\documentclass[letterpaper,11pt]{article}
\usepackage{jheppub} % for details on the use of the package, please
\usepackage{graphicx}
\usepackage{epstopdf}
\usepackage{amsmath, amssymb, float}
\usepackage{longtable}

\usepackage{bm}
\usepackage{bbold}
\usepackage{subcaption}
\usepackage[utf8]{inputenc}
\usepackage[all]{xy}
\usepackage{ytableau}
\ytableausetup{smalltableaux, aligntableaux = center}
\def\fund{{\small{\ydiagram{1}}}}

	\newcommand{\ba}[1]{\begin{align} #1 \end{align} }

	\newcommand{\mc}[1]{\mathcal{ #1} }

	\newcommand{\be}{\begin{eqnarray}}
	\newcommand{\ee}{\end{eqnarray}}
	
	\newcommand{\bn}{\begin{enumerate}}
	\newcommand{\en}{\end{enumerate}}
	\newcommand{\bea}{\begin{eqnarray}}
	\newcommand{\eea}{\end{eqnarray}}

	\def\CB{{\cal B}}

	\def\CE{{\cal E}}

	\def\CI{{\cal I}}

	\def\CL{{\cal L}}
	\def\CM{{\cal M}}
	\def\CN{{\cal N}}
	\def\CO{{\cal O}}

	\def\CR{{\cal R}}
	\def\CS{{\cal S}}
	\def\CT{{\cal T}}

	\def\a{\alpha}

	\def\ch{\chi}
	\def\half{\frac{1}{2}}

	\def\goto{\rightarrow}

	\def\Tr{{\rm Tr}}
	\def\tr{{\rm Tr}}
	
	\def\fp{\mathfrak{p}}
	\def\fq{\mathfrak{q}}
	\def\ft{\mathfrak{t}}
	\def\PE{\textrm{PE}}

\usepackage{pdflscape}

\title{Large Landscape of 4d Superconformal Field Theories from Small Gauge Theories}

\author[a]{Minseok Cho,}
\author[b,c]{Kazunobu Maruyoshi,}
\author[d,e]{Emily Nardoni,}
\author[a]{and Jaewon Song}
\affiliation[a]{Department of Physics, Korea Advanced Institute of Science and Technology\\ 291 Daehak-ro, Yuseong-gu, Daejeon 34141, Republic of Korea}
\affiliation[b]{Faculty of Science and Technology, Seikei University\\ 3-3-1 Kichijoji-Kitamachi, Musashino-shi, Tokyo, 180-8633, Japan}
\affiliation[c]{Nambu Yoichiro Institute of Theoretical and Experimental Physics (NITEP), \\Osaka Metropolitan University}
\affiliation[d]{Kavli Institute for the Physics and Mathematics of the Universe\\ University of Tokyo, Kashiwa, Chiba 277-8583, Japan}
\affiliation[e]{Department of Physics and Astronomy, Vassar College\\ 124 Raymond Avenue, Poughkeepsie, New York 12604, USA}
\emailAdd{cms1308@kaist.ac.kr}
\emailAdd{maruyoshi@st.seikei.ac.jp}
\emailAdd{enardoni@vassar.edu}
\emailAdd{jaewon.song@kaist.ac.kr}

\abstract
{
  We systematically explore the space of renormalization group flows of four-dimensional $\CN=1$ superconformal field theories (SCFTs) triggered by relevant deformations, as well as by coupling to free chiral multiplets with relevant operators. 
    In this way, we classify all possible fixed point SCFTs that can be obtained from certain rank 1 and 2 supersymmetric gauge theories with small amount of matter multiplets, identifying 7,346 inequivalent fixed points which pass a series of non-trivial consistency checks.
  This set of fixed points exhibits  interesting statistical behaviors, including a narrow distribution of central charges $(a, c)$, a correlation between the number of relevant operators and the ratio $a/c$, and trends in the lightest operator dimension versus $a/c$. The ratio $a/c$ of this set is distributed between $0.7228$ and $1.2100$, where the upper bound is larger than that of previously known interacting SCFTs. 
  Moreover, we find a plethora of highly non-perturbative phenomena, such as (super)symmetry enhancements, operator decoupling, non-commuting renormalization group flows, and dualities. 
  We especially identify amongst these fixed points a new SCFT that has smaller central charges $(a, c) = (\frac{633}{2000},\frac{683}{2000})$ than that of the deformed minimal Argyres-Douglas theory, as well as novel Lagrangian duals for certain $\CN=1$ deformed Argyres-Douglas theories.
  We provide a website \url{https://qft.kaist.ac.kr/landscape} to navigate through our set of fixed points. 
}

\begin{document} 
\maketitle

%%%%%%%%%%%%%%%%%%%%%%%%%%%%%%%%%%%%%%%%%%
%%%%%%%%%%%%%%%%%%%%%%%%%%%%%%%%%%%%%%%%%%
\section{Introduction} \label{sec:intro}

The low-energy dynamics of quantum field theories (QFTs) are rich and varied. One of the most fundamental questions in the study of gauge theory is to understand precisely how the asymptotically free Yang-Mills theory develops a mass gap above the vacuum. By contrast, it is also possible that a QFT has no energy gap at low energies, and instead exhibits emergent scale invariance. In this case, the low-energy dynamics is described by a conformal field theory (CFT). Conformal field theories describe fixed points of renormalization group (RG) flows.
It is possible (and common) that distinct quantum field theories can flow under the renormalization group to the same fixed point; if so they are referred to as residing in the same universality class, and they share the same infrared (IR) dynamics. Because there can be multiple QFTs in the same universality class, the set of possible IR phases is smaller than that of all the possible QFTs. 

Classifying the set of all possible conformal field theories would tell us all possible gapless phases of QFTs. This is an extremely challenging problem, where progress is rather restrictive and generally relies on making additional assumptions. 
There are generally two complementary approaches to the classification problem. 
One is to start with the most general constraints imposed by conformal symmetry and unitarity, and then try to find  universal conditions for a CFT to satisfy. This conformal bootstrap program has led to fruitful results, which include classifying the minimal models in two dimensions \cite{Belavin:1984vu}, and providing very accurate determinations of the critical exponents of the 3d Ising model \cite{El-Showk:2012cjh}. Generally, bootstrap methods are very efficient for deriving general constraints---see \cite{Poland:2018epd, Bissi:2022mrs} and the references therein for a summary of recent developments. However, a full classification of CFTs from the most general constraints seems to be intractable. In fact, even for the 2d case---which has infinite-dimensional conformal symmetry constraining the problem---the full classification of CFTs is still a challenging open question.  

Alternatively, one can directly construct a comprehensive set of CFTs with a high degree of supersymmetry. Especially in six dimensions, the classification of $\CN=(1, 0)$ SCFTs \cite{Heckman:2013pva, Heckman:2015bfa} was conjectured based on string theory compactifications. 
However, applying a similar method in four dimensions is extremely challenging. 
Assuming $\CN=2$ supersymmetry in 4d, there have been several approaches for a systematic construction of SCFTs. One fruitful approach is via compactification of 6d $\CN=(2, 0)$ theory on a Riemann surface. 4d $\CN=2$ SCFTs that can be constructed in this way are called class $\CS$ theories \cite{Gaiotto:2009we, Gaiotto:2009hg}. The class $\CS$ construction unveiled a vast landscape of SCFTs that exhibit a plethora of non-perturbative phenomena---such as dualities---and led to the discovery of many new non-Lagrangian theories which have no obvious weak-coupling limit. Such constructions are readily generalized to $\CN=1$ theories \cite{Bah:2012dg, Gaiotto:2015usa, Hanany:2015pfa, Razamat:2016dpl}, whose landscape is even richer. While these geometric constructions yield a very large set of SCFTs, it is unclear whether one can find a string-theoretic realization of all possible 4d SCFTs. Indeed, one can construct SCFTs that are not readily realized via the top-down constructions described above; many of the conformal Lagrangian gauge theories have no known string theory realization. Nevertheless, such constructions provide a large set of SCFTs equipped with various additional tools that allow a quantitative understanding of non-perturbative properties. 

One significant line of progress in the classification of 4d $\CN=2$ SCFTs  is  through the construction/classification of the scale-invariant Coulomb branch geometries. In this way, rank-1 SCFTs ({\it i.e.}~having 1-dimensional Coulomb branch) have been completely classified \cite{Argyres:2015ffa, Argyres:2015gha, Argyres:2016xmc, Argyres:2016xua}. 
On the other hand, all possible Lagrangian $\CN=2$ SCFTs are classified in \cite{Bhardwaj:2013qia}. This latter result is possible because $\CN=2$ supersymmetry completely fixes the Lagrangian once the gauge group and matter hypermultiplets are fixed. For such a gauge theory to be conformal, it is enough to find the choice of matters that gives vanishing one-loop beta functions, thus converting the classification problem to a completely group-theoretical question. 

Our goal in this paper is to make modest progress in the classification of 4d $\CN=1$ SCFTs, via systematic  construction. Unlike the case of $\CN=2$ SCFTs, there has been little progress on classification of $\CN=1$ SCFTs, except for \cite{Razamat:2020pra} wherein all possible $\CN=1$ superconformal gauge theories with vanishing beta functions are classified. (This can be thought of as a generalization of \cite{Bhardwaj:2013qia}, except that one has to ensure that there is a non-trivial conformal manifold \cite{Leigh:1995ep, Green:2010da}.) 
In the current paper, we continue the classification program of \emph{Lagrangian} superconformal theories that was initiated in \cite{Maruyoshi:2018nod}. We first start with a supersymmetric gauge theory that flows to an interacting conformal fixed point, which we call a ``seed theory". All possible $\CN=1$ gauge theories with gauge group $SU/SO/Sp$ that flow to non-trivial fixed points are classified in \cite{Agarwal:2020pol, LargeNClassification}, and can serve as the seed theory.  
From this seed theory, we perform all possible supersymmetry preserving deformations via relevant operators. In addition, we also consider all possible relevant ``flip" deformation, which is to include an additional chiral multiplet and particular superpotential coupling. This procedure generates all possible gauge theories that flow to SCFTs for the fixed gauge group and  matter content. As we will see, the set of such fixed points can be very large. 

In this paper, we focus on simple rank-1 and -2 gauge theories with small amount of matter chiral multiplets. We are able to classify all possible fixed points that can be realized by $SU(2), SU(3), Sp(2), SO(5),$ and $G_2$ gauge theories with ``small" amount of matter fields, in doing so finding more than 7,000 fixed points. With this large set of SCFTs---that we refer to as the landscape throughout this work---we perform various statistical analyses. 

\begin{figure}[t] 
    \centering
    \includegraphics[width=.9\linewidth]{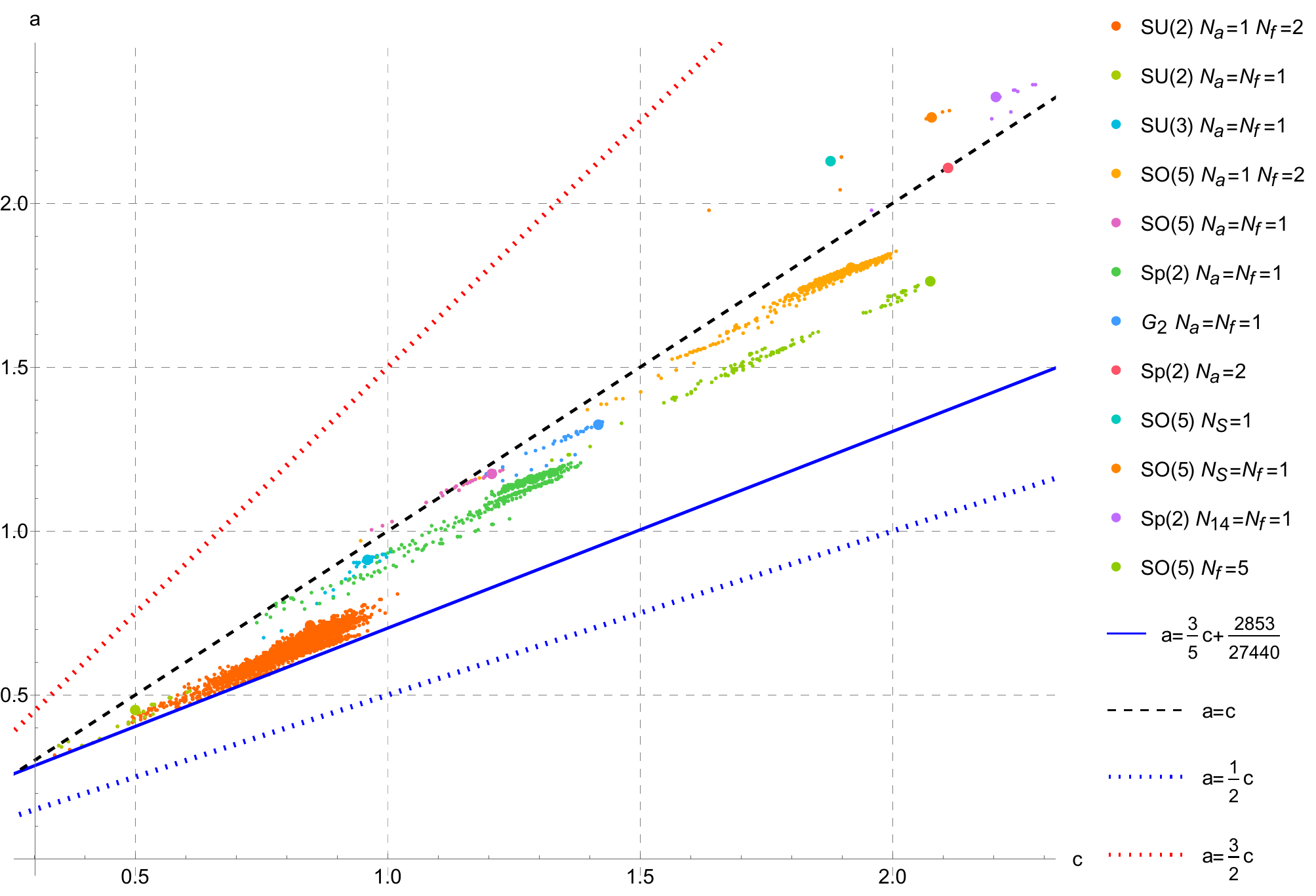}
    \caption{The plot of $a$ versus $c$ for the entire set of fixed points we obtain. Colors label the seed theory and the fixed points that originate from it. The large dots denote the seed theories. The dashed blue/red lines denote unitarity bound, and the black dashed line denotes the $a=c$ slice. We also draw the blue solid line below which no fixed point exists. Other symbols will be explained in Section \ref{subsec:landscape}.}
    \label{fig:acDist1}
  \end{figure}

We first explore the distribution of central charges $a$ and $c$ as well as their ratio $a/c$ in our SCFT landscape---see Figure \ref{fig:acDist1}.
One thing we look for is the minimal 4d $\CN=1$ SCFT, in the following sense. 
For the case of $\CN=2$ SCFT, the minimal value of $c$ for an interacting theory is given as $\frac{11}{30}$ \cite{Liendo:2015ofa}, which is realized by the minimal Argyres-Douglas theory \cite{Argyres:1995jj}. This theory has the value  $a=\frac{43}{120}$, which is conjectured to be the minimal value of $a$ for any interacting $\CN=2$ SCFT. We can ask a similar question for 4d $\CN=1$ SCFT. Unfortunately, there is no rigorous bound known for this case. Analysis from the conformal bootstrap  suggests the bound $c \ge \frac{1}{9} \simeq 0.11$ \cite{Poland:2015mta}, but the minimal value of $c$ that has been realized until now is $\frac{271}{768} \simeq 0.35$, as possessed  by the $\CN=1$ deformed minimal Argyres-Douglas theory $H_0=(A_1, A_2)$ \cite{Xie:2016hny, Buican:2016hnq}. More recently, it was found in \cite{Xie:2021omd} that the $\CN=2$ $(A_1, A_4)$ theory deformed via Coulomb branch operator has slightly lower central charges $(a, c) = (\frac{633}{2000}, \frac{683}{2000}) \simeq (0.3165, 0.3415)$. As we will see, we find that this value is reproduced by RG flow from a Lagrangian $SU(2)$ gauge theory, which further supports the existence of such a fixed point. It will be discussed in detail in section \ref{subsec:deformAD} below.

Unitarity constrains the ratio of central charges of any 4d $\CN=1$ SCFT to be $\frac{1}{2} \le \frac{a}{c} \le \frac{3}{2}$ \cite{Hofman:2008ar}. The lower and upper-limits are saturated by free chiral and vector multiplets respectively. One curious observation is that for all the known interacting $\CN=1$ SCFTs, the observed range is much more restrictive than the unitarity bound, so that lower and upper limits are $\sim$~$0.6$ and $\sim$~$1.2$ respectively. See also \cite{Benini:2015bwz, Bobev:2017uzs} for arguments for the lower bound of $3/5$ from RG flows across dimenions.  
The largest value of $a/c$ of an interacting SCFT that the authors were aware of was that of the Intriligator-Seiberg-Shenker (ISS) model \cite{Intriligator:1995ax}, with $a/c \simeq 1.16$. In this work, we  find an $\CN=1$ SCFT with higher ratio $a/c \simeq 1.21$, breaking this record. We discuss this in detail in section \ref{subsec:acRecord}.

We are also motivated by the AdS/CFT correspondence \cite{Maldacena:1997re, Gubser:1998bc, Witten:1998qj}. For  holographic theories (meaning theories that are dual to weakly-coupled Einstein (super)gravity), the central charges $a$ and $c$ are equal (at large $N$) \cite{Henningson:1998gx}. The difference between the central charges $c-a$ controls the $R_{\mu\nu\rho\sigma}R^{\mu\nu\rho\sigma}$ higher-derivative corrections in the supergravity effective action, so that the famous entropy-viscosity ratio bound $\eta/s \ge 1/4\pi$ \cite{Kovtun:2004de} gets corrected \cite{Buchel:2008vz}. This combination of central charges appears also in the supersymmetric Cardy formula \cite{DiPietro:2014bca}\footnote{~The degeneracies captured by the superconformal index are actually much larger than the one accounted for by the formula of \cite{DiPietro:2014bca}. The high-energy asymptotics of the index turn out to be controlled by $3c-2a$ \cite{Kim:2019yrz, Cabo-Bizet:2019osg}.} and in the universal part of the entanglement entropy \cite{Perlmutter:2015vma}. 
Even though we focus on SCFTs with small gauge group rank, since AdS/CFT should work beyond the large-$N$ limit or the limit of weakly-coupled gravity in the bulk (sometimes referred to as the ``strongest version" of the correspondence), it may shed light on non-perturbative, strongly-coupled, non-Einstein-like gravity that is fully quantum. 
In particular, our analysis may be useful in finding/refining an AdS/CFT version of the Swampland conjectures \cite{Vafa:2005ui}. 

  We also analyze possible correlations between the central charge ratio against the dimension of the lightest operator in the SCFT. It is conjectured that a large $N$ CFT with a parametrically large gap in the higher-spin operator  is dual to  weakly-coupled gravity in the bulk \cite{Heemskerk:2009pn}. The higher-spin gap is also constrained by the difference between the central charges $c-a$ \cite{Camanho:2014apa} for  Einstein gravity in the bulk. Here, we look for any correlation between $a/c$ and the dimension of the lightest \emph{scalar} operator. We do find a curious correlation between the two, which we discuss in section \ref{sec:statistics}.

\begin{figure}[t]
    \centering
    \includegraphics[width=0.95\linewidth]{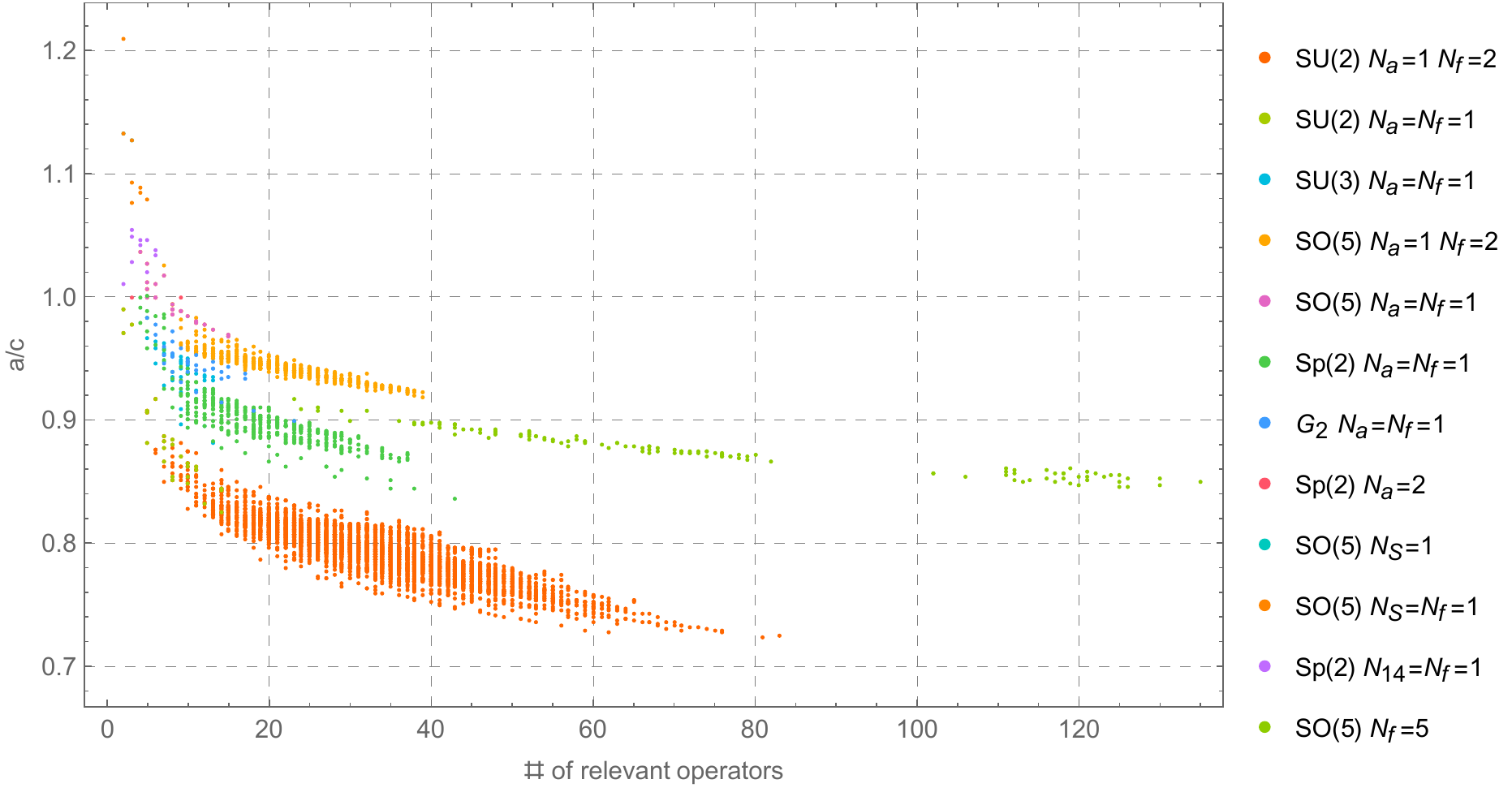}
    \caption{The plot of $a/c$ versus the number of relevant operators for the fixed points obtained in our landscape of  SCFTs. 
    Colors label the `seed theory' that the fixed point originates from, and the large dots denote the seed theories. 
    Other symbols will be explained in section \ref{subsec:landscape}.}
    \label{fig:acVrelops}
\end{figure}

  We also find an interesting correlation between $a/c$ and the number of relevant operators. There is a sense in which the number of relevant operators measures the degrees of freedom of the CFT---this notion has been proposed and explored in \cite{Gukov:2015qea, Gukov:2016tnp}. Here we find a tantalizing connection between $a/c$ versus the number of relevant operators, as demonstrated in Figure \ref{fig:acVrelops}.

  Another motivation for our work comes from the discovery of  UV $\CN=1$ Lagrangian gauge theories that flow to $\CN=2$ ``non-Lagrangian" superconformal fixed points \cite{Gadde:2015xta, Maruyoshi:2016tqk, Maruyoshi:2016aim, Agarwal:2016pjo, Agarwal:2017roi, Benvenuti:2017bpg, Agarwal:2018ejn, Zafrir:2020epd}. We search amongst our fixed points for  those that exhibit similar supersymmetry enhancement in the IR. In doing so we reproduce the previous results of \cite{Maruyoshi:2016tqk, Maruyoshi:2016aim, Agarwal:2016pjo}, but do not identify any additional $\CN=2$ non-Lagrangian SCFTs realized as the end point of an $\CN=1$ RG flow. However, we do find one case that passes all the consistency requirements that can be tested via the superconformal index, and so is seemingly a new rank-1 SCFT that is not present in the classification via  Coulomb branch geometry \cite{Argyres:2015ffa, Argyres:2015gha, Argyres:2016xua, Argyres:2016xmc}. We discuss this example in detail in section \ref{subsec:newN2}.

  We also identify fixed points with identical central charges $a = c$, as motivated by the series of works \cite{Kang:2021lic, Kang:2021ccs, Kang:2022vab, Kang:2022zsl, Kang:2023dsa} that construct $a=c$ SCFTs via gauging of Argyres-Douglas type theories. In view of holography, this could provide an interesting set of cases to look for a family of large-$N$ theories. 

The organization of this paper is as follows. In section \ref{sec:method}, we discuss our method of constructing the large set of fixed points. We emphasize that our method critically relies on the assumption that we correctly identify all the symmetries of the theory. In section \ref{sec:statistics}, we present the results for various gauge groups and matter contents, and perform statistical analysis on the landscape of SCFTs that we find. We look in particular for universal patterns that persists amongst the large set of fixed points. In section \ref{sec:details}, we discuss several key examples and features that we discover in detail. These involve new IR dualities, non-commuting RG flows, possible candidates of supersymmetry enhancement, and so on. 
In section \ref{sec:discussion}, we summarize the results  and conclude with a discussion of future directions. Appendix \ref{app:ops} reviews $\CN=2$ and $\CN=1$ superconformal multiplets and describes our deformation process for a generic non-Lagrangian $\CN=2$ SCFT. In Appendix \ref{app:database}, we elaborate on technical aspects of constructing our database of fixed points.

%%%%%%%%%%%%%%%%%%%%%%%%%%%%
\section{Methods and Strategy} \label{sec:method}
  In this section we introduce the methods and strategy to search for fixed point SCFTs. We will classify all possible interacting SCFTs that can be realized as the end point of a renormalization group flow from 4d $\CN=1$ supersymmetric gauge theories %with rank 1 and 2 gauge groups and a small number of chiral multiplets, 
  via relevant superpotential deformations. While our method does not rely on the existence of a Lagrangian description, we will focus herein on flows amongst Lagrangian gauge theories.

\subsection{Deforming $\mathcal{N}=1$ SCFT} \label{subsec:deform}

  Relevant supersymmetry-preserving superpotential deformations of 4d $\CN=1$ SCFTs arise from chiral multiplets with scalar chiral primaries $\CO$, whose dimension $\Delta$ is related to their $U(1)_R$ charge $R$ as, 
  \ba{
  \delta W = \CO\,,\qquad \Delta(\CO)  = \frac{3}{2}R(\CO) < 3\,.
  }
  This fact, as well as a summary of pertinent $\CN=1$ multiplets, is reviewed in Appendix~\ref{app:ops}.

  In order to obtain all possible flows amongst fixed points of 4d $\CN=1$ gauge theories, we employ a refinement of the deformation procedure introduced in \cite{Maruyoshi:2018nod}. This procedure can be applied to any $\mathcal{N}=1$ SCFT in four dimensions, as long as the local operator spectrum is known. The steps may be summarized as follows. 

  \begin{enumerate}
  \item Start with an SCFT $\CT$, which we will refer to as the seed theory. Identify all the relevant chiral operators of $\CT$ ({\it i.e.}~those having dimension less than $3$, or equivalently $U(1)_R$ R-charge less than $2$), and let us call the set of them as $\CR_{\CT} $. Let us also denote by $\CS_\CT \subset \CR_\CT$  the subset of so-called {\it super-relevant} operators with R-charge less than $4/3$. 
  
  \item Consider the fixed points $\{ \CT_\CO \}$ obtained by the deformation $\delta W = \CO$ for all $\CO \in \CR_\CT$, or the fixed points $\{ {\CT}_{\overline{\CO}} \}$ given by adding an additional gauge singlet field $M$ and the superpotential coupling $\delta W=M \CO$ for all $\CO \in \CS_\CT$.\footnote{~This type of deformation is commonly referred to as ``flipping" the operator $\CO$, or as a flip deformation, and crucially featured in (for example) \cite{Barnes:2004jj, Maruyoshi:2016tqk, Benvenuti:2017lle, Maruyoshi:2018nod}.} 
  
  \item For each of the new fixed point theories obtained in the previous step, check if it has an operator $\CO_d$ that decouples. (We will comment more below on which classes of operator decouplings we can expect to identify.) If so, remove the operator by introducing a flip field $X$ and a superpotential coupling $\delta W = X \CO_d$.
  
  \item For each new fixed point, go back to Step 1 and repeat the procedure. Terminate if there is no new (non-trivial) fixed point. 
  
  \end{enumerate}
  In this way, we find all possible superconformal fixed points that originate from the seed theory $\CT$ and its superpotential deformations. Let us make a few remarks. 
  
  At each step, we employ $a$-maximization \cite{Intriligator:2003jj} and its modification \cite{Kutasov:2003iy} to compute the superconformal R-charges. Upon relevant deformation (including the flip deformations), the superconformal R-charge in the IR is modified compared to the one before deformation. The identification of the superconformal $U(1)_R$ symmetry is essential in our computation, since it allows us to identify the spectrum of relevant (and marginal) operators and to compute the central charges. In order for the $a$-maximization procedure to yield a correct result, one must identify \emph{all} possible $U(1)$ symmetries that can possibly be mixed with $R$. There exist many subtleties in making this identification, since there can be (and often are) accidental symmetries in the IR. In section \ref{subsec:amax} we will outline the procedure to account for such accidental symmetries when possible. However, we emphasize that this is not always possible, and that incomplete identification of accidental  symmetries can sometimes remove a putative (interacting) superconformal fixed point, or miss possible fixed points. 
  
  We perform the $a$-maximization procedure at each step of the RG flow to ensure that we do not miss any relevant operators along the flow. It is not sufficient to simply enumerate all possible relevant operators of the seed CFT $\CT$, as it is possible that some of the  operators that are irrelevant in the seed CFT can become relevant after flow. Such ``dangerously irrelevant'' operators  feature frequently in our analysis.\footnote{~These dangerously irrelevant operators frequently appear in the studies of supersymmetric theories and their dualities, see for instance \cite{Kutasov:1995np, Kutasov:1995ss, Intriligator:1995ax, Agarwal:2014rua, Agarwal:2015vla, Maruyoshi:2016tqk, Maruyoshi:2016aim, Agarwal:2016pjo}. } 
  
  At each putative fixed point we perform various unitarity checks. Firstly we check if the central charges are positive and also satisfy the Hofman-Maldacena bounds $\frac{1}{2} < \frac{a}{c} < \frac{3}{2}$ \cite{Hofman:2008ar}. As we have mentioned, we also check if all the operators in the chiral ring satisfy the unitarity bounds $\Delta > 1$ (or $R>\frac{2}{3}$). Especially, we compute the superconformal index, which can be used to test whether a given theory is a viable unitary SCFT. As we will review in section \ref{subsec:index},  the superconformal index provides a much more stringent test than the ones based on central charges or the chiral ring.

\subsection{$a$-maximization and Central Charges} \label{subsec:amax}
  The central charges $a$ and $c$ of an $\CN=1$ SCFT are related to the 't Hooft anomalies of the superconformal $U(1)_R$ symmetry as \cite{Anselmi:1997am},
    \ba{
    a = \frac{3}{32} (3\tr R^3 - \tr R)\,,\qquad c = \frac{1}{32} (9\tr R^3-5\tr R)\,.\label{eq:ac}
    }
  Recall that the dimension of a scalar chiral primary operator $\CO$ satisfies the unitarity constraint $\Delta(\mc{O}) = \frac{3}{2} R(\mc{O}) \geq 1$, where a free operator saturates the bound.

  For $\CO$ to be a relevant deformation at the $\CT$ fixed point, it should have $R(\CO) < 2$.\footnote{~The strict equality $R(\CO)=2$ is never relevant; recall that a marginal deformation that is neutral under the flavor symmetries is exactly marginal, while a marginal deformation that is not neutral is marginally irrelevant \cite{Green:2010da}.} Upon a relevant deformation in Step 2 in subsection \ref{subsec:deform} by $\delta W = \CO$, the R-symmetry of $\CT$ can be mixed with global $U(1)$ symmetries. Suppose we know the central charges $a$ and $c$, and thus the $U(1)_R$ anomalies at the $\CT$ fixed point. Also suppose $\CT$ has global Abelian symmetries $U(1)_i$ ($i=1,\ldots, \ell$), and we know their anomalies as well as their mixed anomalies with $U(1)_R$. Then in general, the operator $\CO$ is charged under a group $\prod_{\bar{i} \in \alpha} U(1)_{\bar{i}}$, where $\alpha$ is a subset of $\{ 1,\ldots, \ell \}$. 
  Upon deformation $\delta W = \CO$, global symmetries and the R-symmetry get broken or shifted, and one must perform $a$-maximization to determine the combination which is identified with the superconformal R-symmetry in the IR \cite{Intriligator:2003jj}. In particular, the R-symmetry that is preserved in the IR will be a linear combination,
  	\ba{
	R_{{\rm IR}}= R + \sum_{\bar{i} \in \alpha} \epsilon_{\bar{i}} J_{\bar{i}}\,, \label{eq:rir}
	}
  where $R$ and $J_i$ are the generators of  $U(1)_R$ and $U(1)_i$ respectively.  The parameters $\epsilon_i$ are determined by plugging $R_{{\rm IR}}$ into the formula for $a$ in \eqref{eq:ac}, and maximizing with respect to the $\epsilon_i$. 
  
  A special case for which no maximization of $a$ is required occurs when $\CO$ is charged under only one global $U(1)$ symmetry, say $\alpha=1$. Let $R(\CO)$ and $J_1(\CO)$ denote the R-charge and $U(1)$ flavor charge of $\CO$ at the $\CT$ fixed point respectively. By assumption, $R(\CO)$ is less than $2$. If the theory flows to an IR fixed point, it has $R_{\rm IR}(\CO)=2$. This uniquely fixes $\epsilon$, such that the R-symmetry preserved in the IR corresponds to,
  	\ba{
	\delta W = \CO:\quad R_{{\rm IR}} = R+ \epsilon J_1\,,\qquad \epsilon = \frac{2-R(\CO)}{J_1(\CO)}\,.
	\label{RIRJ}
	}
  Note that when $\{ \alpha \} = \emptyset$, we cannot access the IR fixed point by this method since we do not have a candidate R-symmetry available. 
  Throughout our paper, we will only consider the case when the superconformal R-symmetry is accessible via a symmetry that acts on the fundamental fields. Even though it is possible for there to be an emergent superconformal R-symmetry in the IR, we will not consider such cases.
  
  Let us remark that it is certainly possible to have an emergent R-symmetry in the IR that is not visible from the UV description. For example, $SU(N)$ SQCD with $N_f < 2N$ has a relevant quartic operator $\delta W = Q^2 \tilde{Q}^2$. In this case, the quartic superpotential breaks the R-symmetry and there is no $U(1)$ symmetry available. However, using Seiberg duality \cite{Seiberg:1994pq}, this theory gets mapped to $SU(N_f-N)$ SQCD with $N_f$ flavors and $N_f^2$ gauge-singlets $M$ coupled via $W_{\text{mag}} = M q\tilde{q}$, that are dual to the meson $Q \tilde{Q}$. Therefore, the quartic superpotential $\delta W = Q^2 \tilde{Q}^2$ gets mapped to the mass term for the dual mesons, $\delta W_{\text{mag}} = M^2$, which upon RG flow gets integrated out. The end point of the flow becomes simply identical to the fixed point of  $SU(N_f-N)$ SQCD with $N_f$ flavors but without any gauge-singlets \cite{Strassler:2005qs}. In our setup, we (unfortunately) do not have a (useful) dual description to argue for such a scenario. Therefore, we emphasize that our analysis heavily depends upon identifying all possible $U(1)$ symmetries that can be mixed with the R-symmetry.
  
  For the flip-deformation of the form $\delta W = M \CO$, the case where $\CO$ is neutral under all $U(1)$'s is straightforward. Let $J_M$ denote the generator of the $U(1)_M$ under which only $M$ is charged. In this case, $R(M)=\frac{2}{3}$, since $M$ is a free field at the fixed point $\CT$, and  we take $J_M(M)=1$. By assumption, $R(\CO)<\frac{4}{3}$. In the IR, the superpotential enforces $R_{{\rm IR}}(\CO) + R_{{\rm IR}}(M)=2$, uniquely fixing the mixing parameter $\epsilon$ as
    \ba{
	\delta W = M\CO:\quad R_{{\rm IR}}= R + \epsilon J_M\,,\qquad \epsilon = \frac{4}{3} - R(\CO)\,.
	}
  Thus the central charges of the fixed point, $a_{{\rm IR}}$ and $c_{{\rm IR}}$, are simply given by 
    \bea
    a_{{\rm IR}}
    &=&    a + \frac{3}{32} (1-R(\CO))(3R(\CO)^2-6R(\CO)+2)\,, \nonumber \\
    c_{{\rm IR}}
    &=&    c + \frac{1}{32} (1-R(\CO))(9R(\CO)^2-18R(\CO)+4)\,.
    \label{acflip}
    \eea
  When $\CO$ is charged under additional an $U(1)_i$, however, it can also contribute to the superconformal R-symmetry, and so we would need to perform $a$-maximization to obtain the mixing parameter and the IR R-charge. 

  The simple form of the central charges \eqref{acflip} gives a characteristic pattern of the fixed points obtained by the flip flow in $a$-$c$ plane. Namely, the slope of the line between the original fixed point and the IR one in an $a$-$c$ plane is
    \bea
    s =\frac{9R(\CO)^2-18R(\CO)+4}{9R(\CO)^2-18R(\CO)+6}\,.
    \eea
  Since $\frac{2}{3} < R(\CO) < \frac{4}{3}$, this slope is in the narrow range $5/3 < s < 2$. We will observe such lines in several concrete examples in the next section. 

  When there is an accidental symmetry in the IR, this must also be accounted for in the $a$-maximization procedure for the result to be valid. An example of this phenomenon is when a scalar operator $\CO_d$ apparently violates the unitarity bound, $R(\CO_d)<2/3$. In this case, the operator actually becomes a free field with an accidental $U(1)$ symmetry that acts on the IR free field operator. Then, $a$ is modified to reflect the decoupling of the free operator \cite{Kutasov:2003iy}, as
  	\ba{
	a_{\text{new}} = a_{\text{old}} + \frac{3}{32} \left ( \frac{2}{9} - \left( 3  (R(\CO_d) - 1)^3- (R(\CO_d)-1) \right) \right)\,.\label{eq:anew}
	} 
  This leads to Step 3 described in section \ref{subsec:deform}. We account for this type of accidental symmetry and decoupling of operators throughout our analysis. In particular, whenever we find this kind of operator, we add the superpotential term $W = X \CO_d$ to get rid of $\CO_d$ from the chiral ring. Coupling in the dynamical chiral field $X$ (sometimes called a flipping field in the literature) in this way is equivalent to subtracting the contribution of $\CO_d$ from the central charge as in \eqref{eq:anew}. Let us remark that this method of flipping is in a sense enforced from the accidental symmetry of the decoupled operator. On the other hand, we can ``actively" flip an operator $\CO$ with $R(\CO)<4/3$ as in Step 2 even when the operator does not get decoupled. Such an ``active" flip deformation can give rise to many interesting phenomena, such as (super)symmetry enhancement. See for example \cite{Razamat:2017wsk, Maruyoshi:2018nod, Hwang:2020ddr, Hwang:2020wpd}.
  
  It is important to emphasize that the results we present make the crucial assumption that we have correctly identified all of the accidental symmetries present in the IR theory. Generically, however, the $a$-maximization procedure as described above is not the whole story. A typical scenario is that there may be accidental symmetries which are only evident in a dual description of the theory.  As we detail below, sometimes in our analysis the superconformal index suggests an accidental symmetry that we have not accounted for, allowing us to rule out the putative fixed point subject to accounting for this accidental symmetry. In this work, we rely on the consistency of the $a$-maximization procedure and checks of the superconformal index as evidence for the fixed points we enumerate. However to be clear, there are examples (such as SQCD near the top of the conformal window) in which consistency of the $a$-maximization procedure is not enough to guarantee that all accidental symmetries have been accounted for.  Our results are subject to this important caveat.

\subsection{Utilizing the Superconformal Index}
 \label{subsec:index}
  Step 1 of the procedure outlined in section~\ref{subsec:deform} can be performed by means of the superconformal index \cite{Romelsberger:2005eg, Kinney:2005ej}. This is defined by
    \bea \label{eq:Idx}
    \CI(t, y; x) = \Tr (-1)^F t^{3(R+2j_1)}y^{2j_2} x^f\,, 
    \eea
  where $(j_1, j_2)$ are the spins of the Lorentz group $SU(2)_1 \times SU(2)_2$, and  $R$ is the $U(1)_R$ R-charge. When the theory has a global symmetry with Cartan generator $f$, we also include the fugacity $x$ for it. 
  
  The states of an $\CN=1$ SCFT are organized into representations of the superconformal algebra, where only certain short representations make non-vanishing contributions to the index. This structure fruitfully implies that the index can be used to count operators of various types that appear in the spectrum of the theory. 
  To obtain the operator spectrum from the index, it is convenient to consider the following reduced index,
    \bea \label{eq:redIdx}
    \CI_{{\rm red}}
     =     (1 - t^3 y)(1- t^3 y^{-1}) (\CI - 1)\,.
    \eea
 The two  pre-factors in the above equation reduce the contributions from an arbitrary number of derivatives acting on the bottom components of $\CN=1$ multiplets, and the subtraction by 1 removes the identity operator. In general, the expansion of the reduced index $\CI_{\text{red}}$ in terms of $t$ tells us the quantum numbers of the operators in $\CT$, up to the ambiguity of recombination of short multiplets into a long multiplet. In particular:
 \begin{itemize}
 
 \item For the relevant chiral operator with $U(1)_R$ charge $R < 2$  (see \eqref{n1chiral}), there is no ambiguity: the number of such operators can be seen from the coefficient of the $t^{3R} y^0$ term in \eqref{eq:redIdx}, with the flavor charge read off from the flavor fugacity \cite{Beem:2012yn}. 
 
 \item  The marginal operator with $R=2$ contributes to the $t^6$ term, as does the conserved flavor current multiplet but with a minus sign. Therefore the coefficient of the $t^6$ term in \eqref{eq:redIdx} describes the number of the marginal operators minus the number of the conserved currents ({\it i.e.}~the dimension of the flavor symmetry) \cite{Beem:2012yn},
  \begin{align}
  \label{t6}
    \CI_{{\rm red}} \supset \alpha t^6\,,\qquad \alpha = \# (\text{marginal operators}) - \#(\text{conserved currents} )\,.
  \end{align}
 
  \end{itemize}
  Therefore once we know the index of $\CT$, we can perform the Steps 1, 2 and 3 in the procedure. The index of the fixed point theory $\CT_{\CO}$ is then obtained by shifting the flavor $U(1)$ fugacities to take into account the mixing with  $U(1)_R$. The index of $\CT_{\bar{\CO}}$ is obtained by the similar shifting and adding the contribution of $M$. The $U(1)_R$ charges of the relevant operators in $\CT$ are changed accordingly. 
  
  We should note that irrelevant operators with $R>2$ in $\CT$ can become relevant in $\CT_{\CO}$ or $\CT_{\bar{\CO}}$. Such operators are called {\it dangerously irrelevant}. However, there is an ambiguity in counting irrelevant operators in $\CT$---while relevant chiral operators lie in absolutely protected multiplets, chiral multiplets with $R\geq 2$ can recombine to pair up with other multiplets to form long multiplets. Therefore, Step 4 and the repeating Steps can be performed up to this recombination ambiguity. This ambiguity is not present, however,  when the seed theory $\CT$ has a Lagrangian description---for example, when it is connected to a weakly-coupled point by an exactly marginal operator (as in $\CN=2$ $SU(N)$ SQCD with $N_f=2N$ flavors), or when a known Lagrangian gauge theory flows to the $\CT$ fixed point at low energies ({\it {\`a} la} $\CN=1$ $SU(N)$ SQCD in conformal window). This is because all  gauge-invariant operators in $\CT$ and $\CT_{\CO}$ are constructed explicitly from the ``letter" fields, and so we do not miss dangerously irrelevant operators in $\CT_{\CO}$. In the subsequent sections we will mostly restrict to this situation, in which we have at our disposal a Lagrangian gauge theory description.

\paragraph{Testing Unitarity via the Superconformal Index}

The superconformal index provides the following powerful consistency checks on the consistency of a putative SCFT. 

\begin{itemize}
  
\item As explained previously, in a given RG flow  some operators will generically violate the unitarity bound. Such a phenomenon can also be diagnosed from the reduced index: if there is a term $t^{R} \chi_j(y)$ with $R<2+2j$ or a term $(-1)^{2j+1} t^R \chi_j(y)$ with $2+2j \leq R < 6+2j$ where $\chi_j$ is the character of an $SU(2)$ spin-$j$ representation, there is a unitarity violating operator \cite{Evtikhiev:2017heo}. The former constraint with $j=0$ corresponds to the decoupled $\CO_d$ which we discussed around \eqref{eq:anew}. 

\item If the coefficient in front of $(-1)^{2j+1} t^{6 + 2j} \chi_j(y)$ ($j\geq 1$) is positive, the theory is either free or contain some free decoupled part. This is because this term gets positive contribution from spin-$(j+1)$ current, and the higher-spin current necessarily implies there is a free field \cite{Maldacena:2011jn}. The converse is not necessarily true, because there is another multiplet which contributes negatively to this term. 

\item In addition, some of the candidate fixed points have vanishing index.
 This indicates that the theory does not preserve supersymmetry anymore, and therefore should not be included in our set of fixed points.  
  
\end{itemize}
The above conditions should be viewed as sanity checks that the index that has been computed indeed can correspond to an index of a unitary $\CN=1$ SCFT.

\paragraph{Flavor symmetry enhancement and conformal manifolds}

Besides providing unitarity checks, the index can be used to diagnose possible flavor symmetry enhancements at the fixed point. As we noted in \eqref{t6}, the coefficient of the $t^6$ term counts the number of scalar marginal operators minus conserved flavor currents.
If we can count the number of marginal operators in the theory (perhaps directly from the Lagrangian gauge theory description), then we can also identify the number of flavor symmetries. 
Furthermore, due to \eqref{t6} the coefficient of the $t^6$ term can be used to diagnose the existence of a conformal manifold. For instance, if the coefficient $\alpha$ is positive then the minimum dimension of the conformal manifold is  $\alpha$. If all flavor symmetries are broken at a generic point on the conformal manifold, then $\alpha$ precisely captures the actual dimension of the conformal manifold \cite{Leigh:1995ep, Green:2010da}.

\subsection{(Super)Symmetry Enhancements} \label{subsec:SUSYenhancement}
 
  If a fixed point passes the unitarity checks stated above, it preserves at least $\CN=1$ supersymmetry. Moreover, the supersymmetry can be enhanced to $\CN \geq 2$. As explained in \cite{Evtikhiev:2017heo}, the superconformal index provides a series of necessary or sufficient conditions for such supersymmetry enhancement. A sufficient condition is related to the required presence of an extra supercurrent multiplet, while necessary conditions arise from requirements on the ability of $\CN=1$ multiplets to recombine into multiplets of higher supersymmetry. 
  None of the sufficient conditions and the necessary conditions we present for supersymmetry enhancements are both necessary and sufficient. 
 
 The conditions from the index to have  enhanced $\CN=2$ supersymmetry are as follows \cite{Maruyoshi:2016aim,Evtikhiev:2017heo} (see Appendix \ref{app:ops} for a review of the pertinent multiplets in $\CN=1$ and $\CN=2$ SCFTs):
  
  \begin{itemize}
  
  \item {\bf Sufficient condition}: If there are the terms $t^7(y+\frac{1}{y})$ with positive coefficient 1, the fixed point theory has either  $\CN=2$ supersymmetry, or is simply free. There is exactly one sufficient condition for $\CN=1$ to $\CN=2$ enhancement, and it follows from the requirement that the extra supercurrent multiplet is present. Since the extra supersymmetry current multiplet is not absolutely protected, it is not a necessary condition. 
  
  \item {\bf Necessary condition $1$}: Let the fixed point have  global flavor symmetry $F$. Then, the coefficient of the  $t^{4}$ term is greater than or equal to ${\rm dim}(F) - 3 - k$, where $k$ is the number of the left free chiral matter multiplets. This is a condition for there to be enough moment map operators to have supersymmetry enhancement; it follows because the moment map operators in the $\CN=2$ flavor current multiplet contribute as $+t^4$, while left free chiral multiplets are the only multiplets that can contribute as $-t^4$, and at most three $\CN=1$ flavor currents can become $\CN=2$ R-currents lying in the $\CN=2$ stress tensor multiplet. 
  
  \item {\bf Necessary condition $2$}: The $\CN=2$ Coulomb branch operator multiplet ($L \overline{B}_1 [0;0]^{(0; 3R)}_{\frac{3R}{2}}$ in \cite{Cordova:2016emh} or $\CE_{\frac{3R}{2}(0,0)}$ in \cite{Dolan:2002zh}) contributes to the index by $t^{3R}-t^{3R+1} \chi_{\frac{1}{2}}(y)+t^{3R+2}$, where the first term is due to the scalar primary and $R=r/3$ ($r$ being the $\CN=2$ $U(1)_r$ charge) is the $\CN=1$ R-charge \eqref{RJ}, and the other terms are its descendant contribution. 
  For an $\CN=2$ theory, the first term cannot come from any other short multiplets. 
  However, the other two terms can get contributions from other multiplets. Thus, this leads to the condition that if there are terms $n t^{3R}  - m_1 t^{3R+1} \chi_{\frac{1}{2}}(y) + m_2 t^{3R+2} $ for  ${2} \leq 3R < {4}$ where $n, m_1, m_2$ are integer coefficients, then $m_1 \geq n$ and $m_2 \geq n$. 
  
  \item {\bf Necessary condition $3$}: A term $t^{3R}$ with $4 < 3R < 6$ can only get contributions from the following multiplets: $L \overline{B}_1 [0;0]^{(0; 3R)}_{\frac{3R}{2}}$, $L \overline{B}_1 [0;0]^{(0;3R-2)}_{\frac{3R-2}{2}}$, $L \overline{B}_1 [0;0]^{(1;3R-2)}_{\frac{3R}{2}}$.\footnote{~Here, we assume that there is no ``spinning $\CN=2$ chiral multiplets" \cite{Buican:2014qla} so that we ignore any possible contribution from $\CE_{r(0, j)}$-type (or $L \overline{B}_1 [j;0]^{(0;3R)}_{\frac{3R}{2}}$) multiplets with $j>0$.}
  The index for the $L \overline{B}_1 [0;0]^{(0;3R-2)}_{\frac{3R-2}{2}}$ multiplet is given as
  \begin{align}
      t^{3R-2}-t^{3R-1}\chi_{\half}(y)+t^{3R}\,.
  \end{align}
  This contribution includes a pair of terms: $(-1)t^{3R-1}\chi_{\half}(y)$ and $t^{3R}$.
  The index for the $L \overline{B}_1 [0;0]^{(0; 3R)}_{\frac{3R}{2}}$ multiplet, 
  \begin{align}
      t^{3R}-t^{3R+1}\chi_{\half}(y)+t^{3R+2} \ , 
  \end{align}
  and the index for the $L \overline{B}_1 [0;0]^{(1;3R-2)}_{\frac{3R}{2}}$ multiplet,
  \begin{align}
      t^{3R}-t^{3R+1}\chi_{\half}(y)+t^{3R+2}+t^{3R+3}\chi_{\half}(y)-t^{3R+4} \ , 
  \end{align}
  both include a pair of $t^{3R}$ and $(-1)t^{3R+1}\chi_{\half}(y)$ terms.
  
  This yields the following necessary condition: If there is a term $t^{3R}$ with coefficient $n$ and $4 < 3R < 6$, then the coefficient of the term $(-1) t^{3R+1} \chi_{\half}(y)$ plus the coefficients of the terms $ t^{3R-2} \chi_{0}(y)$ is greater than or equal to $n$.

  \end{itemize}
 
One can similarly find conditions from the superconformal index for enhancement of $\CN=1$ to $\CN=3$ supersymmetry \cite{Evtikhiev:2017heo}. In this case, the sufficient condition is that the coefficient in front of $t^7(y+1/y)$ is positive and equal to $2$. We will not write the necessary conditions here since no fixed points in our landscape exhibit $\CN=3$ enhancement, except to comment that one necessary condition is that $a=c$.

%%%%%%%%%%%%%%%%%%%%%%%%%%%%%%%%%%%%%%%%%%%%%%%%%%%%%%%%%%%%%%%%%%%%%%%%%%%%%
\section{Statistics on the Landscape} \label{sec:statistics}

\subsection{Generating a Landscape of SCFTs}\label{subsec:gen}
In this section we describe the result of our analysis. We will pick a set of seed superconformal field theories, that are derived as fixed points of $\CN=1$ supersymmetric gauge theories of rank 1 and 2 simple gauge groups with a small number of chiral multiplets. They can be easily classified as follows. The only allowed gauge groups $G$ (we do not distinguish the global form of the gauge groups in our analysis) are,
\begin{align}
    G = SU(2), \ SU(3), \ Sp(2)=SO(5), \  G_2 \,.
\end{align}
For a given gauge group, the matter content is restricted by gauge anomaly cancellation so that
\begin{align}
    \sum_i \mathcal{A}(\mathbf{R}_i) = 0 \ ,     
\end{align}
where $\mathcal{A}(\mathbf{R}_i)$ denotes the cubic Casimir of the chiral multiplet in the representation $\mathbf{R}_i$ and the sum is over all chiral multiplets in the theory. 
We also impose the condition of asymptotic freedom for the gauge coupling so that the theory flows to an interacting theory in the IR, which implies,
\begin{align}
    b_0 = 3 h^\vee - \sum_i T(\mathbf{R}_i) \ge 0 \ , 
\end{align}
where the sum is again over all charged matter fields and $T(\mathbf{R})$ is the Dynkin index of the representation $\mathbf{R}$. 
In addition, we demand that the R-charges of the charged chiral multiplets  be positive so that there is no Affleck-Dine-Seiberg (ADS) type of runaway superpotential \cite{Affleck:1983mk} generated. When such a superpotential is generated, we do not expect the IR theory to be superconformal.\footnote{~It is possible to have non-positive R-charge but still flow to a non-trivial SCFT. When the F-terms appropriately remove all the would-be unitary violating operators, the gauge-invariant operator spectrum can have proper scaling dimensions \cite{Kutasov:1995np, Kutasov:1995ve, Agarwal:2014rua,Agarwal:2015vla, Gadde:2015xta, Agarwal:2018ejn}.}

For $G = SU(2)$, asymptotic freedom allows us to have spin-1/2, 1, and  3/2 irreducible representations. The number of chiral multiplets are restricted by asymptotic freedom as,
\begin{align}
    n_{1/2} + 4 n_1 + 10 n_{3/2} \le 12 \ , 
\end{align}
where $n_j$ refers to the number of chiral multiplets with spin-$j$ irreducible representations. The Witten $SU(2)$ anomaly restricts $n_{1/2}$ to be even \cite{Witten:1982fp}. We find that there are the following $SU(2)$  gauge theories that flow in the IR to an interacting SCFT:
\begin{align}
    (n_{1/2}, n_1, n_{3/2}) =
    \begin{cases} 
    (8, 0, 0), (10, 0, 0), \\
     (2, 1, 0), (4, 1, 0), (6, 1, 0), (8, 1, 0)^*, \\
     (0, 2, 0), (2, 2, 0), (4, 2, 0)^* ,\\
     (0, 3, 0)^*, \\
     (0, 0, 1) .
    \end{cases}
\end{align}
  Here, the ones with * denote conformal gauge theories with a vanishing 1-loop beta function, but with a non-trivial conformal manifold. The case $(8, 1, 0)$ is $\CN=2$ $SU(2)$ SQCD with 4 fundamental flavors connected via marginal deformation, and $(0, 3, 0)$ is $\CN=4$ $SU(2)$ SYM theory upon appropriate marginal deformation. The one with $(0, 0, 1)$ is the Intriligator-Seiberg-Shenker (ISS) model \cite{Intriligator:1994rx}. The classification of $\CN=1$ SCFTs that can be obtained as a fixed point of classical gauge theories with simple gauge group is reported in \cite{Agarwal:2020pol, Cho:2023koe} and will be further explored in \cite{LargeNClassification}. All possible $\CN=1$ gauge theories with vanishing beta function for the gauge coupling but  non-trivial conformal manifold have been classified in \cite{Razamat:2020pra}. 

  In this analysis, we start with the following set of supersymmetric gauge theories (with vanishing superpotential):
  \begin{align} \label{eq:seedSCFTs}
    \begin{array}{c|c}
    \textrm{gauge group} & \textrm{matter multiplets} \\
    \hline
    SU(2) & \mathbf{adj} \oplus 2 ~ \fund \\
    SU(2) & \mathbf{adj} \oplus 4 ~\fund \\ 
    SU(3) & \mathbf{adj} \oplus  1~(\fund \oplus \overline{\fund}) \\
    SO(5) & \mathbf{10} \oplus  \mathbf{5}  \\
    SO(5) & \mathbf{10} \oplus  2 \cdot \mathbf{5}  \\
    SO(5) & \mathbf{14} \oplus  \mathbf{5}  \\
    SO(5) & \mathbf{14} \oplus 2 \cdot \mathbf{5}  \\
    Sp(2) & \mathbf{14} \oplus 2 \cdot \mathbf{4} \\
    Sp(2) & \mathbf{10} \oplus 2 \cdot \mathbf{4}  \\
    G_2 & \mathbf{adj} \oplus \mathbf{7}
    \end{array}
\end{align}
  These theories flow to non-trivial SCFTs in the IR, sometimes up to decoupled free fields. They play the role of seed SCFTs, from which we construct our SCFT landscape. 
 
  Once we obtain a non-trivial seed SCFT from these gauge theories, we perform the analysis outlined in section \ref{sec:method} to obtain all possible non-trivial fixed points that can be obtained via general deformations. There can be several different descriptions of the same fixed points. For example, one can give a mass  to some of the fields, which get integrated out upon the flow to the IR. Or, we might have a non-trivial duality between two UV descriptions that flow to the same fixed point. Therefore, we need a rationale to distinguish various fixed points. We label inequivalent fixed points in our database as follows: 
  \begin{itemize}
    \item Each fixed point has an integer identifier. 
    \item We construct the database for a given seed SCFT and perform all possible deformations outlined in section \ref{subsec:deform}.
    \item We say that two fixed points are equivalent if they have the same \emph{unrefined} index.\footnote{~Unrefined index refers to the index with the flavor fugacities switched off.} This is because there may be two distinct descriptions of a fixed point with different amount of (non-)manifest global symmetries. 
    \item If there are multiple equivalent fixed points, then we choose the shortest superpotential to represent them, which may exhibit the largest flavor symmetry. Other representations are also included in the database. 
    \item If possible, we identify dual descriptions that can be obtained from another seed SCFT. For a technical reason, this may include mass deformations. (For example, $N_f=1$ theory can be obtained via mass deformation of $N_f=2$ theory.)
  \end{itemize}
  We obtain a vast set (landscape) of fixed points, and it is impossible to present the entire set of results in the paper format. Instead, we present our data on a website: \url{https://qft.kaist.ac.kr/landscape}. We construct a large database that consists of the following information about the (candidate) superconformal field theories:
  \begin{itemize}
    \item gauge group, matter content and superpotential;
    \item central charges $a$, $c$ and their ratio;
    \item R-charges of each chiral multiplet;
    \item superconformal index;
    \item distinct UV (or dual) descriptions when available;
    \item set of relevant and marginal (chiral) operators.
  \end{itemize}
  Typically, since the $a$-maximization procedure eventually reduces to solving a set of quadratic equations, R-charges are generally given by irrational numbers and we give numerical results only. However, sometimes we find rational solutions to the $a$-maximization problem. For such cases, we provide the exact rational R-charges and central charges. 

  We also provide the list of relevant and marginal operators for a given fixed point. However, this result should be taken with care. Quite often, these operators turns out to be either removed from the chiral ring or constrained by non-trivial relations. The true number of independent relevant operators can be counted by reading the coefficients of the superconformal index. For the case of marginal operators, this is tricky since the coefficient of the $t^6$ term in the index gets contributions from both marginal operators (with positive sign) and the conserved flavor currents (with negative sign). If this term is negative, then there must be a flavor symmetry at the fixed point. It can tell us if there is an enhanced symmetry in the IR. If the coefficient of the $t^6$ term is positive, then there must be an exactly marginal deformations of the theory, therefore the fixed point lies on a non-trivial conformal manifold \cite{Leigh:1995ep, Green:2010da, Beem:2012yn}. However, it is often tricky to count the marginal operators and the conserved currents separately due to the quantum correction to the chiral ring.

  For the rest of this section, we perform a statistical analysis of the landscape of SCFTs spawned from each of the seed SCFTs  listed in \eqref{eq:seedSCFTs}. 
  Before we begin, let us explain our conventions for the fixed points. For a given gauge group and set of matter fields, we write down the superpotential to specify the theory. 
  In the superpotential, we suppress all the gauge indices. Also we simply write $q^n$ instead of $q \cdots q$ to save space. Appropriate contraction is always implied. We denote gauge singlet flip fields as $M_i$ where the subscript start with 1 for the first (active) flip deformation, and increases as we introduce more flip fields. Another class of flip fields that we denote as $X_i$ are introduced when we flip an operator that gets decoupled along the RG flow. Finally, the superpotential is written in accordance with the order of relevant deformations we perform. This is important, since ordering of such deformations does matter as we see for example in section \ref{subsec:non-commuting}. We always write the superconformal indices in the reduced form \eqref{eq:redIdx}, so that we remove (trivial) contributions from the descendants and the identity operator.

\subsection{$SU(2)$ $N_f=1$ Adjoint SQCD} \label{sec:su2nf1}
  We start with a seed theory of $SU(2)$ gauge theory with one adjoint $\phi$, one fundamental $q$, and one anti-fundamental $\tilde{q}$ chiral multiplets, without superpotential. This theory flows to an interacting SCFT for which $\tr \phi^2$ gets decoupled. Therefore, we flip this operator by $W=X \tr \phi^2$ to isolate the interacting part. All possible deformations generate the landscape of this theory, which has the following features.
  \begin{itemize}
    \item The maximal number of relevant deformations (modulo flipping free operators): 6.
    \item The number of (inequivalent) fixed points: 26.
    \item The number of pairs of inequivalent fixed points with identical central charges: 0.
    \item The number of fixed points with $a=c$: 0.
    \item The number of fixed points with rational central charges: 12.
  \end{itemize}
This set of theories was previously studied in \cite{Maruyoshi:2018nod}, where 34 distinct fixed points were identified. Here, we obtain only 26 fixed points. 
The difference arises because here we check for consistency at every step during deformation, whereas in the previous work, the consistency check was performed only after all possible deformations were completed. As a result, some seemingly consistent fixed points in \cite{Maruyoshi:2018nod} were obtained by deforming inconsistent fixed points, which our procedure does not capture. However, we do find those extra fixed points in the $SU(2)$ $N_f=2$ landscape that we study in section \ref{sec:su2nf2}. We list all such cases in section \ref{subsec:nf1nf2}.

\begin{figure}[t]
    \centering
     \begin{subfigure}[b]{0.45\textwidth}
         \includegraphics[width=\linewidth]{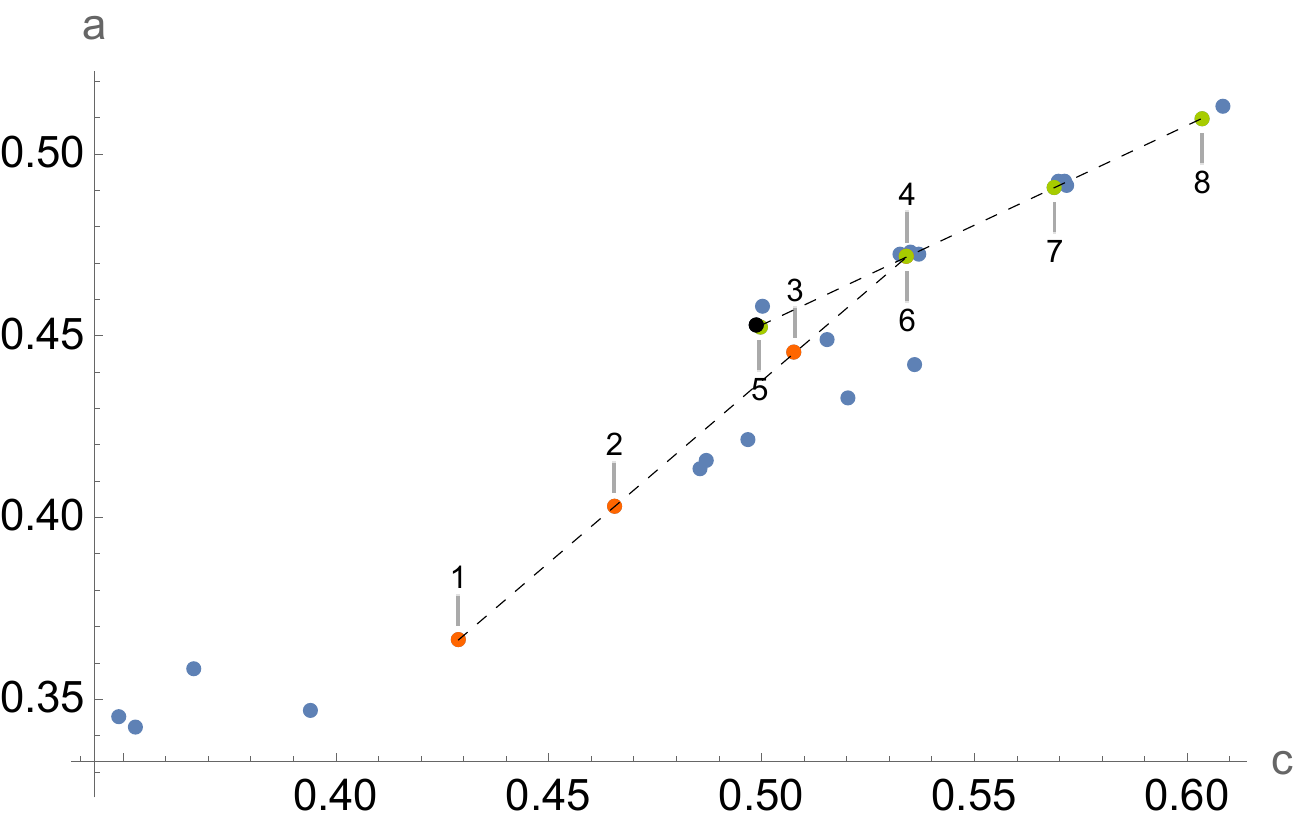}
         \caption{$a$ vs. $c$}
     \end{subfigure}
     \hspace{4mm}
     \begin{subfigure}[b]{0.45\textwidth}
         \includegraphics[width=\linewidth]{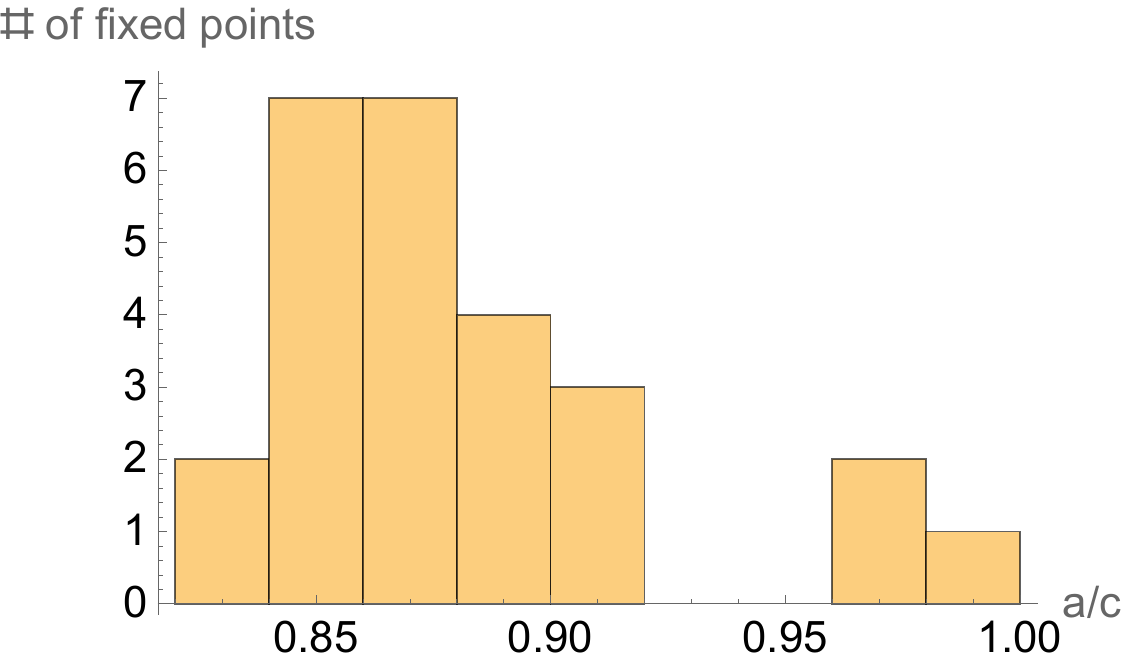}
         \caption{Histogram of $a/c$}
     \end{subfigure}
     \hfill
     \caption{(a) The plot of $a$ versus $c$ for fixed points in the landscape of $SU(2)$ $N_f=1$ adjoint SQCD. The black dot represents the fixed point of $W=X\Tr\phi^2$, which is the seed theory. The colored dots, labeled by an integer $i$, represent fixed points for the $W_i$, and lie on the dashed line. (b) The histogram of $a/c$.}
     \label{fig:SU2adj1nf1ac}
\end{figure}

The distribution of the central charges $a$ and $c$ is plotted in Figure \ref{fig:SU2adj1nf1ac}. The minimal and maximal values of central charges are given by:
\begin{align}
    a_{\min}=\frac{263}{768}\simeq0.3424,\quad a_{\max}\simeq 0.5129,\quad c_{\min}\simeq 0.3488,\quad c_{\max}\simeq 0.6085\,.
\end{align}
The superpotential leading  to the fixed point with the minimal $a$ (\href{https://qft.kaist.ac.kr/landscape/detail.php?id=16}{\#16}) is given by:
\begin{align}
    W=X_1\Tr\phi^2+M_1\phi q q + \phi\tilde{q} \tilde{q} +M_1^2\,.
\end{align}
All fixed points have $a<c$. The minimal and maximal values of $a/c$ are given by:
\begin{align}
    (a/c)_{\min}\simeq 0.8246,\quad (a/c)_{\max}\simeq 0.9895\,.
\end{align}

Interestingly, we find there are two sets fixed points exactly aligned on straight lines, as illustrated in Figure \ref{fig:SU2adj1nf1ac}(a), including:
\begin{itemize}
    \item The fixed points lying on a line $a=c-\frac{1}{16}$:
        \begin{equation}
    \begin{aligned}
        W_4&=X_1\Tr\phi^2+M_1q\tilde{q}+M_1^2+M_2\phi q\tilde{q},\\
        W_3&=X_1\Tr\phi^2+M_1 \phi q^2+M_2\phi q\tilde{q}+M_1^2+q^2\tilde{q}^2\,,\\
        W_2&= X_1\Tr\phi^2+M_1 \phi q^2+M_2\phi q\tilde{q}+M_1M_2+M_1q\tilde{q}, \\
        W_1&=X_1\Tr\phi^2+M_1 \phi q^2+M_2\phi q\tilde{q}+M_1M_2+M_1M_3+q^2\tilde{q}^2\,.
    \end{aligned}
    \end{equation}

    \item The fixed points lying on the line $a=\frac{39}{71}c+\frac{405}{2272}$:
    \begin{equation}
    \begin{aligned}
        W_{5}&=X_1\phi^2+M_1q\tilde{q}+M_1^2, &W_{6}&=W_{5}+M_2\phi q\tilde{q}\,,\\
        W_{7}&= W_{5}+M_2\phi q^2+M_3\phi\tilde{q}^2, &W_{8}&=W_{5}+M_2\phi q^2+M_3\phi\tilde{q}^2+M_4\phi q\tilde{q}\,.\\
    \end{aligned}
    \end{equation}
\end{itemize}
The second line of fixed points arises from a sequence of flipping super-relevant operators from the $W_5$ fixed point. As explained in section \ref{subsec:amax} around equation \eqref{acflip}, when flipping an operator that is not charged under any flavor symmetry, this gives a simple slope on the $a-c$ plane. 

\begin{figure}[t]
    \centering
     \begin{subfigure}[b]{0.45\textwidth}
         \includegraphics[width=\linewidth]{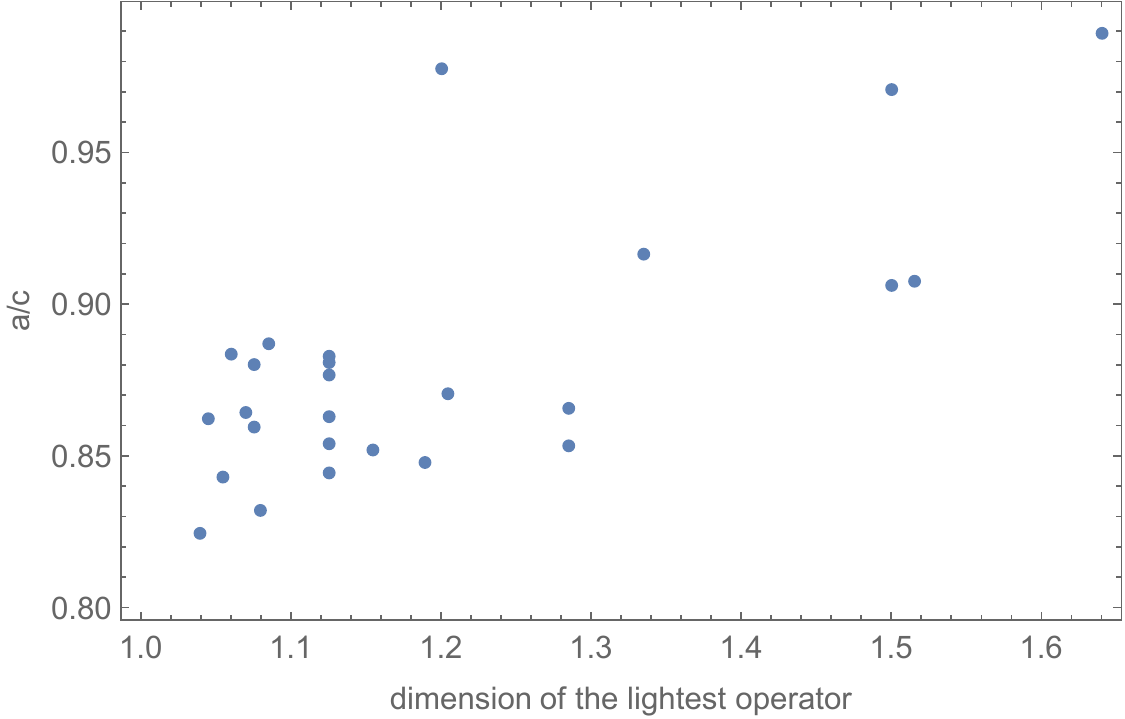}
     \end{subfigure}
     \hspace{4mm}
     \begin{subfigure}[b]{0.45\textwidth}
         \includegraphics[width=\linewidth]{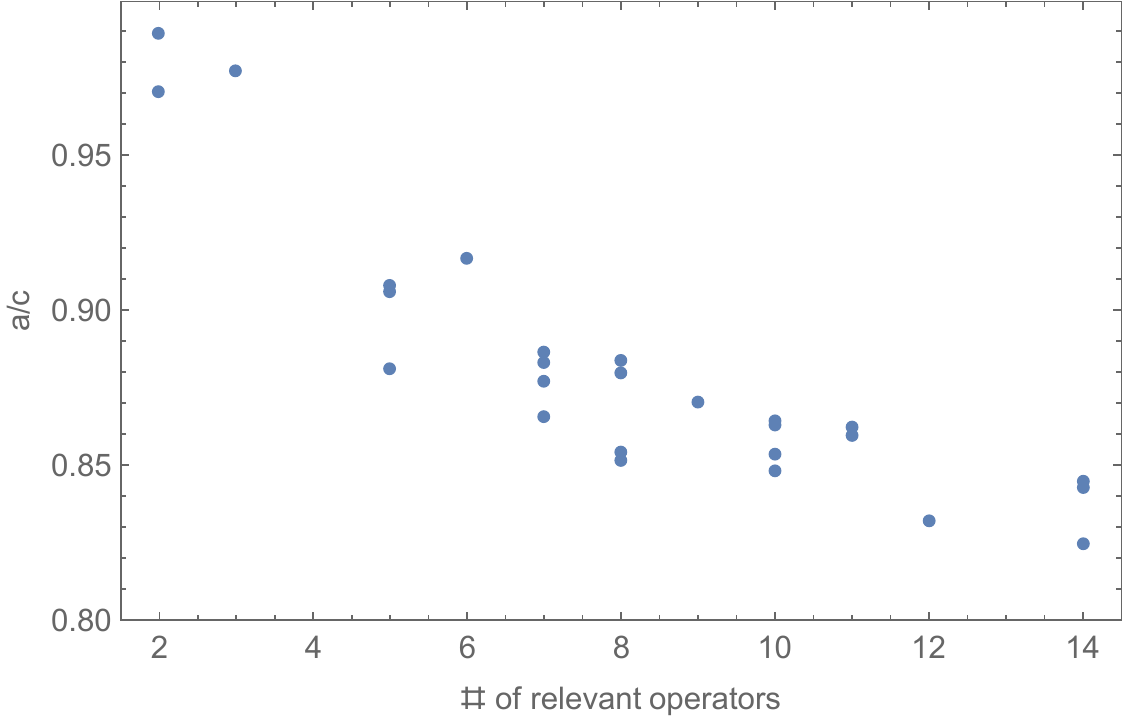}
     \end{subfigure}
     \hfill
     \caption{Left: $a/c$ versus the dimension of the lightest operator. Right: $a/c$ versus the number of relevant operators.}
     \label{fig:SU2adj1nf1ratio}
\end{figure}

We plot $a/c$ versus the dimension of the lightest operator and the number of relevant operators, respectively, in Figure \ref{fig:SU2adj1nf1ratio}. 
We see that the dimension of the lightest chiral operator is positively correlated with the central charge ratio. It suggests that the lightest scalar operator cannot be too light (near the unitarity bound) if it were to be holographic. 
Also, we find that as the number of relevant operators increases, $a/c$ decreases. Loosely speaking, this may mean that holographic theories tend to be more stable against the perturbation, in the sense that they have a smaller number of relevant operators. 

\paragraph{Conformal manifold / Flavor symmetry enhancement}
As described in section \ref{subsec:gen}, we can identify fixed points that have a non-trivial conformal manifold or non-manifest symmetry enhancement, which do not appear in the UV Lagrangian, by examining the coefficient of $t^6$ in the superconformal index. The results are as follows.
\begin{itemize}
    \item The number of fixed points with a positive coefficient in $t^6$ term: 0.
    \item The number of fixed points with no flavor symmetry in the UV but with a negative coefficient in the $t^6$ term: 4.
\end{itemize}
Therefore at least 4 fixed points have enhanced flavor symmetry. We do not find a theory with a non-trivial conformal manifold.

\paragraph{SUSY enhancement}
There are two fixed points that satisfy the sufficient condition for $\CN=2$ supersymmetry enhancement. Moreover, they are the only fixed points that satisfy the necessary conditions for $\CN=2$ supersymmetry as well. They are:
\begin{itemize}
    \item $W=X_1\Tr\phi^2 + M_1\phi q^2+ \phi \tilde{q}^2$. This fixed point corresponds to the $H_0 = (A_1, A_2)$ minimal Argyres-Douglas theory \cite{Argyres:1995jj}, which was found in \cite{Maruyoshi:2016tqk}. The central charges and R-charges of matter fields are given by:
    {\renewcommand\arraystretch{1.4}
        \begin{table}[H]
            \centering
            \begin{tabular}{|c|c|c||c|c|c|c|c|}\hline
                id &$a$ & $c$ & $R_{X_1}$ &  $R_{M_1}$ & $R_{q}$ & $R_{\tilde{q}}$ & $R_\phi$ \\\hline\hline
              \href{https://qft.kaist.ac.kr/landscape/detail.php?id=9}{\#9} & $\frac{43}{120}$& $\frac{11}{30}$ & $\frac{26}{15}$  & $\frac{4}{5}$  & $\frac{8}{15}$ & $\frac{14}{15}$ & $\frac{2}{15}$\\\hline
            \end{tabular}
        \end{table}}
    \item $W=X_1\Tr\phi^2+M_1 q\tilde{q}$. This fixed points corresponds to the $H_1 = (A_1, A_3)$ Argyres-Douglas theory \cite{Argyres:1995xn}, which was found in \cite{Maruyoshi:2016aim}. The central charges and R-charges of matter fields are given by:
    {\renewcommand\arraystretch{1.4}
        \begin{table}[H]
            \centering
            \begin{tabular}{|c|c|c||c|c|c|c|c|}\hline
                id &$a$ & $c$ & $R_{X_1}$ &  $R_{M_1}$ & $R_{q}$ & $R_{\tilde{q}}$ & $R_\phi$ \\\hline\hline
               \href{https://qft.kaist.ac.kr/landscape/detail.php?id=2}{\#2} & $\frac{11}{24}$& $\frac{1}{2}$ & $\frac{14}{9}$  & $\frac{8}{9}$  & $\frac{5}{9}$ & $\frac{5}{9}$ & $\frac{2}{9}$\\\hline
            \end{tabular}
        \end{table}}
\end{itemize}

\subsection{$SU(2)$ $N_f=2$ Adjoint SQCD} \label{sec:su2nf2}
We start with the seed theory given by the IR fixed point of $SU(2)$ supersymmetric gauge theory with one adjoint $\phi$, and four fundamental chiral multiplets $q_{1,2}, \tilde{q}_{1,2}$. The $N_f=2$ landscape turns out to be very large, so we have not yet completely classified the fixed points. There are 20 relevant operators + flip deformations at the seed theory, therefore we in principle have to analyze $2^{20} \sim 10^6$ cases in a crude estimation. Due to  computational complexity and time limitations, we truncated after considering 12 relevant deformations, including the flipping of super-relevant operators. This is enough to include the entire $N_f=1$ landscape we discussed in section \ref{sec:su2nf1}. 

The $N_f=2$ landscape (up to level 12) has the following features.
\begin{itemize}
    \item The maximal number of relevant deformations (modulo flipping free operators): 12+$\a$.
    \item The number of inequivalent fixed points: 5894+$\a$.
    \item The number of pairs of inequivalent fixed points with identical central charges: 26.
    \item The number of fixed points with $a=c$: 0.
    \item The number of inequivalent fixed points with rational central charges: 3428+$\a$.
\end{itemize}
We put $+\a$ to denote that our search of the landscape is not fully accomplished. As we see the number of fixed points is huge, even in this simple setup. 
Two distinct fixed points (having different superconformal indices) may have  identical central charges $a$ and $c$, and we indeed find such cases in this landscape. We provide an example in section \ref{subsec:iden}.

\begin{figure}[t]
    \centering
     \begin{subfigure}[b]{0.45\textwidth}
         \includegraphics[width=\linewidth]{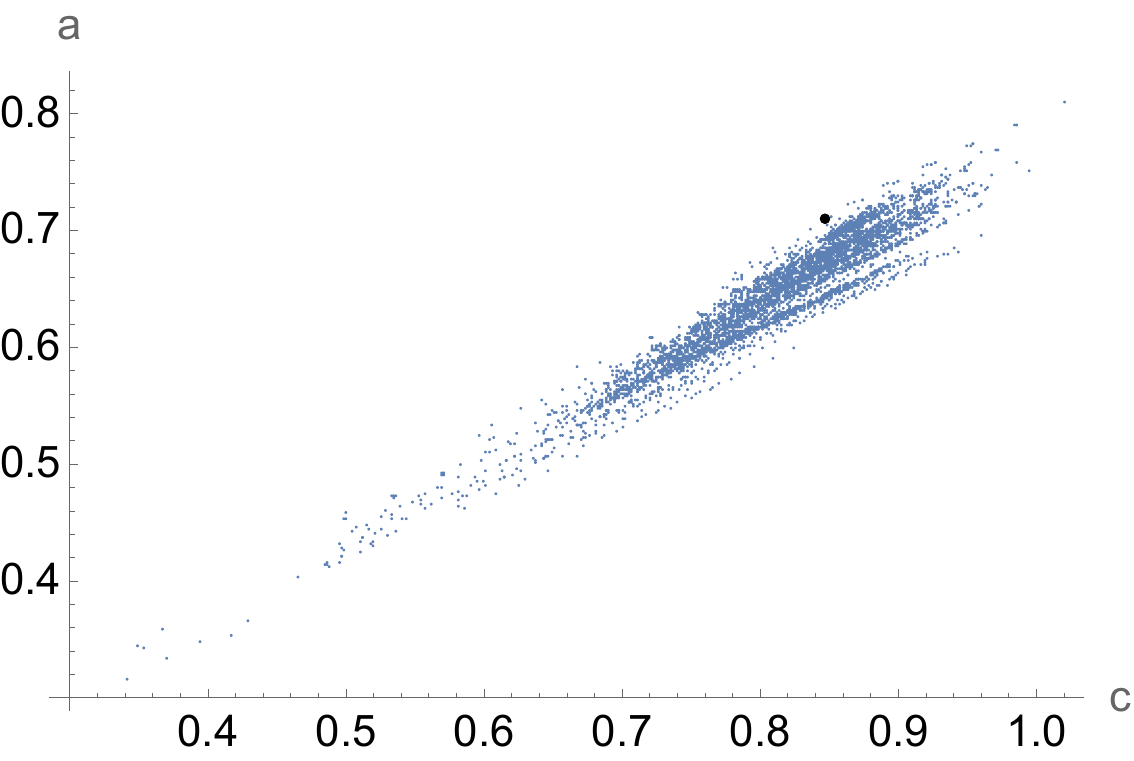}
     \end{subfigure}
     \hspace{4mm}
     \begin{subfigure}[b]{0.45\textwidth}
         \includegraphics[width=\linewidth]{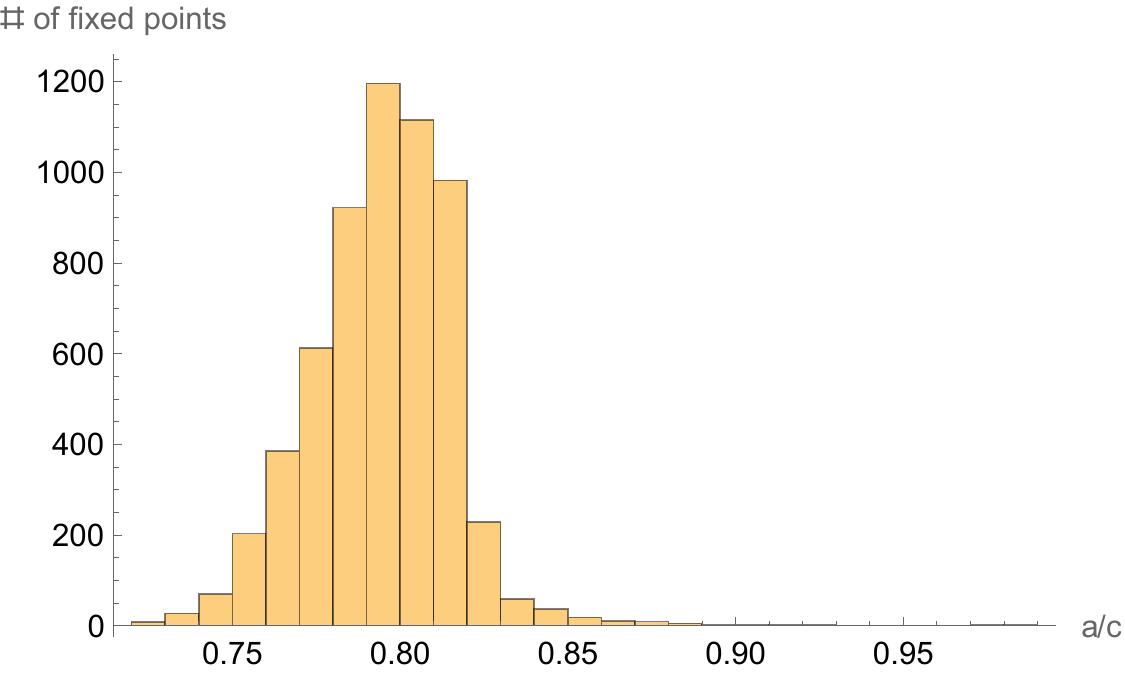}
     \end{subfigure}
     \hfill
     \caption{Left: The plot of $a$ versus $c$ for the fixed points in the landscape of $SU(2)$ $N_f=2$ adjoint SQCD. Right: The histogram of $a/c$}
     \label{fig:SU2adj1nf2ac}
\end{figure}

The distribution of the central charges $a$ and $c$ is plotted in Figure \ref{fig:SU2adj1nf2ac}. 
The minimal and maximal values of central charges are given by,
\begin{align}
    a_{\min}=\frac{633}{2000}\simeq0.3165,\quad a_{\max}\simeq 0.8092,\quad c_{\min}=\frac{683}{2000}\simeq 0.3415,\quad c_{\max}\simeq 1.0197 \, .
\end{align}
We find that all of the fixed points have central charges $a<c$. The minimal and maximal values of $a/c$ are given by
\begin{align}
    (a/c)_{\min}\simeq 0.7228,\quad (a/c)_{\max}\simeq 0.9895\,.
\end{align}
We see that the distribution of central charges is rather narrow and almost linear, but denser near the seed theory and becomes sparser for low central charges. 

\paragraph{A new minimal $\CN=1$ SCFT} 
We find that the superpotential for the fixed point with minimal value of $a$  (\href{https://qft.kaist.ac.kr/landscape/detail.php?id=46107}{\#46107}) is given by
\begin{align}
    W=M_1 q_1 q_2 + M_2 q_1 \tilde{q}_1 + \phi q_1\tilde{q}_2 + M_2\phi q_2^2 + M_1M_2 + \phi \tilde{q}_1^2 + X_1\Tr\phi^2\,.\label{eq:minimal a}
\end{align}
This minimal fixed point turns out to be equivalent to the $\CN=1$ deformation of the $\CN=2$ supersymmetric $(A_1,A_4)$ Argyres-Douglas theory \cite{Eguchi:1996vu, Eguchi:1996ds, Cecotti:2010fi, Xie:2012hs} by a Coulomb branch operator of dimension $\frac{10}{7}$, which was considered in \cite{Xie:2021omd}. 

Since the $(A_1,A_4)$ theory has an $\CN=1$ UV Lagrangian (\href{https://qft.kaist.ac.kr/landscape/detail.php?id=8491}{\#8491}) in terms of $Sp(2)$ $N_f=1$ adjoint SQCD \cite{Maruyoshi:2016aim}, we therefore have a dual description of this fixed point. This indicates a non-trivial dual description for the fixed point of $\CN=1$ deformed AD theory, whose UV descriptions are given by either $SU(2)$ $N_f=2$, or $Sp(2)$ $N_f=1$ adjoint SQCD. We will discuss this example in detail in section \ref{subsec:deformAD}.

\begin{figure}[t]
    \centering
     \begin{subfigure}[b]{0.45\textwidth}
         \includegraphics[width=\linewidth]{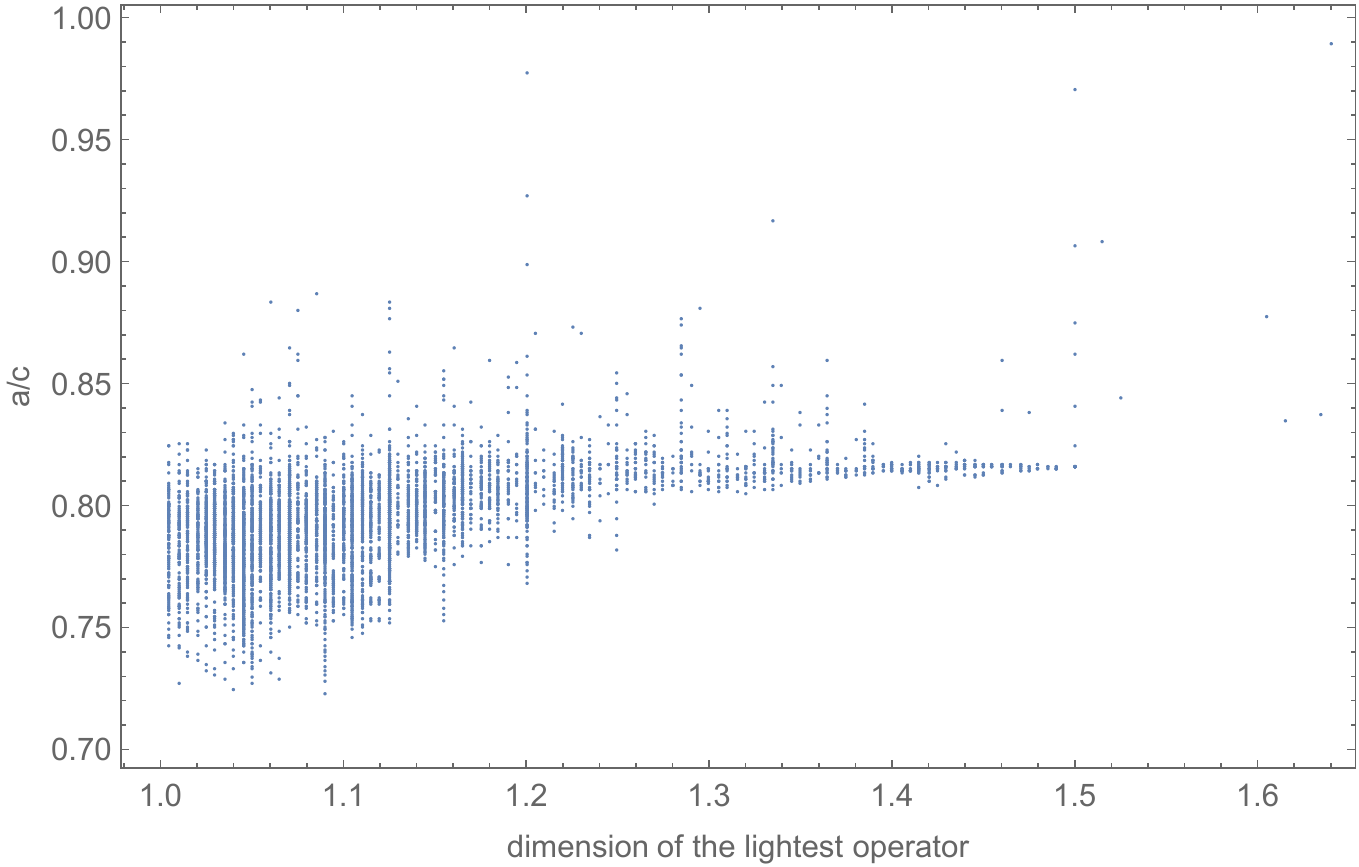}
     \end{subfigure}
     \hspace{4mm}
     \begin{subfigure}[b]{0.45\textwidth}
         \includegraphics[width=\linewidth]{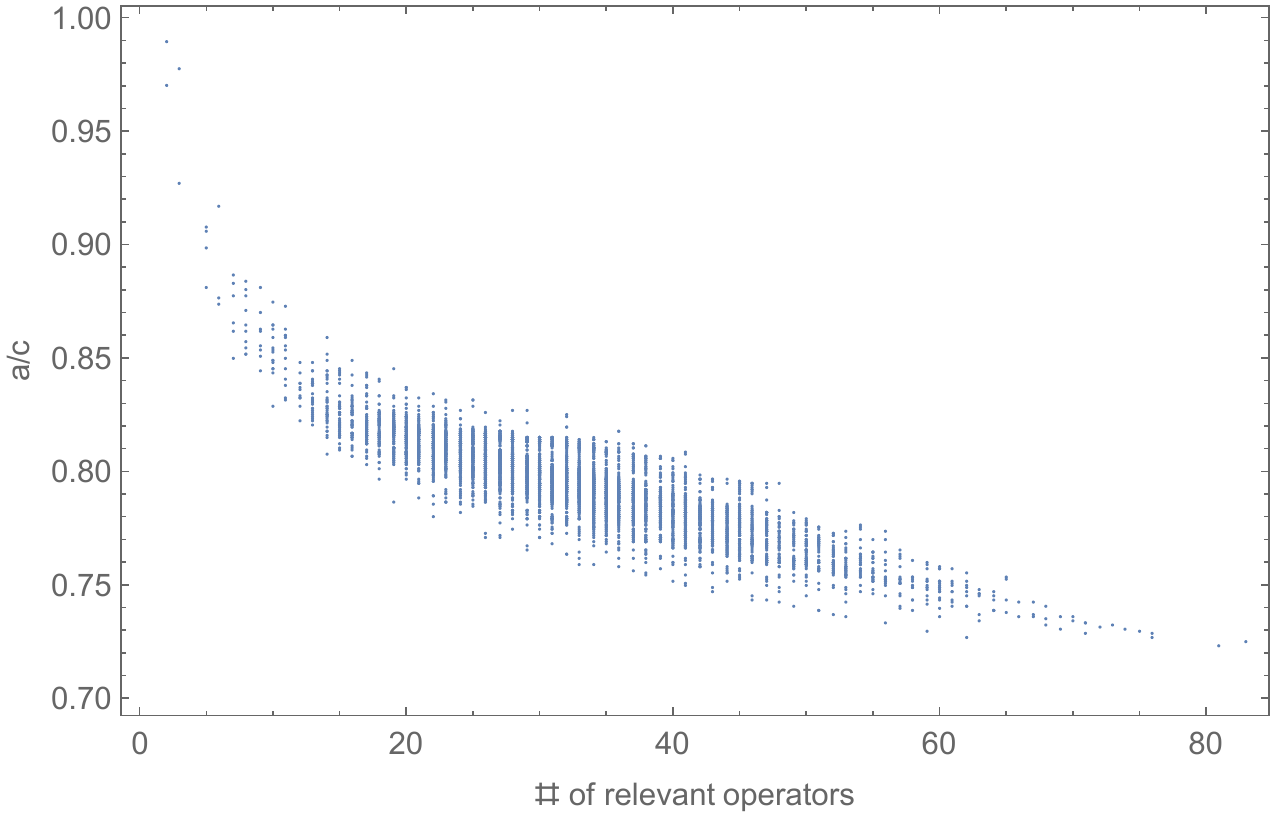}
     \end{subfigure}
     \hfill
     \caption{Left: The plot of $a/c$ versus the dimension of the lightest operator. Right: The plot of $a/c$ versus the number of relevant operators.}
     \label{fig:SU2adj1nf2ratio}
\end{figure}

\paragraph{$a/c$ vs other quantities}
We plot $a/c$ versus the dimension of the lightest operator and the number of relevant operators, respectively, in Figure \ref{fig:SU2adj1nf2ratio}. 
We see a clear pattern that as the dimension of the lightest operator increases, the lower bound for the ratio of the central charges increases. This can be interpreted as follows: When the dimension of the lightest operator is close to 1 (the value for the free field), there can be an approximate higher-spin symmetry \cite{Maldacena:2011jn, Maldacena:2012sf}. In view of the AdS/CFT correspondence, the bulk theory is further away from Einstein gravity, so that $a/c$ is further away from 1. Loosely speaking, the mass gap of the scalar in the bulk seems to constrain the higher-derivative corrections in the bulk. 

As in section \ref{sec:su2nf1}, we find that as the number of relevant operators increases, $a/c$ gets smaller (than 1), which can be thought of as a hint that holographic theories tend to be more ``stable" against relevant perturbations. 
Among the fixed points we have found, the one with the maximal number of relevant operator is the one with the superpotential (\href{https://qft.kaist.ac.kr/landscape/detail.php?id=5969}{\#5969}),
\begin{align}
\begin{split}
    W&=M_1q_1q_2+ M_2\tilde{q}_1\tilde{q}_2+ \phi q_1^2+ M_1\phi ^2+ M_2M_3+ M_4\phi q_2\tilde{q}_1+ M_4\phi q_2^2\\
    &\quad + M_5 \phi  q_2 \tilde{q}_2+ M_6 \phi  q_2^2 + M_7 q_1 \tilde{q}_2+ M_8 \phi  \tilde{q}_1^2+ M_9 \phi  \tilde{q}_1 \tilde{q}_2 \ , 
\end{split}
\end{align}
which has 83 relevant operators. 
The reduced index of this theory is given by
\begin{align}
\begin{split}
    I_{\textrm{red}}&=3 t^{2.08} + 5 t^{2.31} + t^{2.54} + t^{2.77} +  2 t^{3.23} + t^{4.15} + 6 t^{4.16} + 15 t^{4.38} +  15 t^{4.61}\\
    &\quad+ t^{4.62} (3 - \chi_{\half}(y))+ 8 t^{4.85} + 5 t^{5.07} + t^{5.08} +  7 t^{5.31} + 10 t^{5.54} - 5 t^6 + \cdots\,.
\end{split}
\end{align}
We indeed see that there are in total 83 terms with $t^{\a<6}$.

\paragraph{Conformal manifold / Flavor symmetry enhancement}
We identify a number of fixed points that have a non-trivial conformal manifold or accidental symmetry enhancement by examining the coefficient of $t^6$ in the superconformal index:
\begin{itemize}
    \item The number of fixed points with a positive coefficient in $t^6$ term: 664+$\a$.
    \item The number of fixed points with no flavor symmetry in the UV but with a negative coefficient in the $t^6$ term: 1395+$\a$.
\end{itemize}
Therefore, there are at least 664 fixed points with non-trivial conformal manifold and at least 1395 fixed points with flavor symmetry enhancement. 

\paragraph{SUSY enhancement}
We have found three fixed points that satisfy the sufficient and necessary conditions for $\CN=2$ enhancement. They are all identical to the previous results of \cite{Maruyoshi:2016aim, Agarwal:2016pjo} upon removing the decoupled chiral multiplets. 
\begin{itemize}
    \item $W=M_1 q_1q_2 + \phi \tilde{q}_1\tilde{q}_2 + X_1\Tr\phi^2$. This fixed point corresponds to the $H_2 = (A_1, D_4)$ Argyres-Douglas theory. The central charges and R-charges of matter fields are given by:
        {\renewcommand\arraystretch{1.4}
        \begin{table}[H]
            \centering
            \begin{tabular}{|c|c|c||c|c|c|c|c|}\hline
               id & $a$ & $c$ & $R_{X_1}$ & $R_{M_1}$ & $R_{q_i}$ & $R_{\tilde{q}_i}$ & $R_\phi$ \\\hline\hline
               \href{https://qft.kaist.ac.kr/landscape/detail.php?id=54}{\#54}&$\frac{7}{12}$& $\frac{2}{3}$ & $\frac{4}{3}$ & $1$ & $\frac{1}{2}$, $\frac{1}{2}$ & $\frac{5}{6}$, $\frac{5}{6}$ & $\frac{1}{3}$\\\hline
            \end{tabular}
        \end{table}}
        
    \item $W=\phi q_1^2+ q_1 q_2 + X_1\Tr\phi^2 + M_1 \tilde{q}_1\tilde{q}_2$. This fixed point corresponds to the $H_1 = (A_1, A_3)$ Argyres-Douglas theory. Upon integrating out massive chiral multiplets $q_1, q_2$, the Lagrangian is identical to the previous one in section \ref{sec:su2nf1}. The central charges and R-charges of matter fields are given by: 
    {\renewcommand\arraystretch{1.4}
        \begin{table}[H]
            \centering
            \begin{tabular}{|c|c|c||c|c|c|c|c|}\hline
                id &$a$ & $c$ & $R_{X_1}$ &  $R_{M_1}$ & $R_{q_i}$ & $R_{\tilde{q}_i}$ & $R_\phi$ \\\hline\hline
               \href{https://qft.kaist.ac.kr/landscape/detail.php?id=67}{\#67}&$\frac{11}{24}$& $\frac{1}{2}$ & $\frac{14}{9}$  & $\frac{8}{9}$  & $\frac{8}{9}$, $\frac{10}{9}$ & $\frac{5}{9}$, $\frac{5}{9}$ & $\frac{2}{9}$\\\hline
            \end{tabular}
        \end{table}}

    \item $W=\phi q_1^2+\phi q_2^2+M_1\phi \tilde{q}_1^2 + \phi\tilde{q}_2^2 + X_1\Tr\phi^2$. This fixed point also corresponds to the $H_1 = (A_1, D_3)$ Argyres-Douglas theory. Notice that in this description, there is no massive matter field. It is a dual description of the previous one. We discuss the duality in detail in section \ref{subsec:H1dual}. 
    {\renewcommand\arraystretch{1.4}
    \begin{table}[H]
        \centering
        \begin{tabular}{|c|c|c||c|c|c|c|c|}\hline
            id &$a$ & $c$ & $R_{X_1}$ &  $R_{M_1}$ & $R_{q_i}$ & $R_{\tilde{q}_i}$ & $R_\phi$ \\\hline\hline
           \href{https://qft.kaist.ac.kr/landscape/detail.php?id=95}{\#95}&$\frac{11}{24}$& $\frac{1}{2}$ & $\frac{14}{9}$  & $\frac{8}{9}$  & $\frac{8}{9}$, $\frac{8}{9}$ & $\frac{4}{9}$, $\frac{8}{9}$ & $\frac{2}{9}$\\\hline
        \end{tabular}
    \end{table}}
        
    \item $W=\phi q_1^2+ \phi q_2^2+ q_1 \tilde{q}_1+X_1\Tr\phi^2+M_1\phi \tilde{q}_2^2$. This fixed point corresponds to the $H_0$ Argyres-Douglas theory. Upon integrating out the massive chiral multiplets $q_1, \tilde{q}_1$, this gives an identical theory to that discussed in section \ref{sec:su2nf1}. The central charges and R-charges of matter fields are given by:
        {\renewcommand\arraystretch{1.4}
        \begin{table}[H]
            \centering
            \begin{tabular}{|c|c|c||c|c|c|c|c|}\hline
              id & $a$ & $c$ & $R_{X_1}$ &  $R_{M_1}$ & $R_{q_i}$ & $R_{\tilde{q}_i}$ & $R_\phi$ \\\hline\hline
              \href{https://qft.kaist.ac.kr/landscape/detail.php?id=98}{\#98} & $\frac{43}{120}$& $\frac{11}{30}$ & $\frac{26}{15}$  & $\frac{4}{5}$  & $\frac{14}{15}$, $\frac{14}{15}$ & $\frac{16}{15}$, $\frac{8}{15}$ & $\frac{2}{15}$\\\hline
            \end{tabular}
        \end{table}}

    \end{itemize}
    There is no other fixed point in the $N_f=2$ $SU(2)$ adjoint SQCD landscape (that we explored so far) that satisfies the necessary conditions for $\CN=2$ SUSY enhancement.

%%%%%%%%%%%%%%%%%%%%%%%%%%%
\subsection{$SU(3)$ $N_f=1$ Adjoint SQCD}
\begin{figure}[t]
    \centering
     \begin{subfigure}[b]{0.45\textwidth}
         \includegraphics[width=\linewidth]{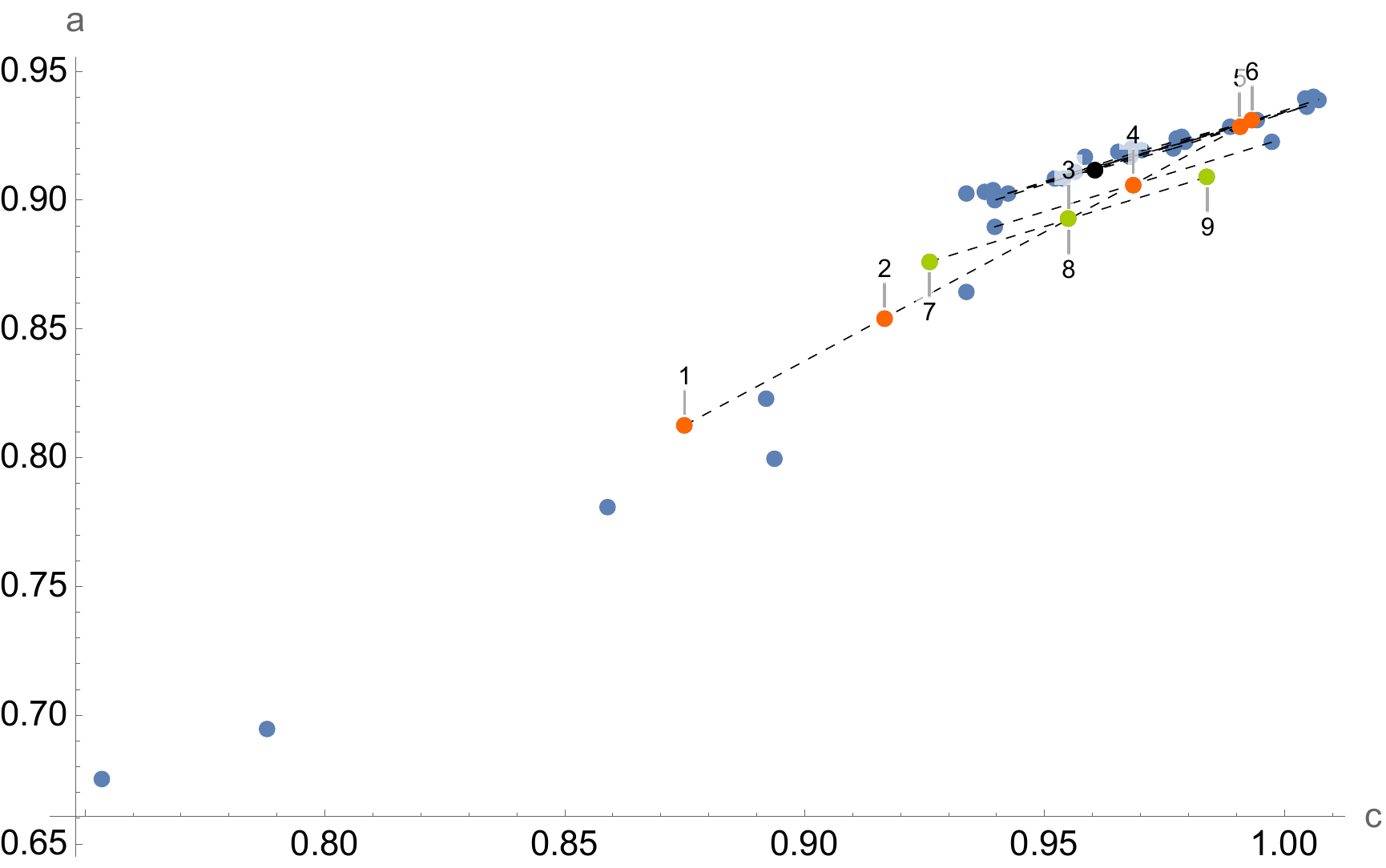}
         \caption{$a$ vs. $c$}
     \end{subfigure}
     \hspace{4mm}
     \begin{subfigure}[b]{0.45\textwidth}
         \includegraphics[width=\linewidth]{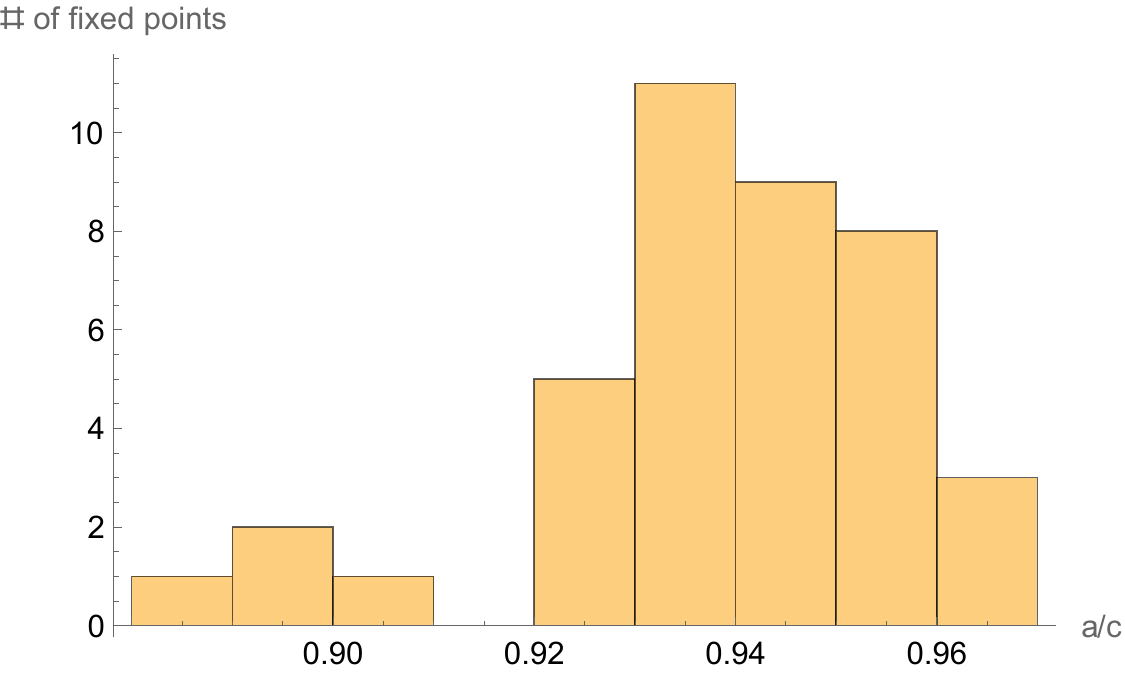}
         \caption{Histogram of $a/c$}
     \end{subfigure}
     \hfill
     \caption{(a) The plot of $a$ versus $c$ for the fixed points in the landscape of $SU(3)$ $N_f=1$ adjoint SQCD. The black dot represents the fixed point with the superpotential $W=X_1\Tr\phi^2+X_2\Tr\phi^3$. The colored dots, labeled by an integer $i$ and each representing the fixed point with the superpotential $W_i$, lie on the dashed lines. (b) The histogram of $a/c$.}
     \label{fig:SU3adj1nf1ac}
\end{figure}

We start with the seed theory given as the fixed point of $SU(3)$ supersymmetric gauge theory with one adjoint $\phi$, one fundamental $q$ and one anti-fundamental $\tilde{q}$ chiral multiplet, without superpotential. The set of fixed points spawned from this theory have the following features.
\begin{itemize}
    \item The maximal number of relevant deformations (modulo flipping decoupled operators): 5.
    \item The number of inequivalent fixed points: 40.
    \item The number of pairs of inequivalent fixed points with identical central charges: 0.
    \item The number of fixed points with $a=c$: 0.
    \item The number of inequivalent fixed points with rational central charges: 29.
\end{itemize}
The distribution of the central charges $a$ and $c$ is plotted in Figure \ref{fig:SU3adj1nf1ac}. 
The minimal and maximal values of central charges are given by,
\begin{align}
\begin{split}
    a_{\min}&=\frac{1383}{2048}\simeq 0.6753,\quad a_{\max}=\frac{361}{384}\simeq 0.9401, \\ c_{\min}&=\frac{1543}{2048}\simeq 0.7534,\quad c_{\max}=\frac{975}{968}\simeq 1.0072\,.
\end{split}
\end{align}
The fixed point with the minimal $a$ in this landscape (\href{https://qft.kaist.ac.kr/landscape/detail.php?id=45786}{\#45786}) is given by the superpotential
\begin{align}
    W=X_1\Tr\phi^2+ X_2\Tr\phi^3+\phi^3\tilde{q}^3+M_1\phi^2q\tilde{q}+M_2^2+X_3 q\tilde{q}\,.
\end{align}

All the fixed points in this landscape have $a<c$. The minimal and maximal values of $a/c$ are given by
\begin{align}
    (a/c)_{\min}=\frac{711}{807}\simeq 0.8810,\quad (a/c)_{\max}=\frac{231}{239}\simeq 0.9665
\end{align}
The fixed point with $(a/c)_{\max}$ has the superpotential
\begin{align}
    W=X_1\Tr\phi^2+X_2\Tr\phi^3+q^2\tilde{q}^2\,.
\end{align}
This fixed point (\href{https://qft.kaist.ac.kr/landscape/detail.php?id=45756}{\#45756}) turns out to be  equivalent to the $\CN=1$ deformation of the $(A_2,A_3)$ Argyres-Douglas theory via Coulomb branch operator of dimension $12/7$. We discuss this example in detail in section \ref{subsec:deformAD}.

We have found that some sets of more than two fixed points are exactly aligned on straight lines, as illustrated in Figure \ref{fig:SU3adj1nf1ac}(a). These are as follows.
\begin{itemize}
    \item The fixed points lying on a line of $a=c-\frac{1}{16}$:
        \begin{equation}
    \begin{aligned}
        W_4&=X_1\Tr\phi^2+X_2\Tr\phi^3+M_1\phi q\tilde{q}+M_1^2\\
        W_{3}&=X_1\Tr\phi^2+X_2\Tr\phi^3+\phi^3\tilde{q}^3+M_1\phi q\tilde{q}+M_1^2\,,\\
        W_{2}&=X_1\Tr\phi^2+X_2\Tr\phi^3+M_1\phi q\tilde{q}+M_2\phi^2q\tilde{q}+M_1M_2+X_3q\tilde{q}\,,\\
        W_1&=X_1\Tr\phi^2+X_2\Tr\phi^3+M_1\phi q\tilde{q}+M_2\phi^2q\tilde{q}+M_1M_2+X_3q\tilde{q}+\phi^3\tilde{q}^3\,,\\
        W_5&=X_1\Tr\phi^2+X_2\Tr\phi^3+\phi q^2\tilde{q}^2+\phi^3q^3+M_1\phi^2q\tilde{q}\,,\\
        W_6&=X_1\Tr\phi^2+X_2\Tr\phi^3+\phi q^2\tilde{q}^2+M_1\phi^2q\tilde{q}\,.
    \end{aligned}
    \end{equation}

    \item The fixed points lying on the line of $a=\frac{33}{58}c + \frac{81}{232}$:
    \begin{equation}
    \begin{aligned}
        W_8&=X_1\Tr\phi^2+X_2\Tr\phi^3+\phi^3\tilde{q}^3+M_1\phi q\tilde{q}+M_1^2\,,\\
        W_7&=W_8 + M_2q\tilde{q},\quad W_9=W_8+M_2\phi^2q\tilde{q}\,.
    \end{aligned}
    \end{equation}
    \item There are 10 more lines with slopes $\frac{123}{244}$, $\frac{150}{271}$, $\frac{33}{58}$, $\frac{177}{298}$\,.
\end{itemize}

\begin{figure}[t]
    \centering
     \begin{subfigure}[b]{0.45\textwidth}
         \includegraphics[width=\linewidth]{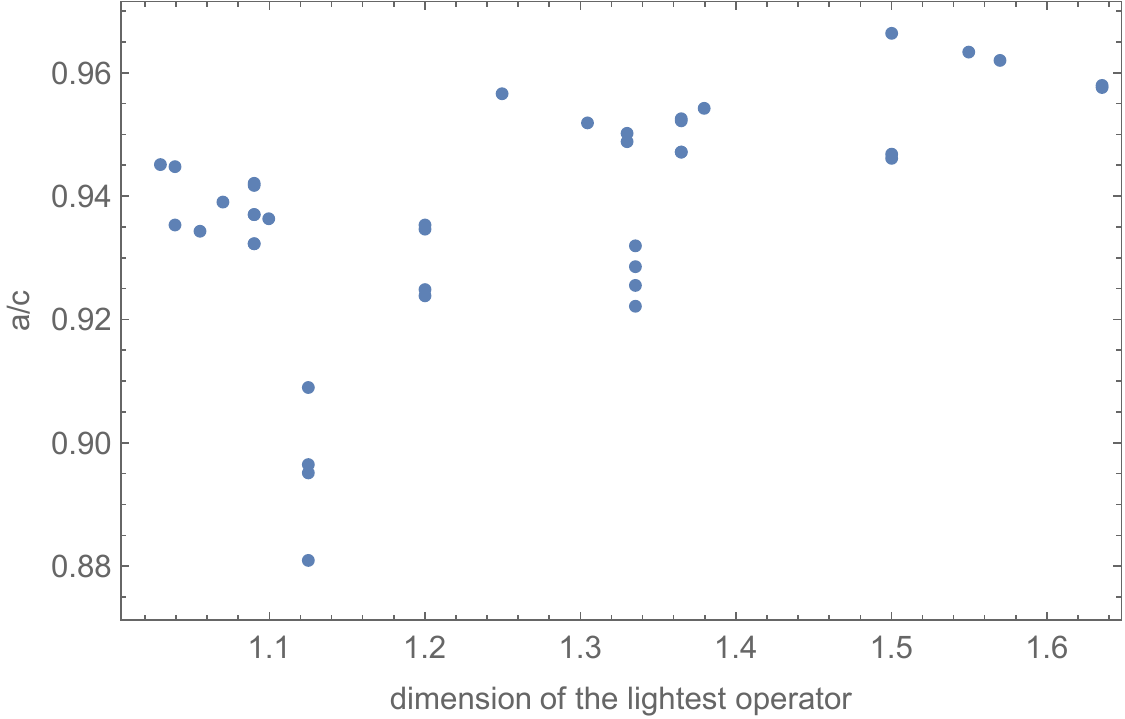}
     \end{subfigure}
     \hspace{4mm}
     \begin{subfigure}[b]{0.45\textwidth}
         \includegraphics[width=\linewidth]{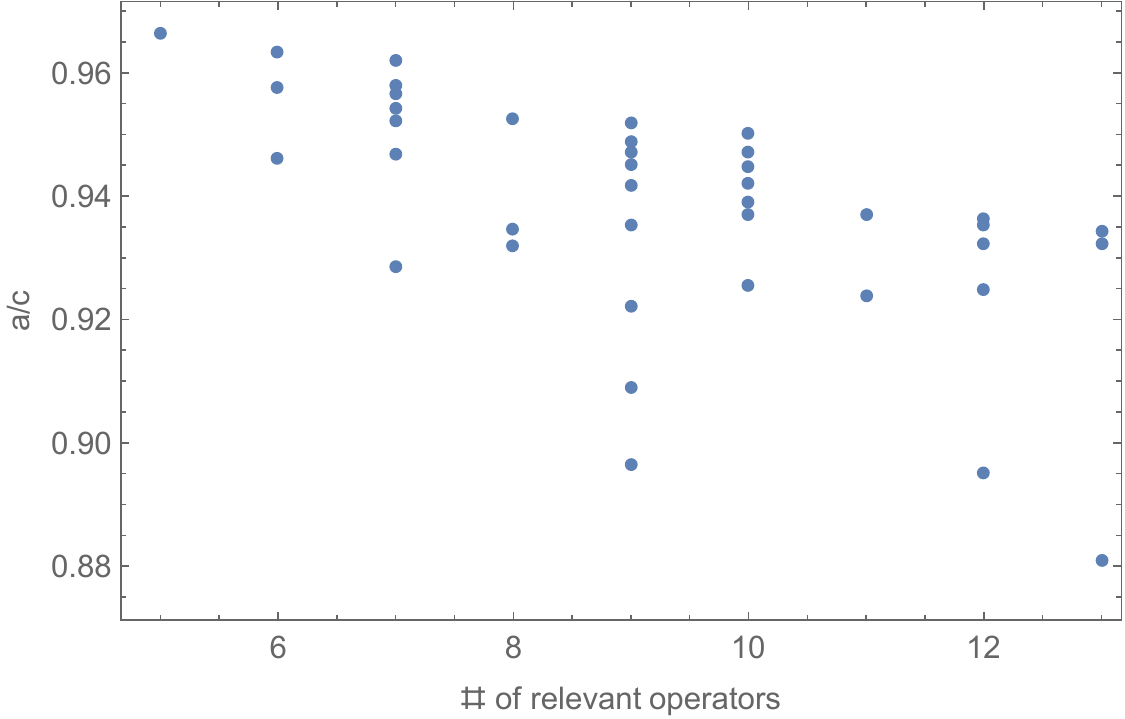}
     \end{subfigure}
     \hfill
     \caption{Left: The plot of $a/c$ versus the dimension of the lightest operator. Right: The plot of $a/c$ versus the number of relevant operators.}
     \label{fig:SU3adj1nf1ratio}
\end{figure}

We have also plotted $a/c$ versus the dimension of the lightest operator and the number of relevant operators, respectively, in Figure \ref{fig:SU3adj1nf1ratio}.
We see a similar correlation between the lightest operator dimension/number of relevant operators vs $a/c$ as in the previous cases. 

\paragraph{Conformal manifold / Flavor symmetry enhancement}
We identify fixed points that have a non-trivial conformal manifold or accidental symmetry enhancement by examining the coefficient of $t^6$ in the superconformal index:
\begin{itemize}
    \item The number of fixed points with a positive coefficient in $t^6$ term: 5.
    \item The number of fixed points which have no flavor symmetry in UV but have a negative coefficient in $t^6$ term: 5.
\end{itemize}
Therefore there are at least 5 fixed points with non-trivial conformal manifold and at least 5 fixed points with symmetry enhancement. 

\paragraph{SUSY enhancement}
There is only one fixed point that satisfies the sufficient condition for $\CN=2$ supersymmetry enhancement, and  also satisfies all the necessary conditions for $\CN=2$ supersymmetry:
\begin{itemize}
    \item $W=X_1\Tr\phi^2+X_2\Tr\phi^3+M_1 \phi q\tilde{q}+M_2q\tilde{q}$. This fixed point corresponds to the $(A_1, A_5)$ Argyres-Douglas theory, which was originally found in \cite{Maruyoshi:2016aim}. 
        {\renewcommand\arraystretch{1.4}
        \begin{table}[H]
            \centering
            \begin{tabular}{|c|c|c||c|c|c|c|c|}\hline
               id & $a$ & $c$ & $R_{X_i}$ &  $R_{M_i}$ & $R_{q_i}$ & $R_{\tilde{q}_i}$ & $R_\phi$ \\\hline\hline
              \href{https://qft.kaist.ac.kr/landscape/detail.php?id=45770}{\#45770} & $\frac{11}{12}$& $\frac{23}{24}$ & $\frac{5}{3}$, $\frac{3}{2}$ & $\frac{5}{6}$ & $\half$ & $\frac{1}{2}$ & $\frac{1}{6}$\\\hline
            \end{tabular}
        \end{table}}
\end{itemize}
We find another fixed point that satisfies the necessary conditions for $\CN=2$ supersymmetry: 
\begin{itemize}
    \item $W=X_1\Tr\phi^2+X_2\Tr\phi^3 + M_1 \phi q\tilde{q} + M_2 \phi^2 q\tilde{q}+ M_1M_2 + X_3 q\tilde{q}+ \phi^3 q^3$.
        {\renewcommand\arraystretch{1.4}
        \begin{table}[H]
            \centering
            \begin{tabular}{|c|c|c||c|c|c|c|c|}\hline
               id & $a$ & $c$ & $R_{X_i}$ &  $R_{M_i}$ & $R_{q_i}$ & $R_{\tilde{q}_i}$ & $R_\phi$ \\\hline\hline
              \href{https://qft.kaist.ac.kr/landscape/detail.php?id=45791}{\#45791} & $\frac{13}{16}$& $\frac{7}{8}$ & $\frac{14}{9}$, $\frac{4}{3}$, $\frac{4}{3}$  & $\frac{10}{9}$, $\frac{8}{9}$ & $\frac{4}{9}$ & $\frac{2}{9}$& $\frac{2}{9}$\\\hline
            \end{tabular}
        \end{table}}
\end{itemize}
At this fixed point, the coefficient of $t^6$ is 0 and no marginal operator exists. This means that there is no global symmetry in view of $\CN=1$ supersymmetry, which is necessary for the supersymmetry enhancement. Consequently, this fixed point lacks a symmetry that could enhance to the $\CN=2$ R-symmetry in  the IR.

\subsection{$SO(5)$ $N_f=1$ Adjoint SQCD}
\begin{figure}[t]
    \centering
     \begin{subfigure}[b]{0.45\textwidth}
         \includegraphics[width=\linewidth]{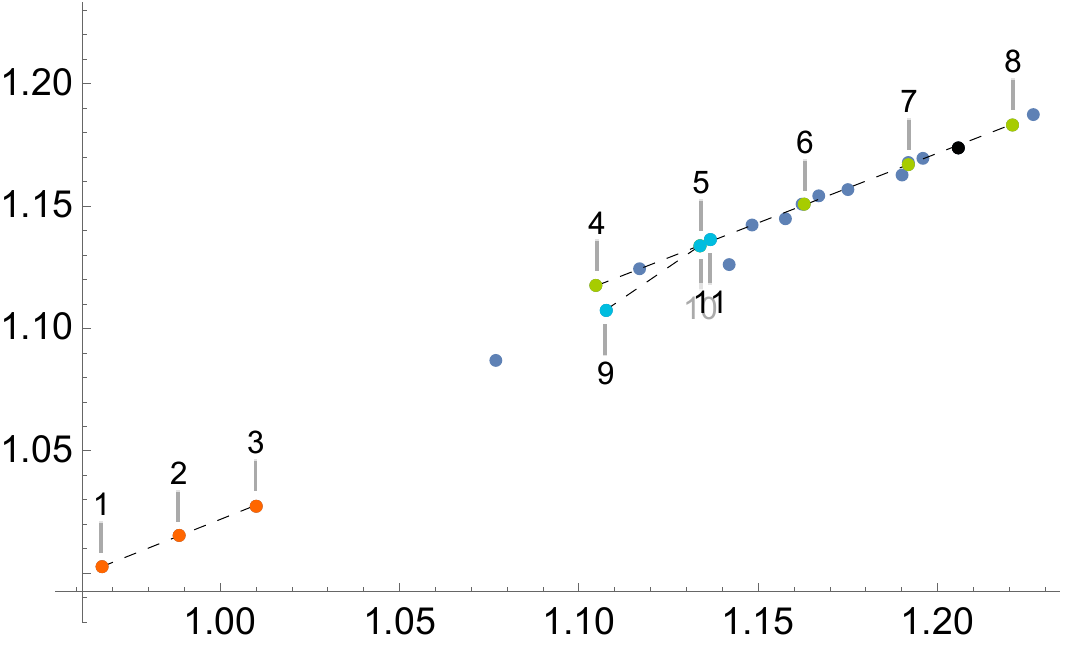}
         \caption{$a$ vs. $c$}
     \end{subfigure}
     \hspace{4mm}
     \begin{subfigure}[b]{0.45\textwidth}
         \includegraphics[width=\linewidth]{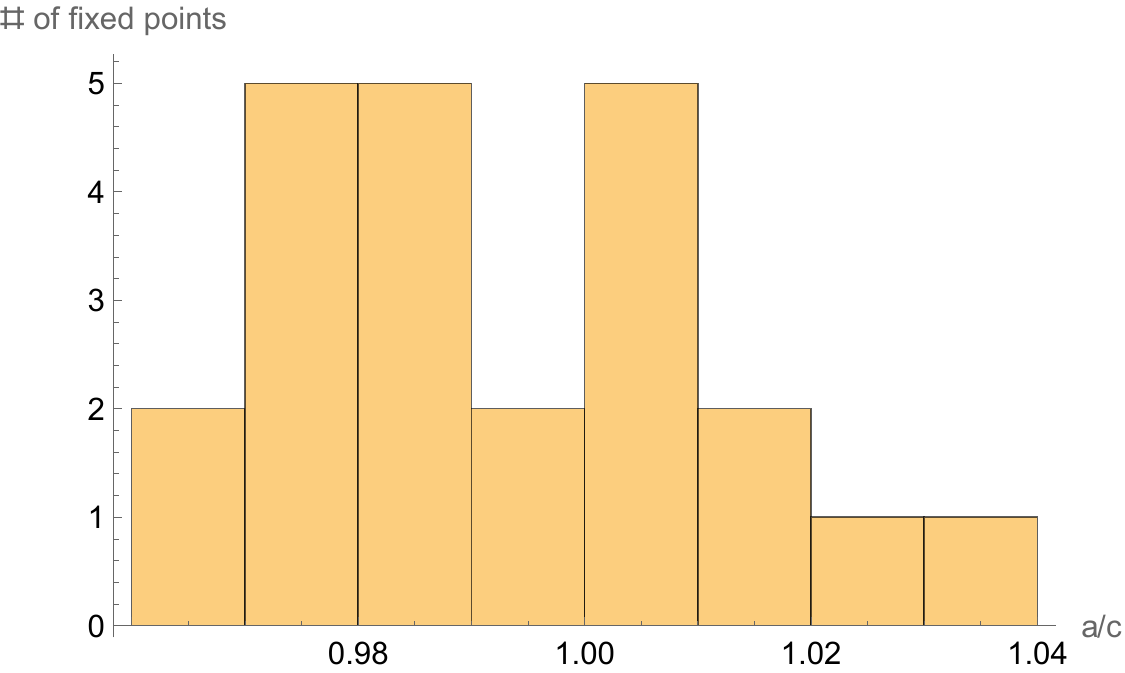}
         \caption{Histogram of $a/c$}
     \end{subfigure}
     \hfill
     \caption{(a) The plot of $a$ versus $c$ for fixed points in the landscape of $SO(5)$ $N_f=1$ adjoint SQCD. The black dot represents the fixed point with the superpotential $W=X_1\Tr\phi^2$. The colored dots, labeled by an integer $i$, represent fixed points with superpotentials $W_i$, and lie on the dashed lines. (b) The histogram of $a/c$.}
     \label{fig:SO5adj1nf1ac}
\end{figure}

The seed theory is the IR fixed point of $SO(5)$ supersymmetric gauge theory with one adjoint ($\phi$, the anti-symmetric tensor in the $\mathbf{10}$-dimensional representation) and one vector ($q$) chiral multiplets, without superpotential.
The landscape generated from this theory has the following features:
\begin{itemize}
    \item The maximal number of relevant deformation (modulo flipping free operators): 4.
    \item The number of inequivalent fixed points: 23.
    \item The number of pairs of inequivalent fixed points with identical central charges: 0.
    \item The number of fixed points with $a=c$: 3.
    \item The number of fixed points with $a>c$: 6.
    \item The number of inequivalent fixed points with rational central charges: 11.
\end{itemize}
The distribution of the central charges $a$ and $c$ is plotted in Figure \ref{fig:SO5adj1nf1ac}. 
The minimal and maximal values of central charges are given by
\begin{align}
    a_{\min}=\frac{393}{392}\simeq 1.0026,\quad a_{\max}\simeq 1.1870,\quad c_{\min}=\frac{379}{392}\simeq 0.9668,\quad c_{\max}\simeq 1.2265\,.
\end{align}
The superpotential leading the fixed point with the minimal $a$ (\href{https://qft.kaist.ac.kr/landscape/detail.php?id=8297}{\#8297}) is given by
\begin{align}
    W=X_1\Tr \phi^2+ \phi^2q^3+ X_2\Tr \phi^4+ M_1\phi^2q\,.
\end{align}
The minimal and maximal values of $a/c$ are given by
\begin{align}
    (a/c)_{\min}=\frac{2127}{2543}\simeq 0.8364,\quad (a/c)_{\max}=\simeq 1.0011\,.
\end{align}

We find three sets of fixed points that are exactly aligned on straight lines, as illustrated in Figure \ref{fig:SO5adj1nf1ac}(a).
\begin{itemize}
    \item The fixed points lying on a line of $a=\frac{69}{118}c+\frac{2889}{6608}$:
        \begin{equation}
    \begin{aligned}
        W_2&=X_1\Tr\phi^2+\phi^2q^3+X_2\Tr\phi^4,\quad W_1=W_2+M_1\phi^2q,\quad W_3=W_2+M_1q^2\,.
    \end{aligned}
    \end{equation}

    \item The fixed points lying on the line of $a=\frac{33}{58}c+\frac{567}{1160}$:
    \begin{equation}
    \begin{aligned}
        W_6&=X_1\Tr\phi^2+M_1\phi^4+M_1\phi^2q& W_5&=W_6+M_2q^2\,,\\
        W_4&=W_6+M_2q^2+M_3\phi^2q\,,& &\\
        W_7&=X_1\Tr\phi^2+\phi^2q^4&W_8&=W_7+M_1\phi^2q^2\,.
    \end{aligned}
    \end{equation}
    \item The fixed points lying on the line of $a=c$ are listed in the last part of this section.
\end{itemize}

\begin{figure}[t]
    \centering
     \begin{subfigure}[b]{0.45\textwidth}
         \includegraphics[width=\linewidth]{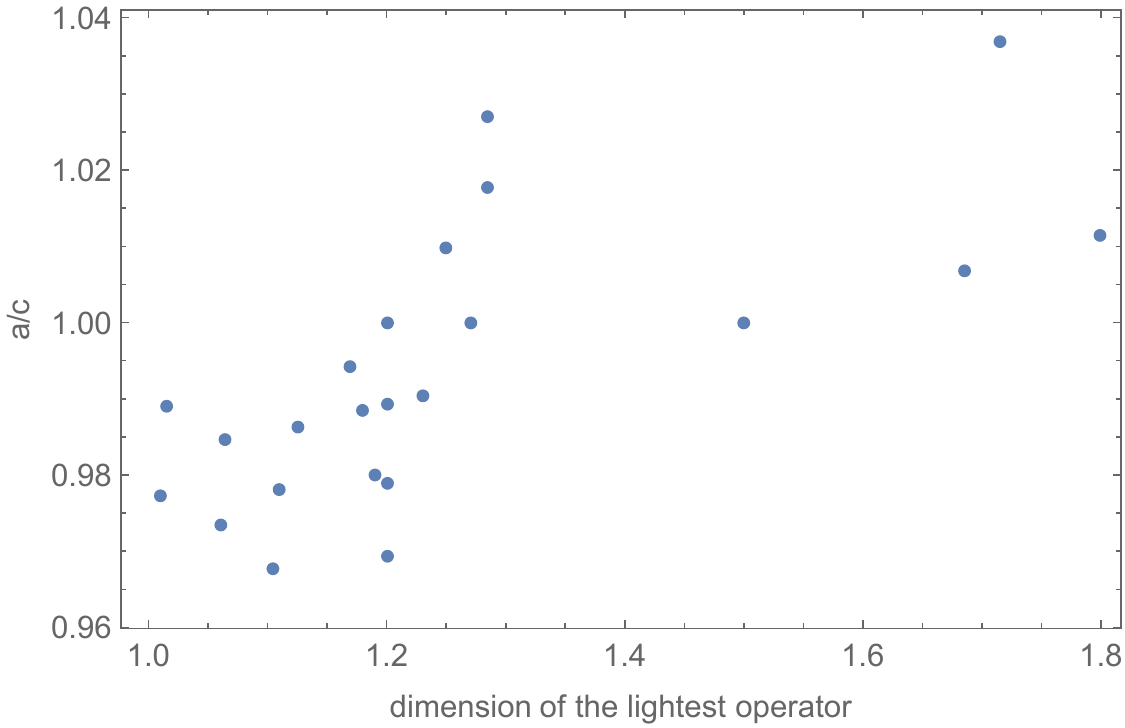}
     \end{subfigure}
     \hspace{4mm}
     \begin{subfigure}[b]{0.45\textwidth}
         \includegraphics[width=\linewidth]{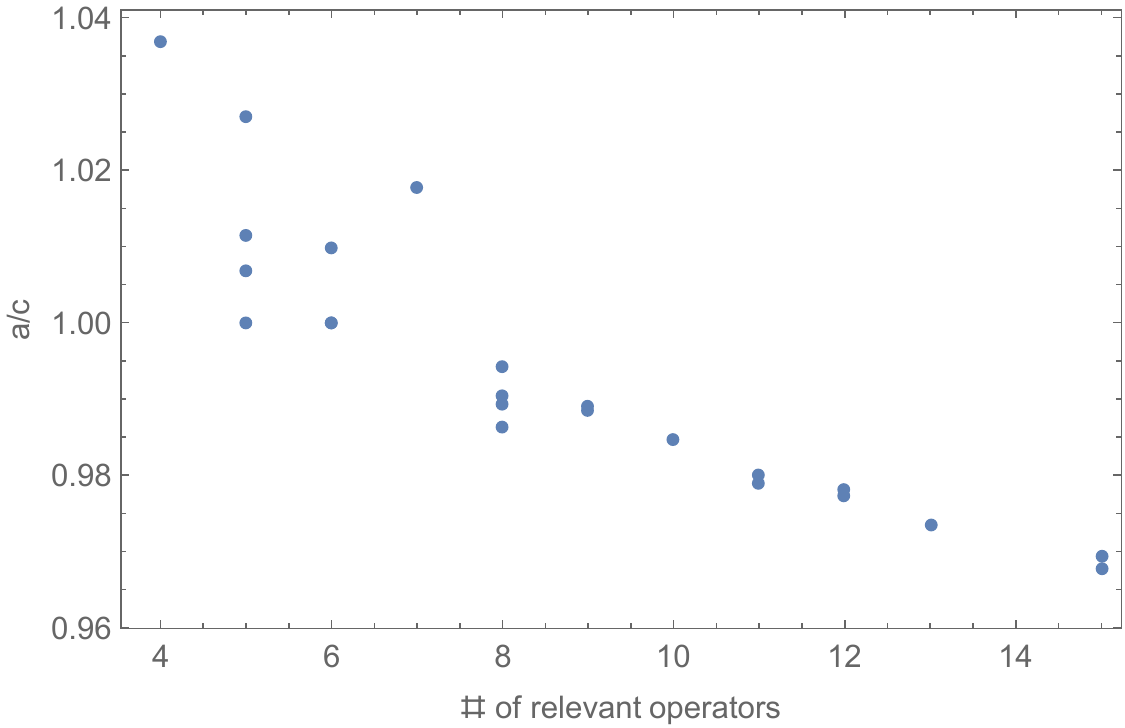}
     \end{subfigure}
     \hfill
     \caption{Left: $a/c$ versus the dimension of the lightest operator. Right: $a/c$ versus the number of relevant operators.}
     \label{fig:SO5adj1nf1ratio}
\end{figure}

We also plot $a/c$ versus the dimension of the lightest operator and the number of relevant operators, respectively, in Figure \ref{fig:SO5adj1nf1ratio}.
We find a similar pattern as in the previous examples. 

\paragraph{Conformal manifold / Flavor symmetry enhancement} 
We identify the fixed points that have a non-trivial conformal manifold or accidental symmetry enhancement by examining the coefficient of $t^6$ in the superconformal index.
\begin{itemize}
    \item The number of fixed points with a positive coefficient in $t^6$ term: 5.
    \item The number of fixed points which have no flavor symmetry in UV but have a negative coefficient in $t^6$ term: 4.
\end{itemize}
Hence, there are at least 5 fixed points with a non-trivial conformal manifold and at least 4 fixed points with flavor symmetry enhancement. 

\paragraph{SUSY enhancement}
There is no fixed point that satisfies the sufficient or necessary conditions for the supersymmetry enhancement in this landscape. 

\paragraph{$a=c$ theory} There are 3 fixed point with $a=c$:
\begin{itemize}
    \item $W=X_1\Tr\phi^2 + \phi^8 + X_2 q^2+ M_1\phi^2 q$.
    {\renewcommand\arraystretch{1.4}
        \begin{table}[H]
            \centering
            \begin{tabular}{|c|c|c||c|c|c|c|}\hline
               id & $a$ & $c$ & $R_{X_i}$ & $R_{M_1}$ & $R_{q}$ & $R_\phi$ \\\hline\hline
              \href{https://qft.kaist.ac.kr/landscape/detail.php?id=8293}{\#8293} & $\frac{567}{512}$& $\frac{567}{512}$ & $\frac{3}{2}$, $\frac{3}{2}$ & $\frac{5}{4}$ & $\frac{1}{4}$  & $\frac{1}{4}$\\\hline
            \end{tabular}
        \end{table}}

    \item $W=X_1\Tr\phi^2 + M_1q^2+M_2\phi^2q$.
    {\renewcommand\arraystretch{1.4}
        \begin{table}[H]
            \centering
            \begin{tabular}{|c|c|c||c|c|c|c|}\hline
              id &  $a$ & $c$ & $R_{X}$ & $R_{M}$ & $R_{q}$ & $R_\phi$ \\\hline\hline
             \href{https://qft.kaist.ac.kr/landscape/detail.php?id=8289}{\#8289} &  $1.1367$& $1.1367$ & $1.5774$ & $1.2680$ & $0.366025$  & $0.2113$\\\hline
            \end{tabular}
        \end{table}}

    \item $W=X_1\Tr\phi^2 + M_1\Tr\phi^4 +M_1\phi^2 q + M_2q^2$.
    {\renewcommand\arraystretch{1.4}
        \begin{table}[H]
            \centering
            \begin{tabular}{|c|c|c||c|c|c|c|}\hline
              id & $a$ & $c$ & $R_{X_1}$ & $R_{M_i}$ & $R_{q}$ & $R_\phi$ \\\hline\hline
              \href{https://qft.kaist.ac.kr/landscape/detail.php?id=8298}{\#8298} & $\frac{567}{500}$ & $\frac{567}{500}$ & $\frac{8}{5}$ & $\frac{6}{5}$, $\frac{6}{5}$ & $\frac{2}{5}$  & $\frac{1}{5}$ \\\hline
            \end{tabular}
        \end{table}}
\end{itemize}

\subsection{$SO(5)$ $N_f=2$ Adjoint SQCD}
Start with the seed theory given by the IR fixed point of $SO(5)$ supersymmetric gauge theory with one adjoint ($\phi$, anti-symmetric tensor, $\textbf{10}$-dimensional representation) and two vector ($q$, $\textbf{5}$-dimensional representation) chiral multiplets, without superpotential.
The set of fixed points created out of the $SO(5)$ $N_f=1$ adjoint SQCD is included in this $N_f=2$ landscape via mass deformation. 
The landscape obtained from this theory has the following features:
\begin{itemize}
    \item The maximal number of relevant deformation (modulo flipping free operators): 10.
    \item The number of inequivalent fixed points: 698.
    \item The number of pairs of inequivalent fixed points with identical central charges: 1.
    \item The number of fixed points with $a=c$: 4.
    \item The number of fixed points with $a>c$: 8.
    \item The number of inequivalent fixed points with rational central charges: 329.
\end{itemize}
We see that the number of fixed points is quite large in this setup. 

\begin{figure}[t]
    \centering
     \begin{subfigure}[b]{0.45\textwidth}
         \includegraphics[width=\linewidth]{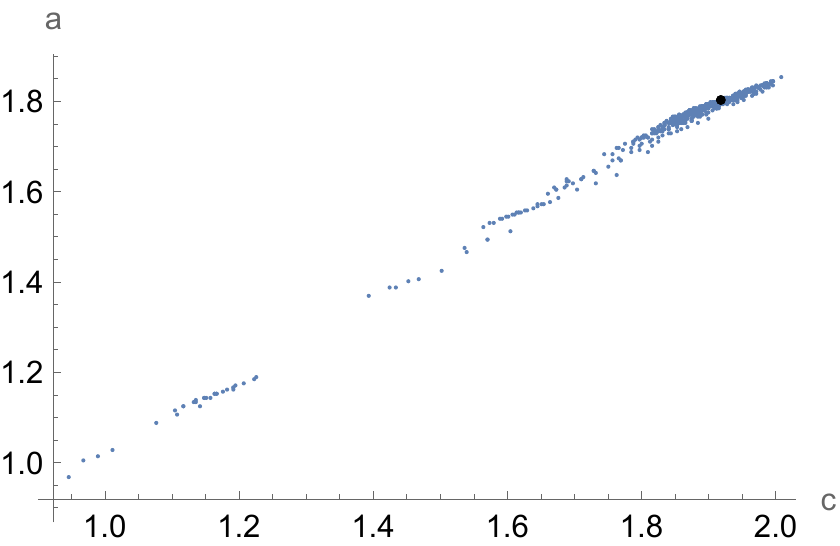}
         \caption{$a$ vs. $c$}
     \end{subfigure}
     \hspace{4mm}
     \begin{subfigure}[b]{0.45\textwidth}
         \includegraphics[width=\linewidth]{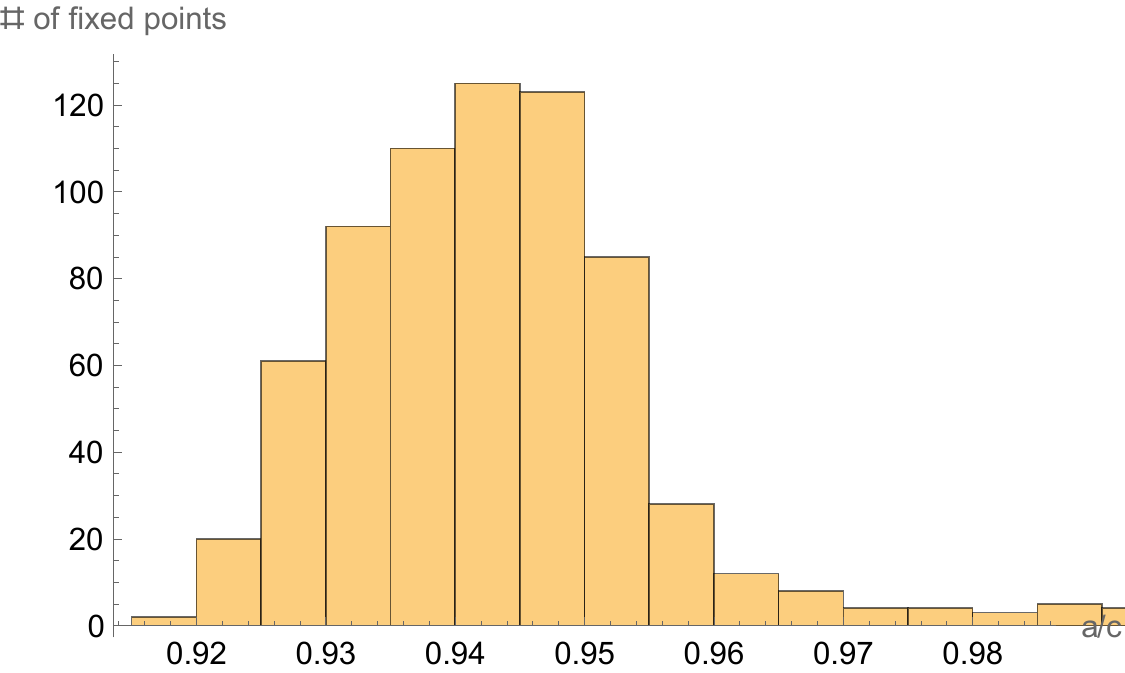}
         \caption{Histogram of $a/c$}
     \end{subfigure}
     \hfill
     \caption{(a) The plot of $a$ versus $c$ for fixed points in the landscape of $SO(5)$ $N_f=2$ adjoint SQCD. The black dot represents the fixed point without superpotential. (b) The histogram of $a/c$.}
     \label{fig:SO5adj1nf2ac}
\end{figure}

The distribution of the central charges $a$ and $c$ is plotted in Figure \ref{fig:SO5adj1nf2ac}. 
Once again, we see a narrow distribution of central charges, and the distribution is denser near the seed CFT. 
The minimal and maximal values of central charges are given by
\begin{align}
    a_{\min}=\frac{15867}{16384}\simeq 0.9684,\quad a_{\max}\simeq 1.8525,\quad c_{\min}=\frac{15483}{16384}\simeq 0.9450,\quad c_{\max}\simeq 2.0069. 
\end{align}
The superpotential leading to the fixed point with the minimal $a$ (\href{https://qft.kaist.ac.kr/landscape/detail.php?id=45037}{\#45037}) is given by
\begin{align}
    W=M_1\phi^2 q_1+ M_1 q_2^2+ q_1^2+ X_1\phi^2+X_2\phi^4\,.
\end{align}
The minimal and maximal values of $a/c$ are given by
\begin{align}
    (a/c)_{\min}=\frac{3669}{3994}\simeq 0.9186,\quad (a/c)_{\max}=\frac{393}{379}\simeq 1.0369\,.
\end{align}

\begin{figure}[t]
    \centering
     \begin{subfigure}[b]{0.45\textwidth}
         \includegraphics[width=\linewidth]{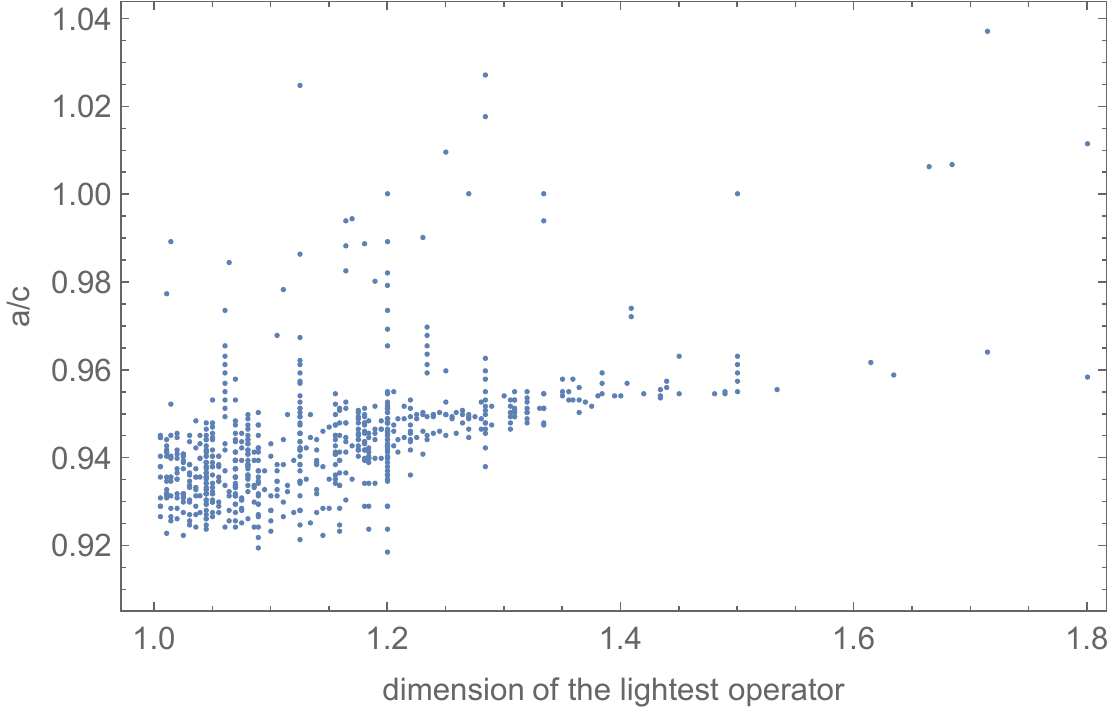}
     \end{subfigure}
     \hspace{4mm}
     \begin{subfigure}[b]{0.45\textwidth}
         \includegraphics[width=\linewidth]{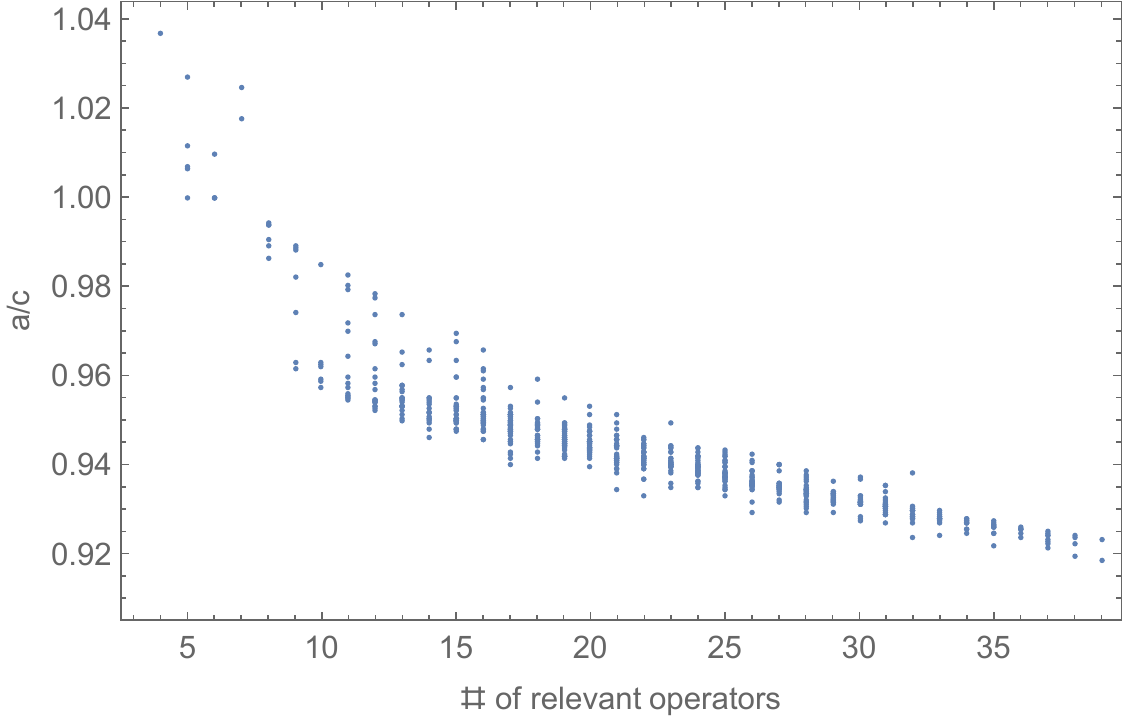}
     \end{subfigure}
     \hfill
     \caption{Left: $a/c$ versus the dimension of the lightest operator. Right: $a/c$ versus the number of relevant operators.}
     \label{fig:SO5adj1nf2ratio}
\end{figure}

We also plot $a/c$ versus the dimension of the lightest operator and the number of relevant operators, respectively, in Figure \ref{fig:SO5adj1nf2ratio}.

\paragraph{Conformal manifold / Flavor symmetry enhancement} We identify the fixed points that have a non-trivial conformal manifold or accidental symmetry enhancement by examining the coefficient of $t^6$ in the superconformal index:
\begin{itemize}
    \item The number of fixed points with a positive coefficient in $t^6$ term: 207.
    \item The number of fixed points which have no flavor symmetry in UV but have a negative coefficient in $t^6$ term: 62.
\end{itemize}
Therefore, at least 207 fixed points have non-trivial conformal manifolds and at least 62 fixed points have emergent flavor symmetry in the IR. 

\paragraph{SUSY enhancement}
There is no fixed point that satisfies the sufficient or necessary conditions for  supersymmetry enhancement.

\paragraph{$a=c$ theory} 
There are 4 fixed point with $a=c$. The first three fixed points are equivalent to the three $a=c$ fixed points that appeared in the $SO(5)$ $N_f=1$ landscape upon integrating out the massive field:
\begin{itemize}
    \item (\href{https://qft.kaist.ac.kr/landscape/detail.php?id=45038}{\#45038}) $W=q_1^2+X_1\phi^2+M_1\phi^2q_2+\phi^8+X_2 q_2^2,\quad a=c=\frac{567}{512}$.

    \item (\href{https://qft.kaist.ac.kr/landscape/detail.php?id=45004}{\#45004}) $W=q_1^2+X_1\phi^2+M_1\phi^2q_2+M_2 q_2^2,\quad a=c=1.1367$.
    
    \item (\href{https://qft.kaist.ac.kr/landscape/detail.php?id=45203}{\#45203}) $W=q_1^2+X_1\phi^2+M_1\phi^2q_2+M_2\phi^4+M_1q_2^2,\quad a=c=\frac{567}{500}$.

    \item (\href{https://qft.kaist.ac.kr/landscape/detail.php?id=45075}{\#45075}) $W=M_1\phi^2 q+ q_1^3q_2+q_2^2+X_1\phi^2+X_2 q_1^2,\quad a=c=\frac{245}{216}$.
\end{itemize}

\subsection{$SO(5)$ with One Traceless Symmetric and One Vector}
\begin{figure}[t]
    \centering
     \begin{subfigure}[b]{0.45\textwidth}
         \includegraphics[width=\linewidth]{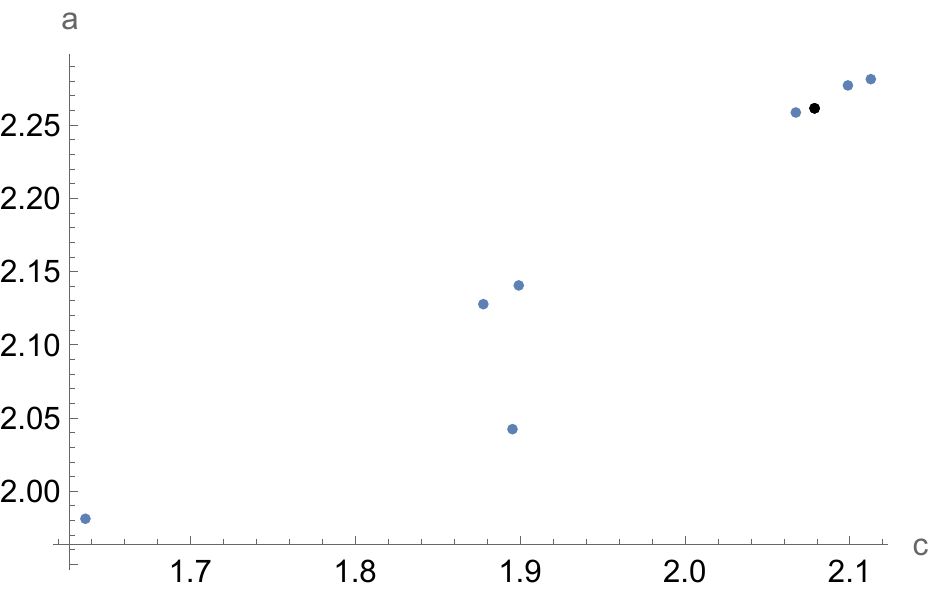}
         \caption{$a$ vs. $c$}
     \end{subfigure}
     \hspace{4mm}
     \begin{subfigure}[b]{0.45\textwidth}
         \includegraphics[width=\linewidth]{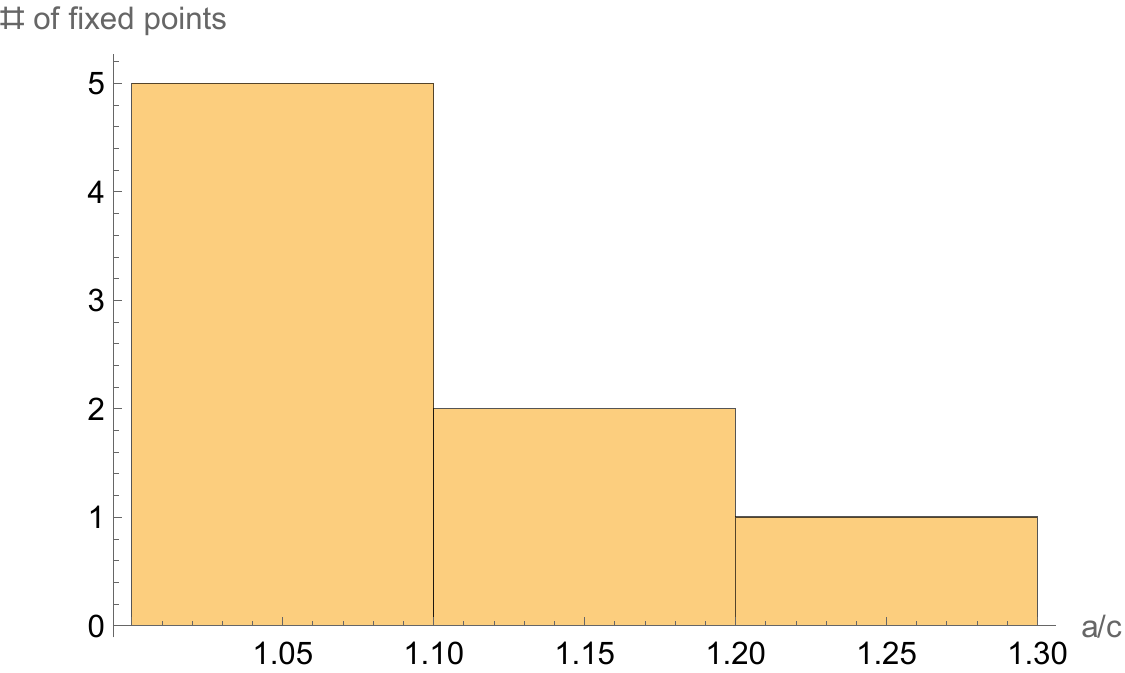}
         \caption{Histogram of $a/c$}
     \end{subfigure}
     \hfill
     \caption{(a) The plot of $a$ versus $c$ for fixed points in the landscape of $SO(5)$ theory with $N_S=N_f=1$. The black dot represents the seed theory. (b) The histogram of $a/c$.}
     \label{fig:SO5S1nf1ac}
\end{figure}

Consider the seed theory to be the IR fixed point of $SO(5)$ supersymmetric gauge theory with one traceless symmetric tensor ($S$, $\mathbf{14}$-dimensional representation) and one vector ($q$) chiral multiplets, without superpotential.
The set of fixed points spawned from this seed theory has the following features:
\begin{itemize}
    \item The maximal number of relevant deformation (modulo flipping free operators): 2.
    \item The number of inequivalent fixed points: 8.
    \item The number of pairs of inequivalent fixed points with identical central charges: 0.
    \item The number of fixed points with $a=c$: 0.
    \item The number of inequivalent fixed points with rational central charges: 6.
\end{itemize}
In this case we obtain a small set of non-trivial fixed points. 
The distribution of the central charges $a$ and $c$ is plotted in Figure \ref{fig:SO5S1nf1ac}. 
The minimal and maximal values of the central charges are given by
\begin{align}
    a_{\min}=\frac{507}{256}\simeq 1.9805,\quad a_{\max}\simeq 2.2808,\quad c_{\min}=\frac{419}{256}\simeq 1.6367,\quad c_{\max}\simeq 2.1128\,.
\end{align}
The superpotential leading the fixed point with the minimal $a$ (\href{https://qft.kaist.ac.kr/landscape/detail.php?id=10701}{\#10701}) is given by
\begin{align}
    W=M_1S^2 + M_1^2\,.\label{eq:largest ratio}
\end{align}

We see that all fixed points in this set have $a>c$. The minimal and maximal values of $a/c$ are given by
\begin{align}
    (a/c)_{\min}=\frac{98}{91}\simeq 1.0769,\quad (a/c)_{\max}=\frac{507}{419}\simeq 1.2100\,.
\end{align}
The fixed point with the superpotential (\ref{eq:largest ratio}) has a large value of $a/c\simeq 1.21$. For 4d $\CN=1$ SCFTs, the bound of central charge ratio is given by $\frac{1}{2}\leq \frac{a}{c}\leq \frac{3}{2}$ \cite{Hofman:2008ar}. However, no known example has $a/c$ close to $3/2$. The previously known largest value of $a/c$ was $1.158$, which is realized by the ISS model \cite{Intriligator:1994rx}, which is an $SU(2)$ theory with the spin-$3/2$ representation. Our $SO(5)$ gauge theory with a symmetric traceless tensor and a vector  sets a new record. We discuss this theory again in section \ref{subsec:acRecord}.

\begin{figure}[t]
    \centering
     \begin{subfigure}[b]{0.45\textwidth}
         \includegraphics[width=\linewidth]{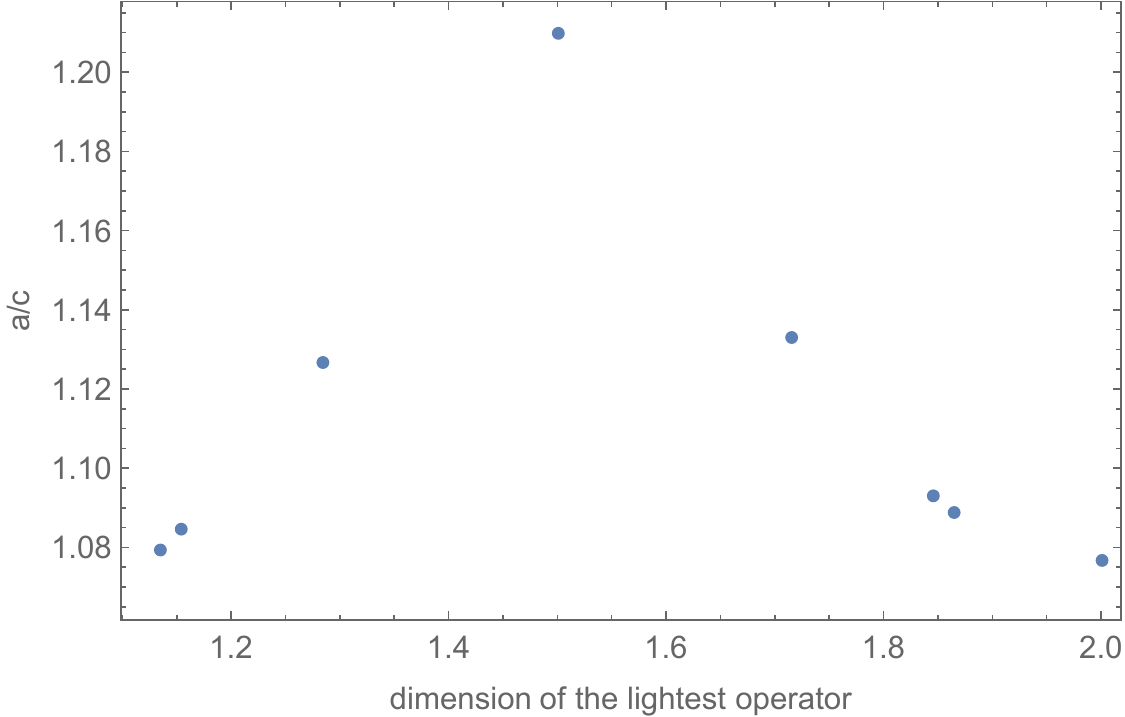}
     \end{subfigure}
     \hspace{4mm}
     \begin{subfigure}[b]{0.45\textwidth}
         \includegraphics[width=\linewidth]{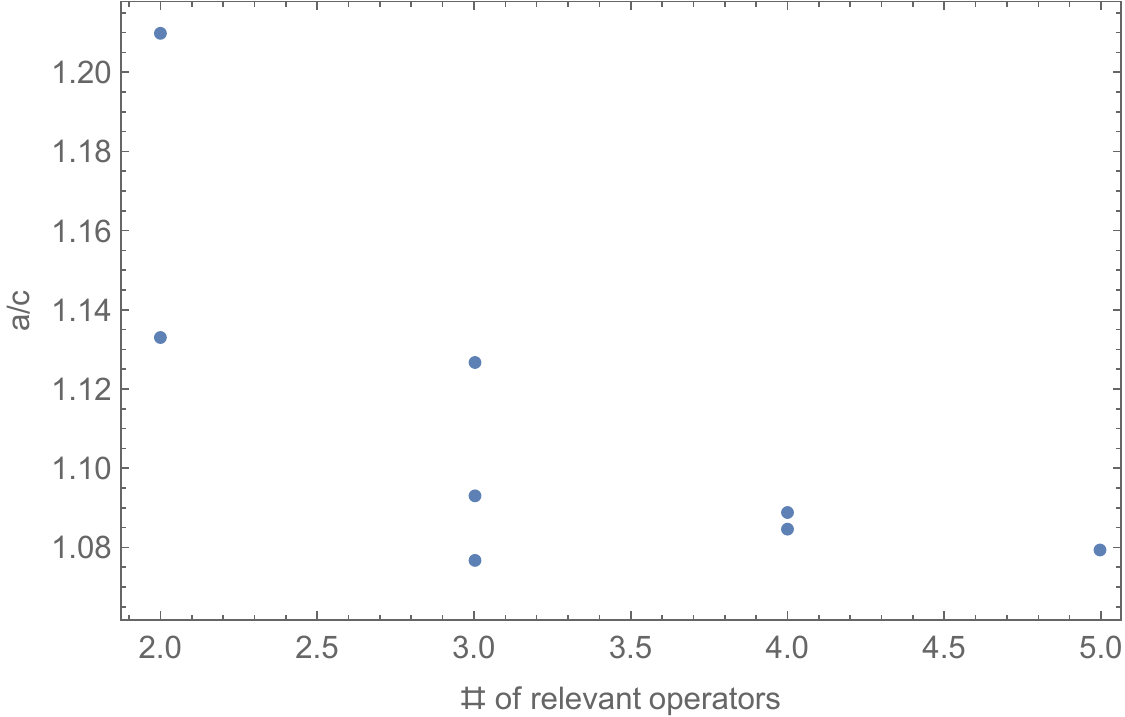}
     \end{subfigure}
     \hfill
     \caption{Left: $a/c$ versus the dimension of the lightest operator. Right: $a/c$ versus the number of relevant operators.}
     \label{fig:SO5S1nf1ratio}
\end{figure}

We also plot $a/c$ versus the dimension of the lightest operator and the number of relevant operators, respectively, in Figure \ref{fig:SO5S1nf1ratio}.

\paragraph{Conformal manifold / Flavor symmetry enhancement} We can identify some fixed points that have a non-trivial conformal manifold or accidental symmetry enhancement by examining the coefficient of $t^6$ in the superconformal index.
\begin{itemize}
    \item The number of fixed points with a positive coefficient in $t^6$ term: 1.
    \item The number of fixed points which have no flavor symmetry in UV but have a negative coefficient in $t^6$ term: 0.
\end{itemize}

\paragraph{SUSY enhancement}
There is no fixed point that satisfies the sufficient for supersymmetry enhancement. However, there is one fixed point satisfying the necessary conditions for $\CN=2$ enhancement:
\begin{itemize}
    \item $W=S^3+X_1 q^2$.
        {\renewcommand\arraystretch{1.4}
        \begin{table}[H]
            \centering
            \begin{tabular}{|c|c|c||c|c|c|}\hline
               id & $a$ & $c$ & $R_{X_1}$ &$R_{q}$ & $R_S$ \\\hline\hline
              \href{https://qft.kaist.ac.kr/landscape/detail.php?id=10699}{\#10699} & $\frac{49}{24}$& $\frac{91}{48}$ & $\frac{4}{3}$  & $\frac{1}{3}$ & $\frac{2}{3}$\\\hline
            \end{tabular}
        \end{table}}
\end{itemize} 
At this fixed point, the coefficient of $t^6$ is 0 and no marginal operator exists. This means that there is no flavor symmetry in view of $\CN=1$ supersymmetry. Consequently, this fixed point lacks a symmetry that could enhance to the $\CN=2$ R-symmetry in IR.

\subsection{$SO(5)$ with One Traceless Symmetric and Two Vectors}
We start with the seed theory given by $SO(5)$ supersymmetric gauge theory with one traceless symmetric ($S$, $\mathbf{14}$-dimensional representation) and two vector ($q$) chiral multiplets, without superpotential. This theory is a conformal gauge theory with vanishing 1-loop beta function for the gauge coupling, but has a non-trivial conformal manifold \cite{Razamat:2020pra}.
The set of fixed points obtained by deforming this theory have the following features:
\begin{itemize}
    \item The maximal number of relevant deformations (modulo flipping free operators): 3.
    \item The number of inequivalent fixed points: 9.
    \item The number of pairs of inequivalent fixed points with identical central charges: 0.
    \item The number of fixed points with $a=c$: 0.
    \item The number of inequivalent fixed points with rational central charges: 7.
\end{itemize}
It turns out that eight out of nine fixed points, except for the seed theory, are equivalent to the fixed points in the landscape of the $N_f=1$ case. Therefore there is no genuinely new theory in this set except for the seed theory, which is given as:
\begin{itemize}
    \item $W=0$.
        {\renewcommand\arraystretch{1.4}
            \begin{table}[H]
                \centering
                \begin{tabular}{|c|c|c||c|c|}\hline
                   id & $a$ & $c$ & $R_{q_i}$ & $R_S$ \\\hline\hline
                  \href{https://qft.kaist.ac.kr/landscape/detail.php?id=10705}{\#10705} & $\frac{19}{8}$& $\frac{9}{4}$  & $\frac{2}{3}$, $\frac{2}{3}$ & $\frac{2}{3}$\\\hline
                \end{tabular}
            \end{table}}
\end{itemize}

\subsection{$Sp(2)$ with One 14 and Two Fundamentals} \label{subsec:Sp2nf2}
\begin{figure}[t]
    \centering
     \begin{subfigure}[b]{0.45\textwidth}
         \includegraphics[width=\linewidth]{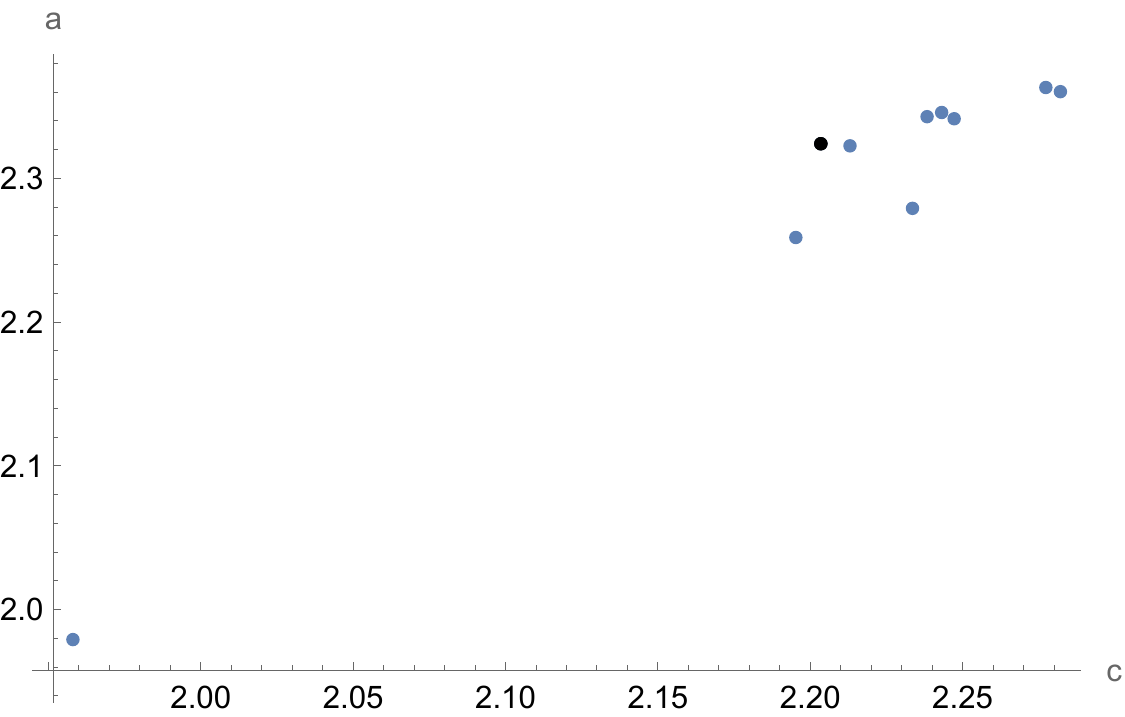}
         \caption{$a$ vs. $c$}
     \end{subfigure}
     \hspace{4mm}
     \begin{subfigure}[b]{0.45\textwidth}
         \includegraphics[width=\linewidth]{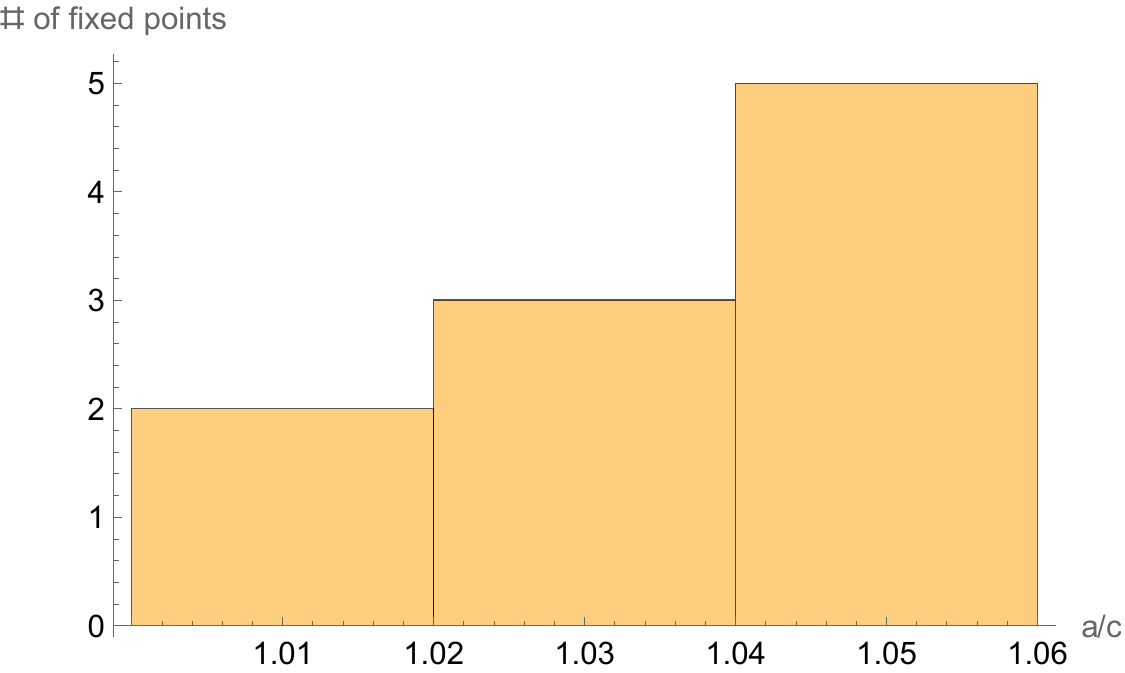}
         \caption{Histogram of $a/c$}
     \end{subfigure}
     \hfill
     \caption{(a) $a$ versus $c$ for the fixed points in the landscape of $Sp(2)$ theory with $N_{\mathbf{14}}=N_{\mathbf{4}}=1$. The black dot represents the seed theory. (b) The histogram of $a/c$.}
     \label{fig:SO5S1sp1ac}
\end{figure}

We start with the seed theory given by the IR fixed point of $Sp(2)=Spin(5)$ supersymmetric gauge theory with one $\mathbf{14}$-dimensional representation ($S$) and two fundamental ($Q_{1,2}$, $\mathbf{4}$-dimensional) chiral multiplets, without superpotential.
The set of fixed points obtained via deforming this seed theory theory has following features:
\begin{itemize}
    \item The maximal number of relevant deformations (modulo flipping decoupled operators): 3.
    \item The number of inequivalent fixed points: 10.
    \item The number of pairs of inequivalent fixed points with identical central charges: 0.
    \item The number of fixed points with $a=c$: 0.
    \item The number of inequivalent fixed points with rational central charges: 6.
\end{itemize}
We plot the distribution of the central charges $a$ and $c$ is in Figure \ref{fig:SO5S1sp1ac}. 
The minimal and maximal values of the central charges are given by
\begin{align}
\begin{split}
    a_{\min}&=\frac{95}{48}\simeq 1.9792,\quad a_{\max}\simeq 2.3637,\quad \\ 
    c_{\min}&=\frac{47}{24}\simeq 1.9583,\quad c_{\max}=\frac{18695}{8192}\simeq 2.821 \ . 
\end{split}
\end{align}
The fixed point theory with the minimal $a$ (\href{https://qft.kaist.ac.kr/landscape/detail.php?id=45748}{\#45748}) is given by the superpotential
\begin{align}
    W=S^3 + X_1 Q_1 Q_2\,.
\end{align}
We see that all fixed points have $a>c$. The minimal and maximal values of $a/c$ are given by
\begin{align}
    (a/c)_{\min}=\frac{95}{94}\simeq 1.0106,\quad (a/c)_{\max}\simeq 1.0547\,.
\end{align}
 \begin{figure}[t]
    \centering
     \begin{subfigure}[b]{0.45\textwidth}
         \includegraphics[width=\linewidth]{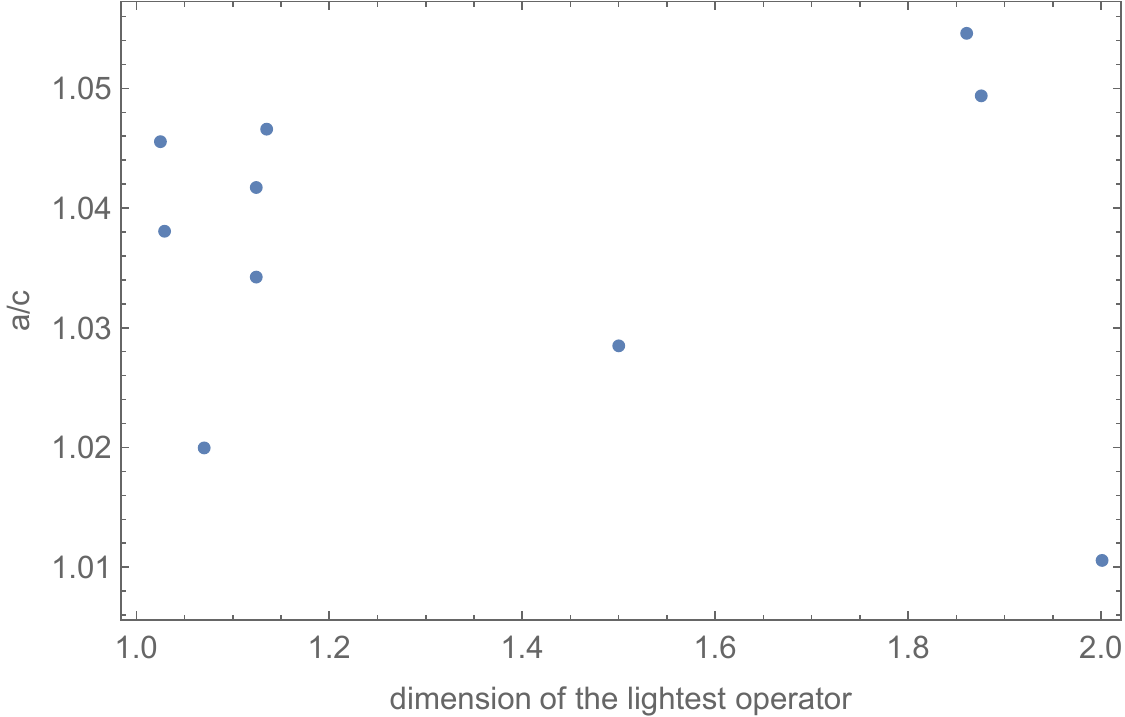}
     \end{subfigure}
     \hspace{4mm}
     \begin{subfigure}[b]{0.45\textwidth}
         \includegraphics[width=\linewidth]{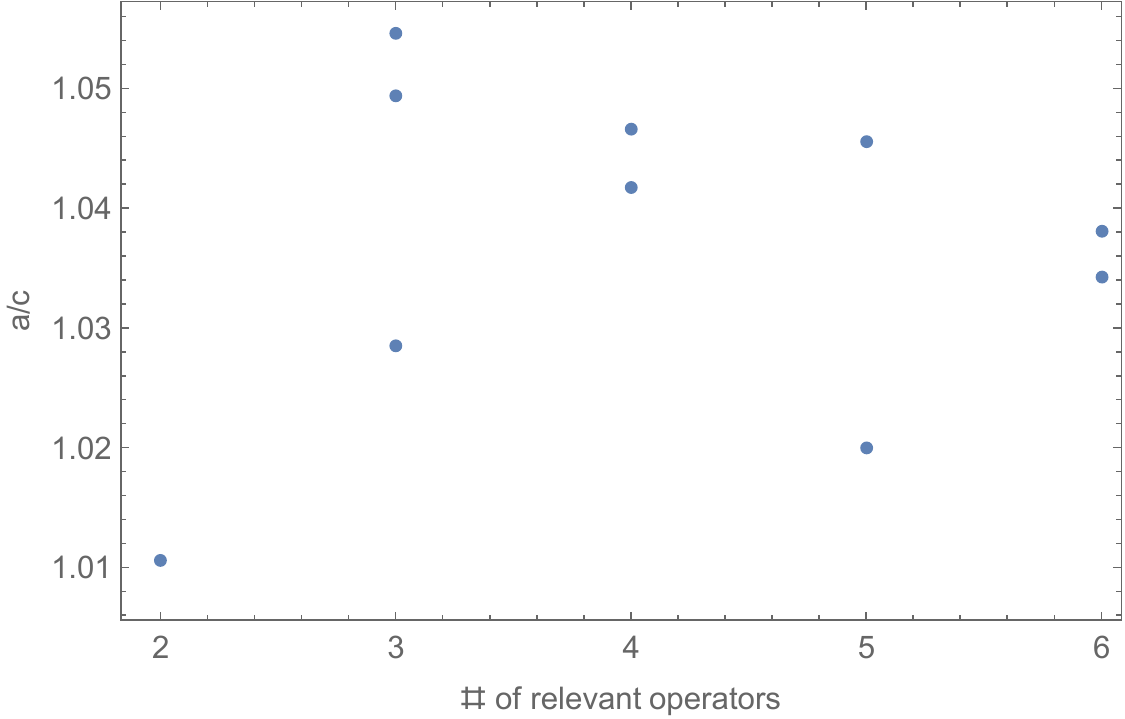}
     \end{subfigure}
     \hfill
     \caption{Left: $a/c$ versus the dimension of the lightest operator. Right: $a/c$ versus the number of relevant operators.}
     \label{fig:SO5S1sp1ratio}
\end{figure}
We also plot $a/c$ versus the dimension of the lightest operator and the number of relevant operators, respectively, in Figure \ref{fig:SO5S1sp1ratio}.

\paragraph{Conformal manifold / Flavor symmetry enhancement} 
We find all of the fixed points in this set have a vanishing coefficient of $t^6$ in the superconformal index. Therefore there is no clear sign of a conformal manifold nor of symmetry enhancement in this set. 

\paragraph{SUSY enhancement}
  There is no fixed point that satisfies the sufficient condition for  supersymmetry enhancement. However, there is one fixed point that satisfies the necessary conditions for  $\CN=2$ enhancement whose superpotential is given as,
  \begin{align}
    W=S^3+X_1 Q_1 Q_2 \ . 
  \end{align}
  Its central charges and R-charges are given as follows:
    {\renewcommand\arraystretch{1.4} 
        \begin{table}[H]
            \centering
            \begin{tabular}{|c|c|c||c|c|c|}\hline
               id & $a$ & $c$ & $R_{X_1}$ & $R_{Q_{i}}$ & $R_S$ \\\hline\hline
               \href{https://qft.kaist.ac.kr/landscape/detail.php?id=45748}{\#45748} & $\frac{95}{48}$& $\frac{47}{24}$ & $\frac{4}{3}$ & $\frac{1}{3}$, $\frac{1}{3}$  & $\frac{2}{3}$\\\hline
            \end{tabular}
        \end{table}
    }
  \noindent This theory is identical to the one with the maximal $a/c$ we discussed above. We will discuss  this fixed point further in section \ref{subsec:newN2} as a candidate  new $\CN=2$ SCFT.

\subsection{$Sp(2)$ $N_f=1$ Adjoint SQCD}
\begin{figure}[t]
    \centering
     \begin{subfigure}[b]{0.45\textwidth}
         \includegraphics[width=\linewidth]{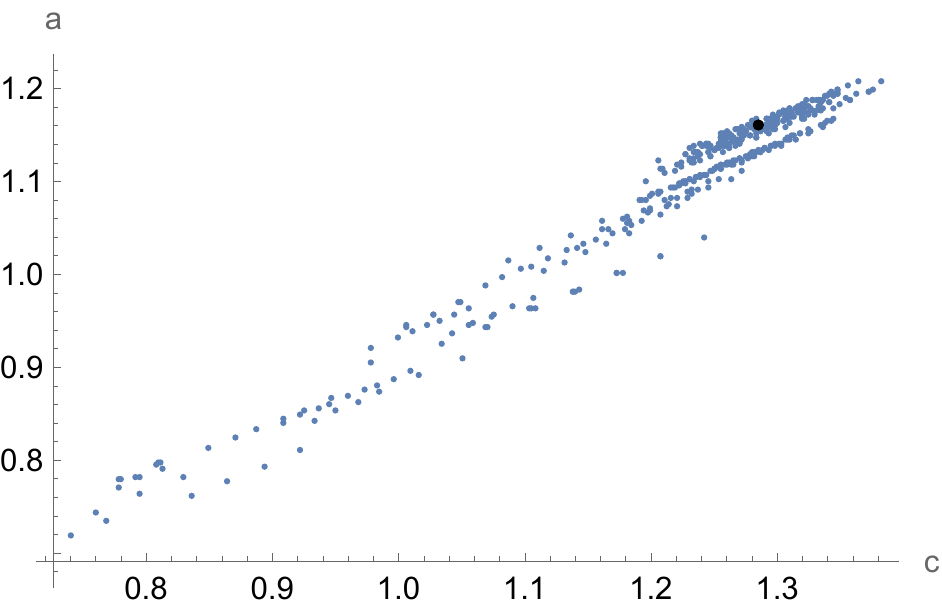}
         \caption{$a$ vs. $c$}
     \end{subfigure}
     \hspace{4mm}
     \begin{subfigure}[b]{0.45\textwidth}
         \includegraphics[width=\linewidth]{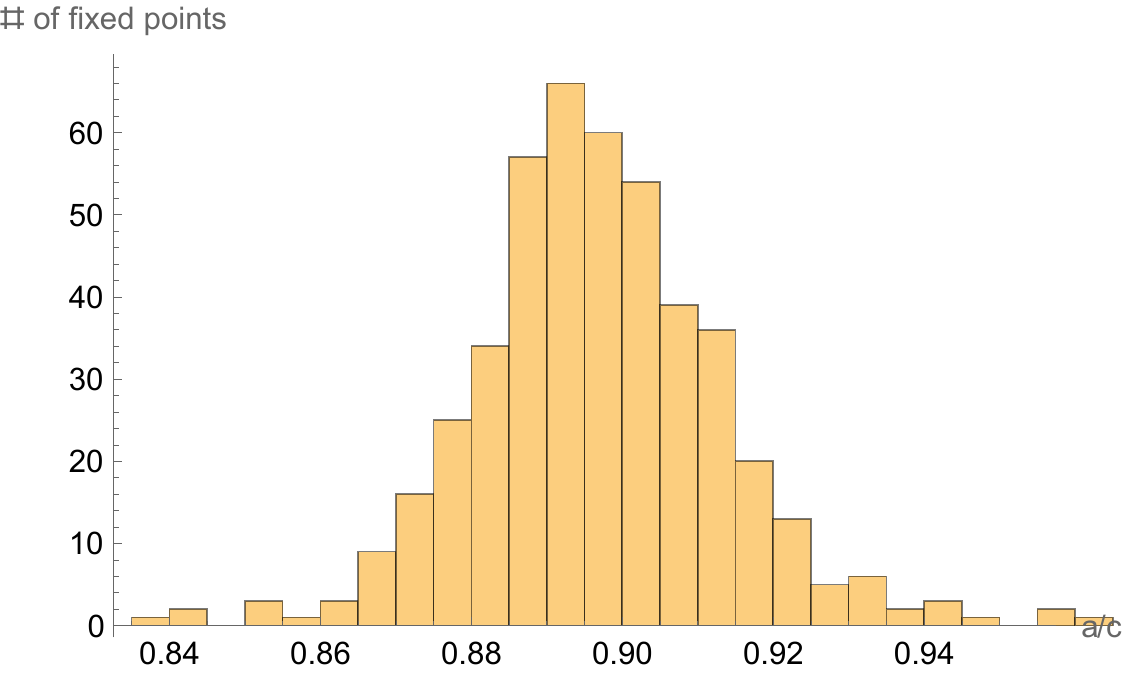}
         \caption{Histogram of $a/c$}
     \end{subfigure}
     \hfill
     \caption{(a) The plot of $a$ versus $c$ for fixed points in the landscape of $Sp(2)$ $N_f=1$ adjoint SQCD. The black dot represents the fixed point without superpotential. (b) The histogram of $a/c$.}
     \label{fig:Sp2adj1nf1ac}
\end{figure}

\begin{figure}[t]
    \centering
     \begin{subfigure}[b]{0.45\textwidth}
         \includegraphics[width=\linewidth]{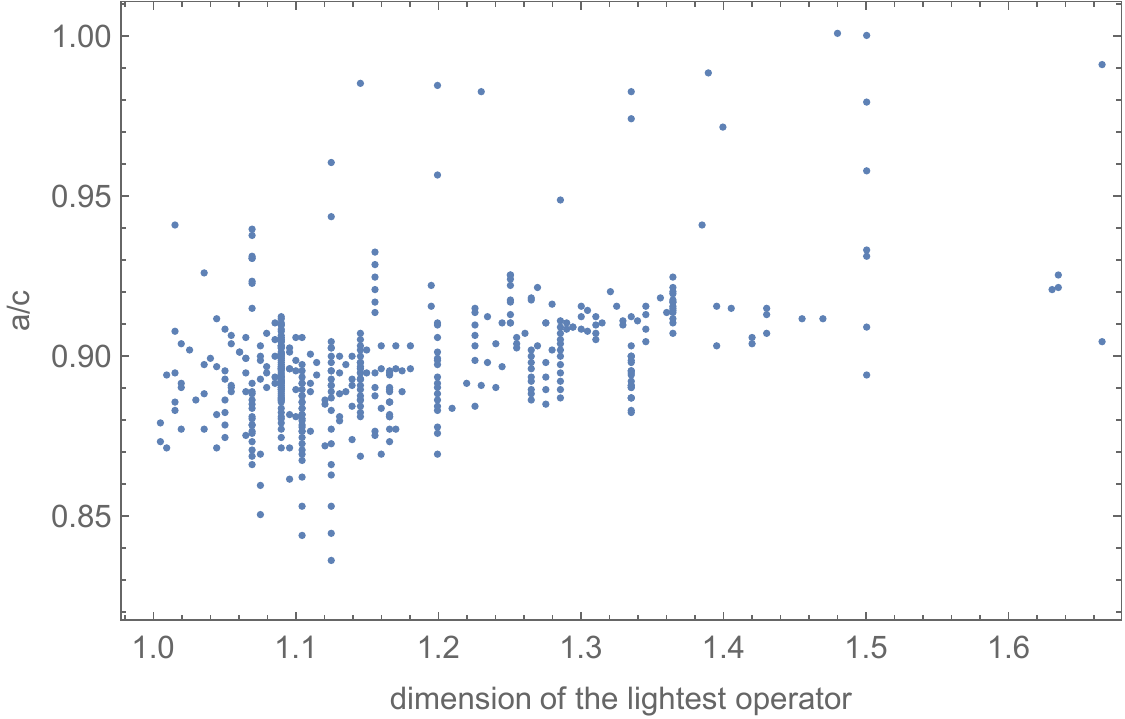}
     \end{subfigure}
     \hspace{4mm}
     \begin{subfigure}[b]{0.45\textwidth}
         \includegraphics[width=\linewidth]{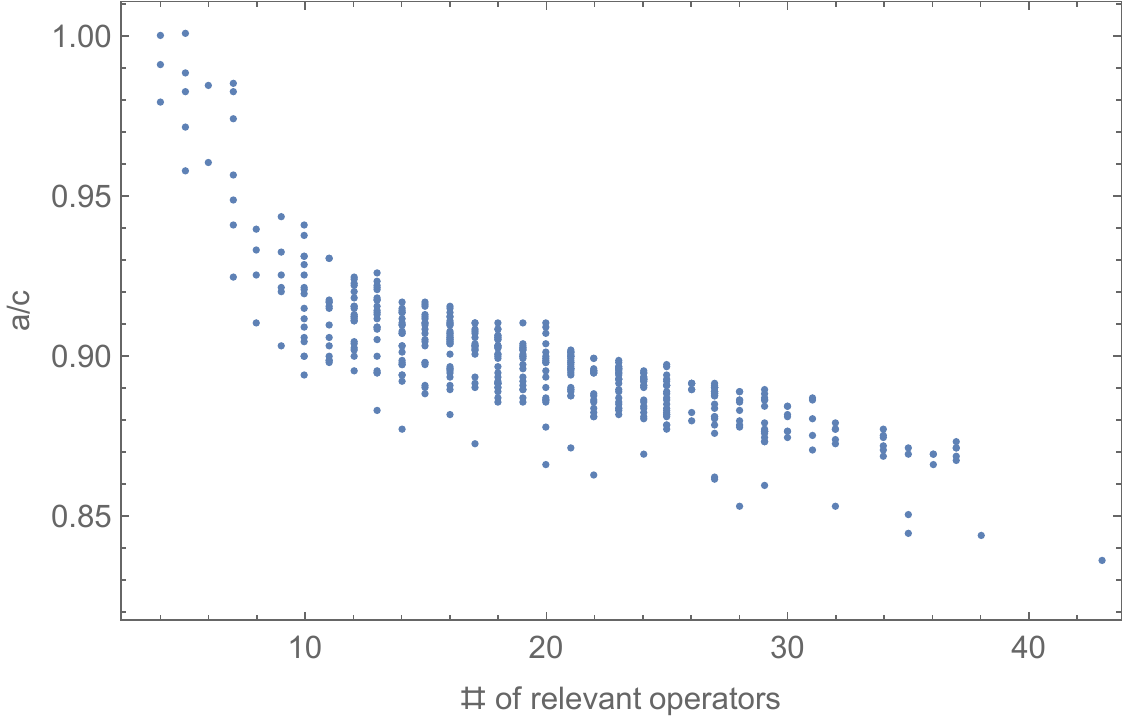}
     \end{subfigure}
     \hfill
     \caption{Left: $a/c$ versus the dimension of the lightest operator. Right: $a/c$ versus the number of relevant operators.}
     \label{fig:Sp2adj1nf1ratio}
\end{figure}

We start with the seed theory given by the IR fixed point of $Sp(2)$ supersymmetric gauge theory with one adjoint ($\phi$, $\mathbf{10}$-dimensional) and two fundamental ($Q_{1,2}$) chiral multiplets, without superpotential.  
The set of fixed points obtained from this theory have the following features:
\begin{itemize}
    \item The maximal number of relevant deformations (modulo flipping decoupled operators): 11.
    \item The number of inequivalent fixed points: 470.
    \item The number of pairs of inequivalent fixed points with identical central charges: 0.
    \item The number of fixed points with $a=c$: 1.
    \item The number of fixed points with $a>c$: 1.
    \item The number of inequivalent fixed points with rational central charges: 272.
\end{itemize}
The distribution of the central charges $a$ and $c$ is plotted in Figure \ref{fig:Sp2adj1nf1ac}. 
The minimal and maximal values of the central charges are given by
\begin{align}
    a_{\min}=\frac{4313}{6000}\simeq 0.7188,\quad a_{\max}\simeq 1.2082,\quad c_{\min}=\frac{2219}{3000}\simeq 0.7397,\quad c_{\max}\simeq 1.3827 \ . 
\end{align}
The fixed point with the minimal central charge $a$ (\href{https://qft.kaist.ac.kr/landscape/detail.php?id=8644}{\#8644}) is given by the superpotential
\begin{align}
    W=X_1\Tr\phi^2+ M_1\phi Q_1 Q_2 + M_2 \phi^2 Q_1^2 + M_1M_2+ \phi Q_2^2+ X_2\Tr\phi^4+ X_3 M_1\,.
\end{align}
The minimal and maximal values of $a/c$ are given by
\begin{align}
    (a/c)_{\min}=\frac{2127}{2543}\simeq 0.8364,\quad (a/c)_{\max} \simeq 1.0011\,.
\end{align}
We also plot $a/c$ versus the dimension of the lightest operator and the number of relevant operators, respectively, in Figure \ref{fig:Sp2adj1nf1ratio}.
We see a similar pattern as in the previous examples. 

\paragraph{Conformal manifold / Flavor symmetry enhancement} 
The fixed points that have a non-trivial conformal manifold or accidental symmetry enhancement can be found by examining the coefficient of $t^6$ in the superconformal index, with the following results.
\begin{itemize}
    \item The number of fixed points with a positive coefficient in $t^6$ term: 104.
    \item The number of fixed points which have no flavor symmetry in UV but have a negative coefficient in $t^6$ term: 106.
\end{itemize}
Therefore, we have at least 104 fixed points with non-trivial conformal manifold and at least 106 fixed points with flavor symmetry enhancement. 

\paragraph{SUSY enhancement}
We find only one fixed point that satisfies the sufficient condition or necessary condition for $\CN=2$ supersymmetry enhancement:
\begin{itemize}
    \item $W=X_1\Tr\phi^2+M_1 \phi Q_1^2 + \phi Q_2^2 + X_2\Tr\phi^4+M_2\phi^3 Q_1^2$. The fixed point corresponds to the $\CN=2$ $(A_1,A_4)$ Argyres-Douglas theory, which was found in \cite{Maruyoshi:2016aim}. 
        {\renewcommand\arraystretch{1.4}
        \begin{table}[H]
            \centering
            \begin{tabular}{|c|c|c||c|c|c|c|}\hline
                id&$a$ & $c$ & $R_{X_i}$ &  $R_{M_i}$ & $R_{Q_i}$ & $R_\phi$ \\\hline\hline
             \href{https://qft.kaist.ac.kr/landscape/detail.php?id=8491}{\#8491} & $\frac{67}{84}$& $\frac{17}{21}$ & $\frac{38}{21}$, $\frac{34}{21}$ & $\frac{20}{21}$, $\frac{16}{21}$ & $\frac{10}{21}$, $\frac{20}{21}$  & $\frac{2}{21}$\\\hline
            \end{tabular}
        \end{table}}
\end{itemize}

\paragraph{$a=c$ theory}
We find one fixed point with equal central charges $a=c$, which is given by:
\begin{align}
     W=X_1 \Tr \phi^2 + \phi Q_1^2+X_2\Tr\phi^4+\phi^2 Q_2^4\,.
\end{align}
{\renewcommand\arraystretch{1.4}
        \begin{table}[H]
            \centering
            \begin{tabular}{|c|c|c||c|c|c|}\hline
               id& $a$ & $c$ & $R_{X_i}$ & $R_{Q_i}$ & $R_\phi$ \\\hline\hline
              \href{https://qft.kaist.ac.kr/landscape/detail.php?id=8251}{\#8251} & $\frac{6237}{8000}$& $\frac{6237}{8000}$ & $\frac{9}{5}$, $\frac{8}{5}$& $\frac{19}{20}$, $\frac{9}{20}$  & $\frac{1}{10}$\\\hline
            \end{tabular}
        \end{table}}

\subsection{$G_2$ $N_f=1$ Adjoint SQCD}
\begin{figure}[t]
    \centering
     \begin{subfigure}[b]{0.45\textwidth}
         \includegraphics[width=\linewidth]{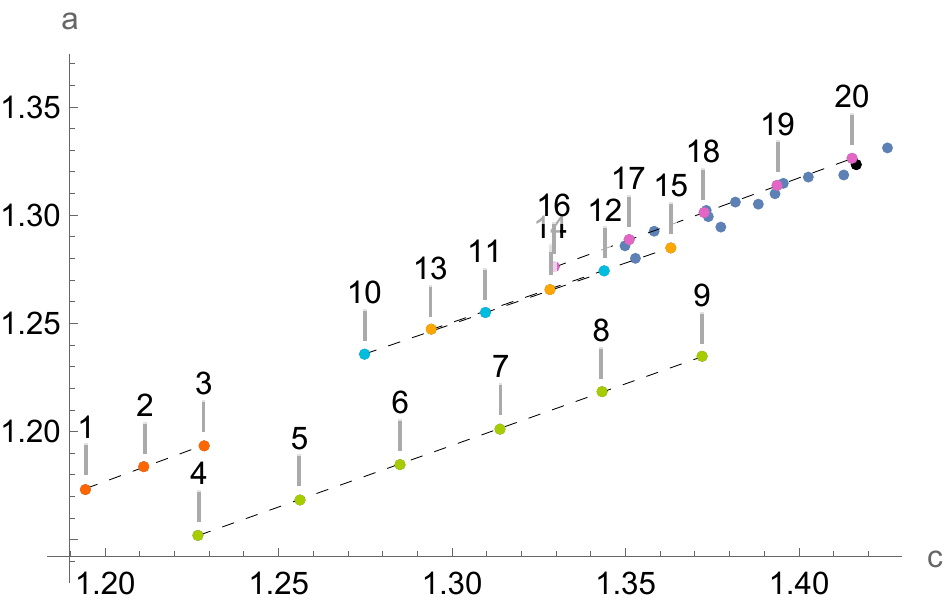}
         \caption{$a$ vs. $c$}
     \end{subfigure}
     \hspace{4mm}
     \begin{subfigure}[b]{0.45\textwidth}
         \includegraphics[width=\linewidth]{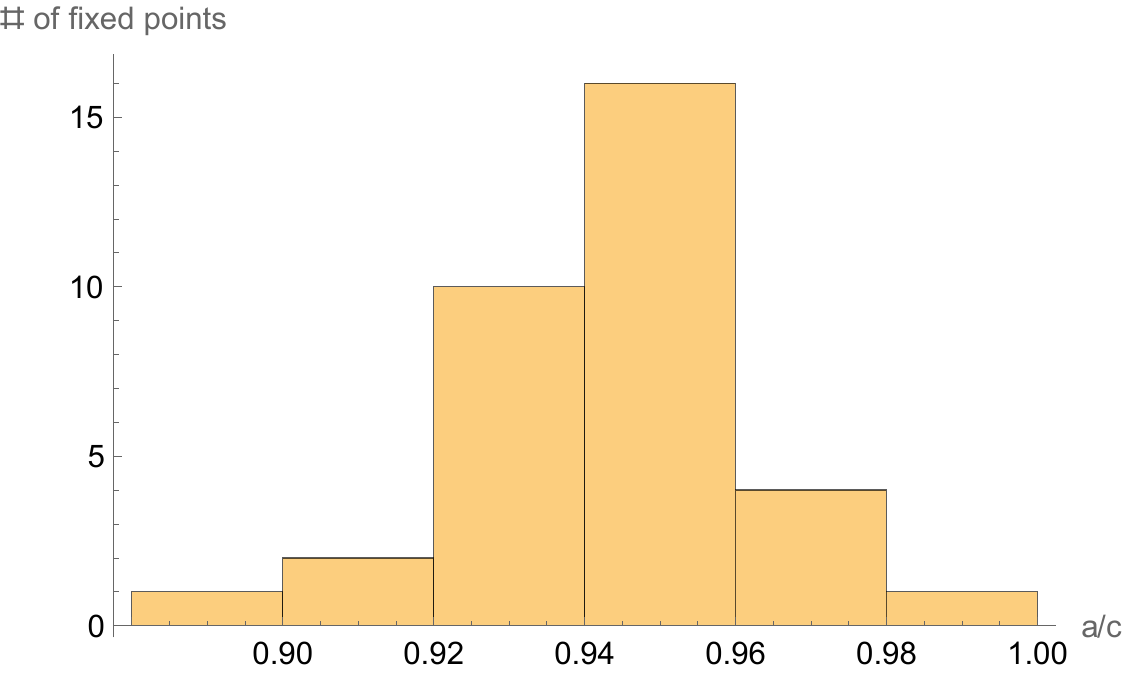}
         \caption{Histogram of $a/c$}
     \end{subfigure}
     \hfill
     \caption{(a) $a$ versus $c$ for the fixed points in the landscape of $G_2$ $N_f=1$ adjoint SQCD. The black dot corresponds to the seed theory. The colored dots, labeled by an integer $i$, represent the fixed points with superpotentials $W_i$, lie on the dashed lines. (b) The histogram of $a/c$.}
     \label{fig:G2adj1nf1ac}
\end{figure}

Let us start with the seed theory given by the IR fixed point of $G_2$ supersymmetric gauge theory with one adjoint ($\phi$, $\mathrm{14}$-dimensional) and one fundamental ($q$, $\mathrm{7}$-dimensional) chiral multiplet, without superpotential. For this theory, $\tr\phi^2$ gets decoupled along the flow, so we remove this by flipping. 
The set of fixed points spawned from this theory via deformations have the following features:
\begin{itemize}
    \item The maximal number of relevant deformations (modulo flipping free operators): 4.
    \item The number of inequivalent fixed points: 34.
    \item The number of pairs of inequivalent fixed points with identical central charges: 0.
    \item The number of fixed points with $a=c$: 0.
    \item The number of inequivalent fixed points with rational central charges: 22.
\end{itemize}
We plot the distribution of the central charges $a$ and $c$ in Figure \ref{fig:G2adj1nf1ac}. 
The minimal and maximal values of the central charges are given by
\begin{align}
    a_{\min}=\frac{144}{125}\simeq 1.152,\quad a_{\max}\simeq 1.3311,\quad c_{\min}=\frac{43}{36}\simeq 1.1944,\quad c_{\max}\simeq 1.4257 \ . 
\end{align}
The fixed point with the minimal central charge $a$ (\href{https://qft.kaist.ac.kr/landscape/detail.php?id=8351}{\#8351}) is given by the superpotential
\begin{align}
    W=X_1\phi^2+ \phi^9q+X_2q^2+M_1\phi^2q^2+M_2\phi^3q\,.
\end{align}
We find all the fixed points in this set have $a<c$. 
The minimal and maximal values of $a/c$ are given by
\begin{align}
    (a/c)_{\min}=\frac{2469}{2744}\simeq 0.8998,\quad (a/c)_{\max}=\frac{169}{172}\simeq 0.9826\,.
\end{align}

In this set, we have found five sets of fixed points exactly aligned on straight lines, as illustrated in Figure \ref{fig:G2adj1nf1ac}(a). They are given as follows: 
\begin{itemize}
    \item The fixed points lying on the line of $a=\frac{13}{22}c+\frac{247}{528}$:
    \begin{align}
        W_2=X_1\phi^2+\phi^3q^3+X_2\phi^6,\quad W_1=W_2+M_1\phi^3q,\quad W_3=W_2+M_1q^2\,.
    \end{align}
    
    \item The fixed points lying on a line of $a=\frac{33}{58}c+\frac{1053}{2320}$:
        \begin{equation}
    \begin{aligned}
        W_6&=X_1\phi^2+\phi^9 q + X_2 q^2, &W_5&=W_6+ M_1\phi^2q^2\,,\\
        W_4&= W_6+ M_1\phi^2q^2+M_2\phi^3q, &W_7&=W_6+M_1\phi^4q^2\,,\\
        W_8&=W_6 +M_1\phi^4q^2+M_2\phi^6, &W_9&=W_6+M_1\phi^4q^2+M_2\phi^6+M_3\phi^3q^3\,.
    \end{aligned}
    \end{equation}

    \item The fixed points lying on the line of $a=\frac{39}{71}c+\frac{19467}{36352}$:
    \begin{align}
        W_{11}=X_1\phi^2 + q^4 + M_1\phi^3 q,\quad W_{10}=W_{11}+M_2\phi^6,\quad W_{12}=W_{11}+M_2\phi^2q^2\,.
    \end{align}

    \item The fixed points lying on the line of $a=\frac{39}{71}c+\frac{9747}{18176}$:
    \begin{align}
        W_{14}=X_1\phi^2 + q^4,\quad W_{13}=W_{14}+M_1\phi^6,\quad W_{15}=W_{14}+M_1\phi^2q^2\,.
    \end{align}

    \item The fixed points lying on the line of $a=\frac{69}{118}c+\frac{1647}{3304}$:
    \begin{equation}
    \begin{aligned}
        W_{19}&=X_1\phi^2+\phi^5q^3, &W_{18}&=W_{19}+ M_1\phi^6\,,\\
        W_{17}&= W_{19}+ M_1\phi^6 + M_2 q^2, &W_{16}&=W_{19}+ M_1\phi^6 + M_2 q^2 + M_3\phi^3 q\,,\\
        W_{20}&=W_{19}+M_1\phi^2q^2\,.
    \end{aligned}
    \end{equation}
\end{itemize}
The fixed points that lives on the same line are connected by flipping of an operator with no flavor symmetry, as we have discussed in section \ref{subsec:amax}.

\begin{figure}[t]
    \centering
     \begin{subfigure}[b]{0.45\textwidth}
         \includegraphics[width=\linewidth]{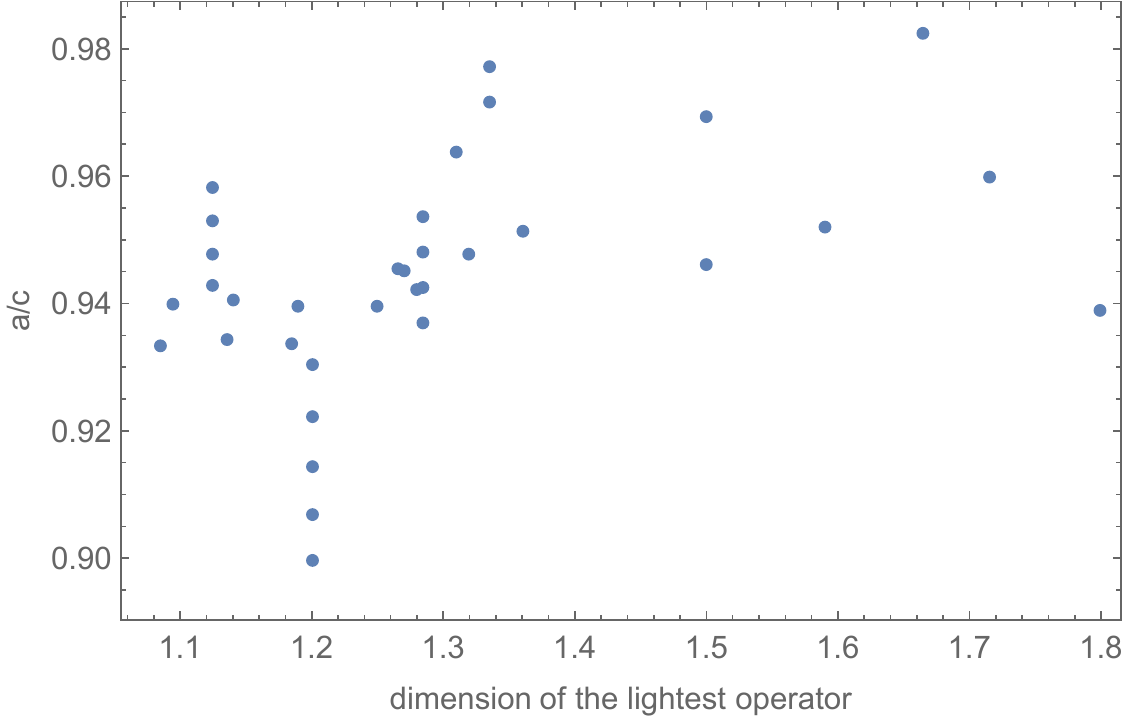}
     \end{subfigure}
     \hspace{4mm}
     \begin{subfigure}[b]{0.45\textwidth}
         \includegraphics[width=\linewidth]{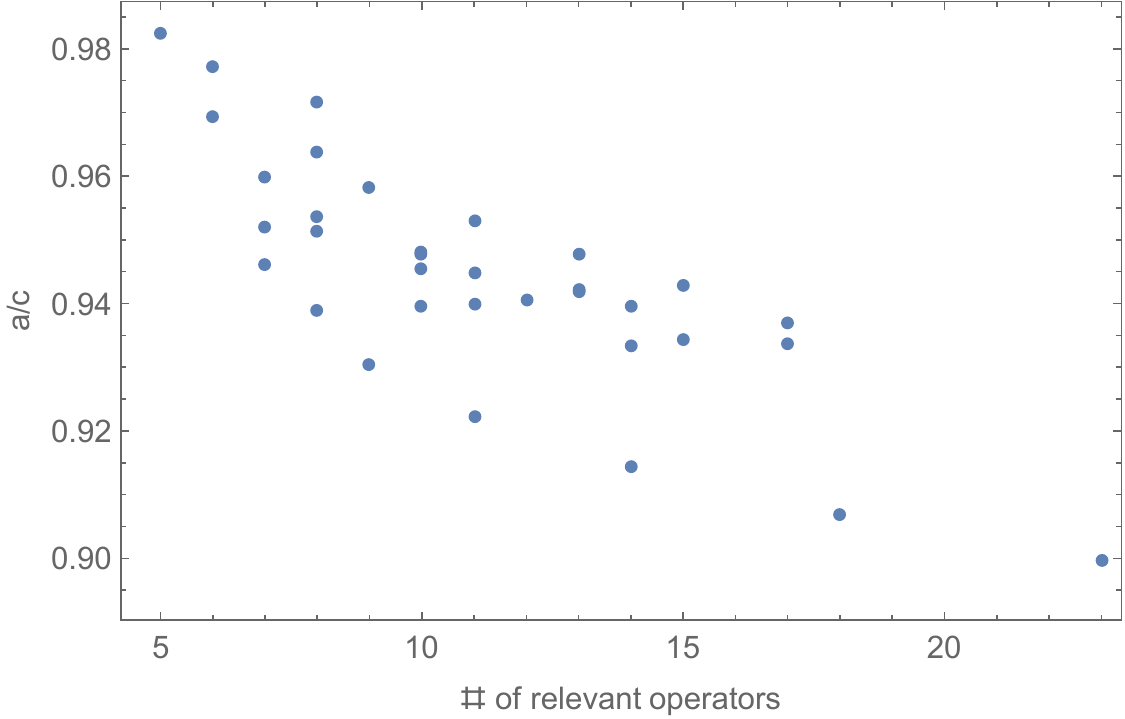}
     \end{subfigure}
     \hfill
     \caption{Left: $a/c$ versus the dimension of the lightest operator. Right: $a/c$ versus the number of relevant operators.}
     \label{fig:G2adj1nf1ratio}
\end{figure}

We also plot $a/c$ versus the dimension of the lightest operator and the number of relevant operators, respectively, in Figure \ref{fig:G2adj1nf1ratio}.
Even with a small number of fixed points, we see a similar overall pattern analogous to the previous examples with ample fixed points. 

\paragraph{Conformal manifold / Flavor symmetry enhancement} 
Here are the number of fixed points with non-vanishing coefficient of $t^6$ in the superconformal index:
\begin{itemize}
    \item The number of fixed points with a positive coefficient in $t^6$ term: 12.
    \item The number of fixed points which have no flavor symmetry in UV but have a negative coefficient in $t^6$ term: 4.
\end{itemize}
This means that there are at least 12 fixed points with non-trivial conformal manifolds, and at least 4 fixed points with emergent flavor symmetry in the IR. 

\paragraph{SUSY enhancement}
We find there is no fixed point in this set that satisfies the sufficient or necessary conditions for the supersymmetry enhancement.

\subsection{Patterns in the Combined SCFT Landscape} \label{subsec:landscape}
In this subsection, we combine all the data points we obtained so far to search for universal properties in the landscape of superconformal field theories. 

\begin{figure}[t]
    \centering
    \includegraphics[width=.9\linewidth]{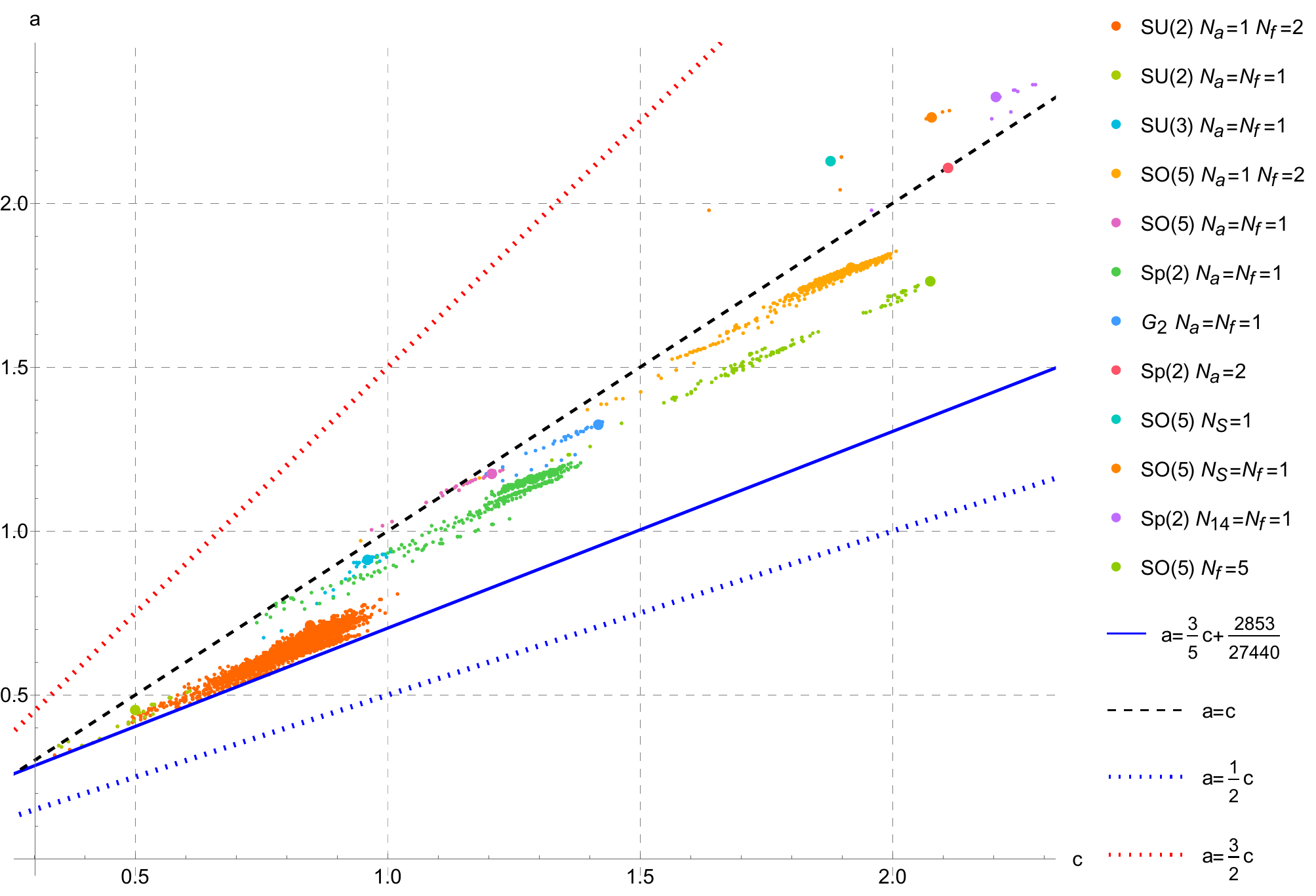}
    \caption{The plot of $a$ versus $c$ for the entire set of fixed points we obtained. The large dots are the seed theories. Here $N_f$, $N_a$, and $N_S$ represents the number of fundamental, adjoint, and rank-2 symmetric chiral multiplets, respectively. $N_{14}$ is the number of chiral multiplets in \textbf{14} representations in $Sp(2)$. The dashed line denotes $a=c$ slice, and the blue solid line denotes a line below which no fixed point exists.}
    \label{fig:acplots}
  \end{figure}

\paragraph{$a$ vs $c$}
  Let us combine all the non-trivial fixed points we obtained from the most general deformations of various gauge groups and matter fields. In addition to the set of fixed points we covered in this section, we also include the fixed points of deformed $Sp(2)$ gauge theory with two \textbf{10} chiral multiplets, $SO(5)$ gauge theory with one \textbf{14} chiral multiplet, and $SO(5)$ $N_f=5$ SQCD. For the case of $Sp(2)$ with two \textbf{10} multiplets, it contains only two fixed points, with the seed theory having identical central charges $a=c$. Also the case of $SO(5)$ with one \textbf{14} contains only two fixed points, both with central charges $a>c$. We plot $a$ versus $c$ for all of these fixed points in Figure \ref{fig:acplots}.

  Let us make a remark for the case of $SU(2)$ $N_f=2$ adjoint SQCD and $SO(5)$ $N_f=5$ SQCD. The full search of the landscape is not yet completed for these cases, because there are too many relevant deformations. Therefore, the fixed points shown in the figure represent only a subset of the full SCFT landscape obtained from these gauge theories. Specifically, for the case of $SO(5)$ $N_f=5$ SQCD, we only examined fixed points consisting up to three superpotential terms, which is a very small subset.

  In Figure \ref{fig:acplots}, we also plot the lower bound line (blue colored) which has slope of $3/5$. It is interesting to note that this slope is identical to the lowest bound for the change of $a/c$ via flipping an operator, as we discussed around equation \eqref{acflip}. 
  We see that the fixed points lie in a rather narrow region within the allowed unitarity bounds of $\half \le \frac{a}{c}  \le \frac{3}{2}$. 
  Generally, for a given seed theory ($\CT$), we observe that the corresponding descendant fixed points (or the landscape of $\CT$-fixed points) are narrowly distributed with a slope of approximately (greater than) $3/5$. We also notice that each of them are widely distributed along the slope with small thickness. 

  Most of our fixed points have $a<c$, but there are numerous cases with $a>c$. In particular, the set of fixed points originating from $SO(5)$ gauge theory with one rank-2 symmetric ($\mathbf{14}$-dimensional, which is also traceless) chiral multiplet all have central charges $a>c$. 

\begin{figure}[t]
    \centering
    \includegraphics[width=0.5\linewidth]{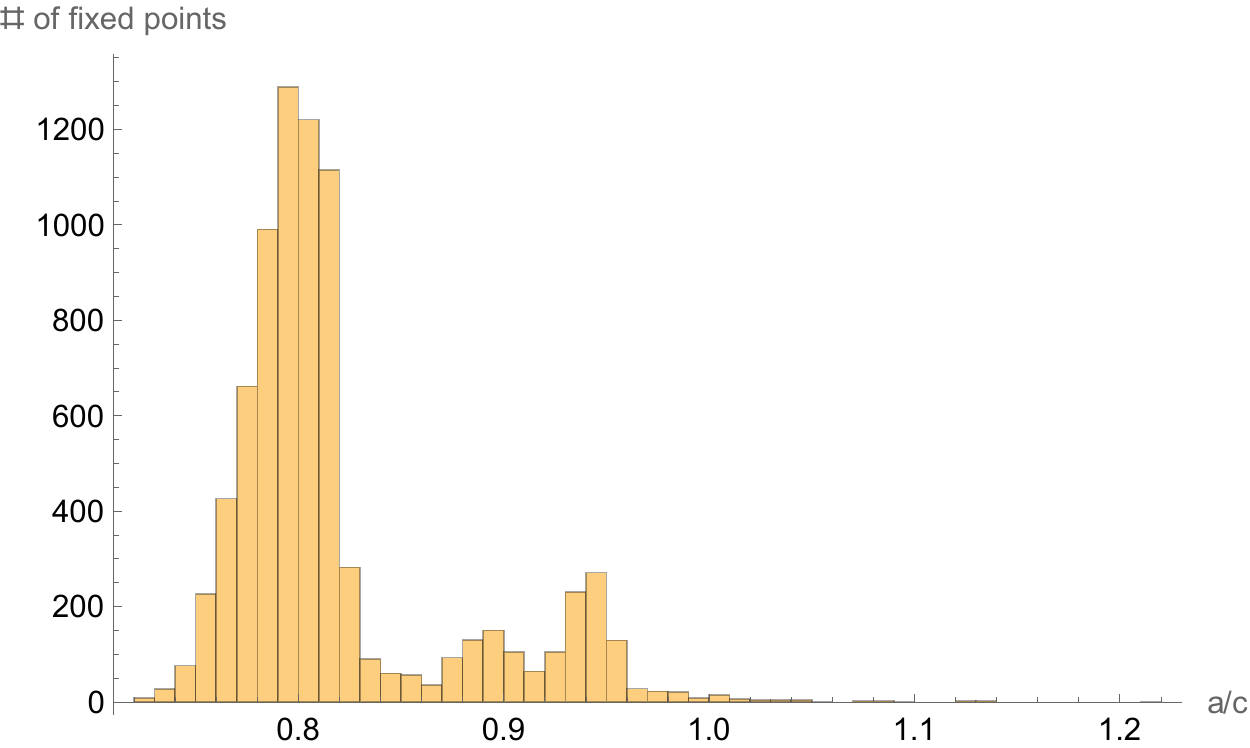}
    \caption{The histogram of $a/c$ for all fixed points in Figure \ref{fig:acplots}.}
    \label{fig:combining histogram}
\end{figure}

We also plot the histogram of $a/c$ for all fixed points we have found in Figure \ref{fig:combining histogram}.
Among the fixed points we find, the lowest and the largest value of $a/c$ are,
\begin{align}
    \left(\frac{a}{c} \right)_{\textrm{min}} \simeq 0.7228 \ , \quad \left(\frac{a}{c} \right)_{\textrm{max}} \simeq 1.210 \ , 
\end{align}
so that the distribution is much narrower than the Hofman-Maldacena bound \cite{Hofman:2008ar}. Notice that the upper bound is lower than that of $\CN=2$ SCFTs, which is $5/4$. 

\paragraph{$a/c$ vs the dimension of the lightest operator}

\begin{figure}[t]
    \centering
    \includegraphics[width=0.95\linewidth]{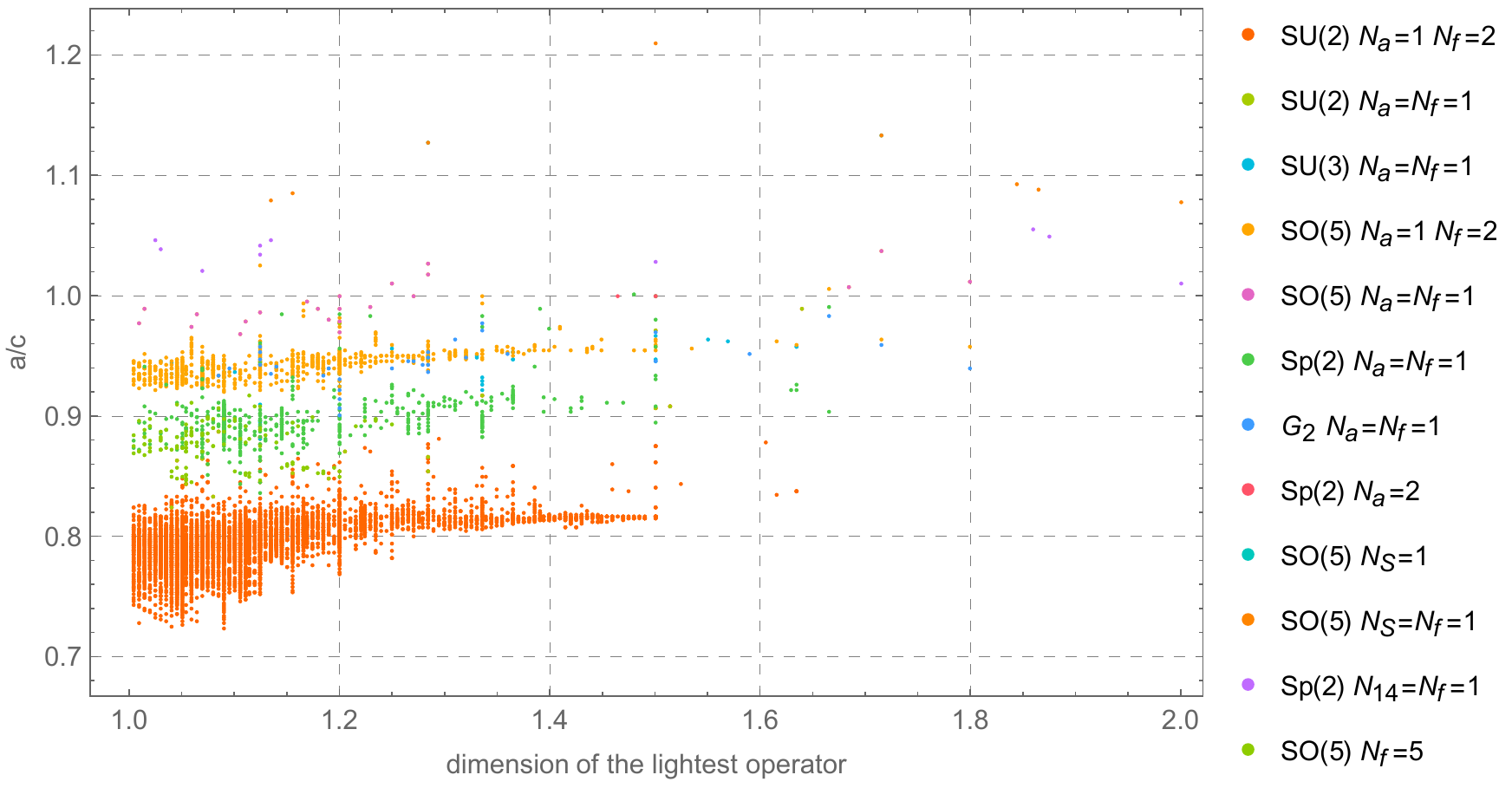}
    \caption{The plot of $a/c$ versus the dimension of the lightest operator in a given fixed point in the landscape. We differentiate the fixed points via their seed theories, which are characterized by the gauge group and the matter field. Here $N_f, N_a, N_S, N_{\mathbf{14}}$ denotes the number of fundamental, adjoint, rank-2 symmetric tensor, and the number of dimension $\mathbf{14}$ chiral multiplets respectively.}
    \label{fig:combining dimtoratio}
  \end{figure}

From our database of superconformal fixed points, we look for a universal pattern of CFT data. One of the correlations we find is between the central charge ratio $a/c$ and the scaling dimension of the lightest scalar operator in the chiral ring. We depict them in Figure \ref{fig:combining dimtoratio}. 
In this figure we distinguish the set of fixed points that originate from different seed theories. Most fixed points are clustered between scaling dimensions 1 and 1.6. 
We find that the set of fixed points originating from different seed theories do not overlap significantly.

  What we find in this plot is that when the dimension of the lightest scalar operator gets larger, the distribution of $a/c$ is more restrictive. On the other hand, when the dimension gets close to $1$, which is the unitarity bound, the $a/c$ distribution tends to become wider. One possible interpretation of this phenomenon is that when the dimension gets close to $1$, the theory will have an approximate higher-spin conserved current \cite{Maldacena:2011jn, Maldacena:2012sf}, which in turn becomes higher-spin gauge fields in the AdS bulk (even though we are in the extremely ``stringy" regime of small $N$). The difference of central charge $c-a$ controls higher-derivative corrections to  Einstein gravity for a theory that is dual to the weakly-coupled holographic dual. When there is a light higher-spin field, there is no reason for this correction to be suppressed. Therefore $a/c$ can significantly deviate from 1. It is well-known that the gap in the higher-spin operator is responsible for local, Einstein-like gravity in the bulk \cite{Heemskerk:2009pn, Camanho:2014apa}. Here we find that there is a curious correlation between the central charges and the \emph{scalar} operator gap. It would be interesting to investigate if our observation persists in a larger, in particular larger-$N$, set of conformal field theories. 

\begin{figure}[t]
    \centering
    \includegraphics[width=0.95\linewidth]{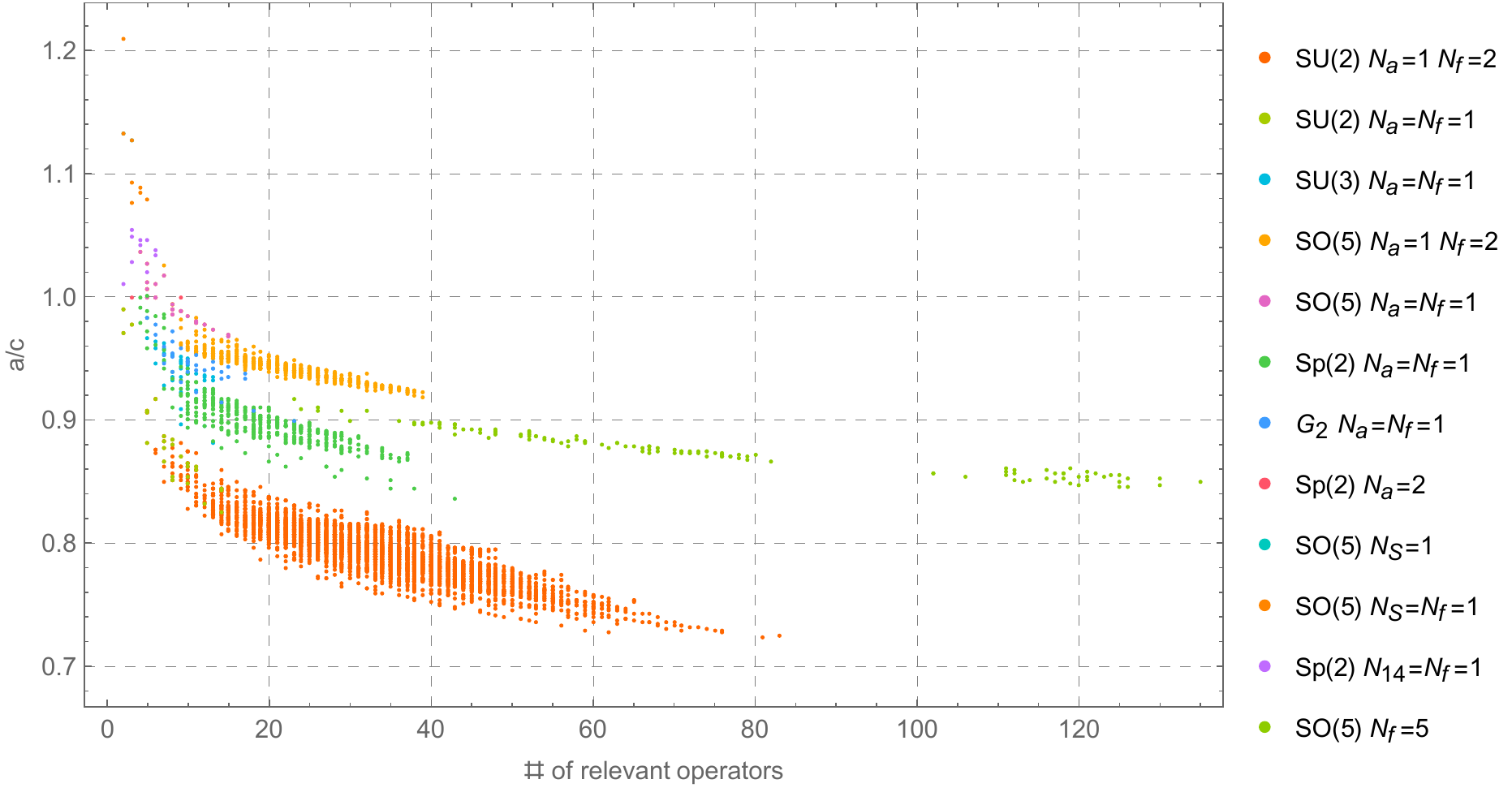}
    \caption{The plot of $a/c$ versus the number of relevant operators for the fixed points in the landscape. We differentiate the fixed points via their seed theories, which are characterized by the gauge group and the matter field. Here $N_f, N_a, N_S, N_{\mathbf{14}}$ denotes the number of fundamental, adjoint, rank-2 symmetric tensor, and the number of dimension $\mathbf{14}$ chiral multiplets respectively.}
    \label{fig:combining reltoratio}
\end{figure}

\paragraph{$a/c$ versus the number relevant operators}
  We also find correlations between the ratio of central charges $a/c$ and the number of relevant operators (scalar, $R<2$) for a given fixed point. 
  In Figure \ref{fig:combining reltoratio}, we plot $a/c$ versus the number of relevant operators for the fixed points. We distinguish the fixed points obtained from different seed theories by color. 
We find that the fixed points originating from the same seed theory tend to cluster together and the ones with different seed theories do not significantly overlap with each other.

  Most curiously, we see there is a general downward trend in the $a/c$ ratio as the number of relevant operator increases. One interpretation of this is as follows. The number of relevant operators can be loosely thought of as a degree of instability of the system. This is because a system has many more ways to be perturbed via relevant deformations. Then what we have here is that $a/c$ can be also thought of as a measure of (in)stability of a given CFT. A similar pattern also exists in the theories of class $\CS$ \cite{Gaiotto:2009hg, Gaiotto:2009we}. For these theories, as we add punctures to the UV curve, the $a/c$ ratio decreases. This is because as we add punctures, we add more flavor symmetry, and in turn more hypermultiplets than vector multiplets. For an $\CN=2$ SCFT, the only relevant operators come from the moment map of flavor symmetries. So at least for this class of theories, our observation  that $a/c$ decreases as the number of relevant operators increases is borne out. 

%%%%%%%%%%%%%%%%%%%%%%%%%%%%%%%%%%%%%%%%%%%%%%%%%%%%%%%%%%%%%%%%%%%%
\section{Detailed Case Studies} \label{sec:details}
  In this section we study in detail  certain fixed points of the gauge theories which we have summarized in the previous section, and associated remarkable phenomena, {\it e.g.} dualities, supersymmetry enhancement, and non-commuting RG flows. 

%%%%%%%%%%%%%%%%%%%%%%%%%%%%%
\subsection{Dual Descriptions of the $H_1$ Argyres-Douglas Fixed Point} \label{subsec:H1dual} 
  The $H_1$ theory is a rank-1 $\mathcal{N}=2$ SCFT with flavor symmetry $SU(2)$. Its central charges are given by $a =\frac{11}{24}$, $c=\frac{1}{2}$. The chiral operators of the $H_1$ theory are listed in Table \ref{tab:H1def2}, where $\mathcal{O}_{\text{CB}}$ denotes the Coulomb branch operator, which is the primary of the $L\bar{B}_1[0,0]_{\Delta}^{(0;2\Delta)}$ multiplet with scaling dimension $\Delta = 4/3$; $\mathcal{O}'_{\text{CB}}$ is the level-$2$ descendent of the same multiplet; and $\mu_i$ is the moment map operator, which is the primary of the $B_1\bar{B}_1[0,0]^{(2;0)}_{2}$ multiplet whose top component corresponds to the conserved $SU(2)$ flavor current. In the Table \ref{tab:H1def2} we have denoted the chargess of $\mathcal{N}=1$ R-symmetry embedded in the $\CN=2$ algebra and of $\mathcal{N}=1$ flavor symmetry \eqref{RJ}. The former is identified with the IR R-symmetry of the gauge theory which we will discuss shortly.

  The superconformal index for the $H_1$ theory is given by \cite{Maruyoshi:2016aim}
    \begin{align}
    \begin{split}
    \CI_{{\rm red}}
    &=    t^{\frac{8}{3}} x^\frac{4}{3} - t^{\frac{11}{3}} \chi_2(y) x^{\frac{1}{3}} + 3 t^4 x^{-1} + t^{\frac{14}{3}} x^{-\frac{2}{3}} + t^{\frac{16}{3}} x^{\frac{8}{3}} 
             - 4 t^6 - t^{\frac{19}{3}} \chi_2(y) x^{\frac{5}{3}} 
               \\
    &  + t^7 \chi_2(y) x^{-1}  + t^{\frac{22}{3}} x^{\frac{2}{3}} + t^8 (x^{4} + x + 5 x^{-2}) - t^{\frac{26}{3}} x^{\frac{4}{3}} - t^9 \chi_2(y) (1+x^3) + \dots \label{eq:indexh1}
    \end{split}
   \end{align}
  The first, third and fourth terms correspond to the operators $\CO_{\text{CB}}$, $\mu_i$ and $\CO_{\text{CB}}'$ respectively. 
  The fifth term denotes the operator $(\CO_{\text{CB}})^2$. 
  The  coefficienet of the $t^6$ term indicates that there are the conserved currents of $SU(2)_f \times U(1)_J$. 
       
    \begin{table}[t!]
    	\centering
    	\begin{tabular}{|c||c|c|}
    		\hline
    		 operators & $R$    & $J$    \\  \hline \hline
		 $\CO_{\text{CB}} = M$ & $\frac{8}{9}$ & $\frac{4}{3}$   \\ \hline
		 $\CO_{\text{CB}}' = X$ & $\frac{14}{9}$ & $-\frac{2}{3}$  \\ \hline
		 $\mu_i = \CO_i$ & $\frac{4}{3}$ & $-1$  \\ \hline
		 $(\CO_{\text{CB}})^2$ & $\frac{16}{9}$  & $\frac{8}{3}$  \\ \hline \hline
		 %$M$ & $\frac{10}{9}$  &  $- \frac{4}{3} $  \\ 
		  $q, \tilde{q}$  &  $\frac{5}{9}$ & $- \frac{2}{3} $  \\ \hline
		 $\phi$  &  $\frac{2}{9}$ &  $\frac{1}{3}$  \\
		\hline 
    	\end{tabular}
    	\caption{Quantum numbers of operators and fields at the $H_1$ fixed point. The top section consists of the relevant $\CN=1$ chiral operators in the $H_1$ spectrum. The fields $q,\tilde{q},\phi$ in the bottom section are the fields of $SU(2)$ $N_f=1$ adjoint SQCD.% with their representations under the $SU(2)$ gauge symmetry listed in parenthesis. 
     } 
    	\label{tab:H1def2} 
    \end{table}

  Below, we concentrate on the two different Lagrangian descriptions that flow to this $H_1$ theory. We begin by discussing the simpler theory, which fits in the analysis of $SU(2)$ $N_f=1$ adjoint SQCD introduced in section~\ref{sec:su2nf1}.  

\paragraph{Description from $SU(2)$ $N_f=1$ adjoint SQCD}
  We consider $SU(2)$ $N_f=1$ adjoint SQCD with the superpotential (\href{https://qft.kaist.ac.kr/landscape/detail.php?id=2}{\#2}),
	\ba{
	W_{H_1} = X \tr \phi^2 + M q \tilde{q}\,,
 \label{eq:suph1}
	}
  which preserves $SU(2)_f \times U(1)_J$ global symmetry where the fundamental and anti-fundamental multiplets $q$ and $\tilde{q}$ form the doublet of $SU(2)_f$. The superpotential \eqref{eq:suph1} drives the theory to the $H_1$ fixed point\footnote{~This is a simplified version of the flow considered in \cite{Maruyoshi:2016aim}. There, one starts from $SU(2)$ $N_f=4$ SQCD and considers nilpotent Higgsing for the principal embedding of $SU(4)\subset SO(8)$, which preserves only a $U(1)$ global symmetry at the Lagrangian level. The resulting superpotential is
	\ba{
	W = M_1 \tr \phi^2 q\tilde{q} + M_2 \tr \phi q\tilde{q} + M q\tilde{q}.\label{eq:suph12}
	}
  Starting from the $W=0$ theory, it is clear that only the last term is needed to drive the theory to the $H_1$ fixed point. %This flow is summarized in Figure \ref{fig:H1}. 
  The simplified superpotential \eqref{eq:suph1} was also considered in \cite{Benvenuti:2017lle}.}.
  The charges of the fields are summarized in the table \ref{tab:H1def2}.
  
  Let us see this by studying the chiral ring of the theory from the perspective of the superpotential \eqref{eq:suph1}. The F-terms are given by,
    \bea
    F_{M}
     =     q \tilde{q}\,, ~~~
    F_\phi
     =     2 X \phi, ~~~
    F_q
     =    M \tilde{q}\,, ~~~
    F_{\tilde{q}}
     =     M q\,, ~~~
    F_X
     =     \tr \phi^2\,.
     \label{FA3}
    \eea
  The middle two equations imply that $M \cdot \CO_{i}^{k} \sim X  \cdot \CO_{i}^k \sim 0$ for all $i$, where we defined $\CO_{i}=\{ \tr \phi \tilde{q} \tilde{q}, \tr \phi q q, \tr \phi q \tilde{q} \}$. The last equation sets all $\CO_{i}^k$ to vanish for $k \geq 2$. Furthermore, the moduli space of $X$ is uplifted by quantum effects\footnote{~This point is also made in \cite{Benvenuti:2017lle}.}. This can be seen from the fact that giving a vev to $X$ reduces the theory to $SU(2)$ $N_f = 1$ SQCD with superpotential $W =M  q \tilde{q}$. The low energy effective superpotential is thus $W_{{\rm eff}} = (\Lambda^5/\CM) + M \CM$, where $\CM =q \tilde{q}$ is the meson field \cite{Affleck:1983mk}. The F-term condition with respect to $M$ sets $\CM=0$ which gives a singular answer. Therefore quantum mechanically $X$ cannot take an expectation value.

  To summarize, the operators in the $\CN=1$ chiral ring and the identification with those of the $H_1$ theory are given by
    \bea
    M = \CO_{{\rm CB}}, ~~X = \CO'_{{\rm CB}}, ~~ \CO_i=\mu_i\,.
    \eea
  In fact, one can check the R-charges of the operators match with those of the $H_1$ theory. They satisfy the chiral ring relations 
    \bea
    M \cdot \CO_{i}
     \sim     0, ~~~~~
    \CO_2 \CO_{-2} \sim \left( \CO_0 \right)^2.
    \eea
  The second relation represents $\mathbb{C}^2/Z_2$, the minimal nilpotent orbit of $SU(2)_f$. Note that the flip field $X$ remains in the chiral ring, as can also be seen from the superconformal index given above in \eqref{eq:indexh1}.
  
\paragraph{Description from $SU(2)$ $N_f=2$ adjoint SQCD}
  Another theory which flows to the $H_1$ theory is $SU(2)$ gauge theory with the chiral multiplets as in the Table \ref{tab:a1d3}, and superpotential (\href{https://qft.kaist.ac.kr/landscape/detail.php?id=95}{\#95})\footnote{~We note that the second superpotential has a quartic term, which means this is not UV finite. However, the UV completion is easy to obtain. For example, the quartic interaction is obtained by adding the additional adjoint chiral $\Phi$ and the superpotential terms $M \Tr \phi \Phi + \Tr \Phi q_2 q_2 + \Tr \Phi^2$ and integrating $\Phi$ out.
    }
    \bea
    W
     =     \Tr \tilde{\phi} q_1q_1 + \tilde{M} \Tr \tilde{\phi} q_2 q_2 + Y \Tr \tilde{\phi}^2\,.
     \label{supNf2}
    \eea
  The IR R-charges are also given in the Table \ref{tab:a1d3}. 
  The $SU(2)_f$ flavor symmetry is manifest in the Lagrangian level.
   
    \begin{table}[t]
    	\centering
    	\begin{tabular}{|c|c|c|c|c|}
    		\hline
    		 fields & $SU(2)_g$ & $SU(2)_f$  &  $R$ & $J$ \\
    		\hline \hline
    		$q_1$ & $\mathbf{2}$ & $\mathbf{3}$  & $\frac{8}{9}$ & $-\frac{1}{6}$ \\ \hline
    		$q_2$ & $\mathbf{2}$ & $\mathbf{1}$ &  $\frac{4}{9}$ & $-\frac{5}{6}$   \\ \hline
    		$\tilde{\phi}$ & \rm{adj} & $\mathbf{1}$  & $\frac{2}{9}$ & $\frac{1}{3}$  \\ \hline
		$\tilde{M}$ &$\mathbf{1}$ & $\mathbf{1}$ & $\frac{8}{9}$ & $\frac{4}{3}$  \\ \hline
		$Y$ & $\mathbf{1}$ & $\mathbf{1}$ & $\frac{14}{9}$ & $-\frac{2}{3}$   \\ \hline
    	\end{tabular}
    	\caption{The charges of the fields of $SU(2)$ $N_f=2$ adjoint SQCD with the superpotential \eqref{supNf2}.
        } 
    	\label{tab:a1d3}
    \end{table}

  Let us consider the chiral ring of the theory.
  The gauge invariant operators are
    \bea &&
    \tilde{M}, ~~~
    Y, ~~~
    \hat{\CO}_s^{k} 
     = \tr \tilde{\phi}^k (q_1)^i (q_1)_i, ~~~
    \hat{\CO}_i^{' k}
     =     \tr \tilde{\phi}^k (q_1)^j (q_1)^l \epsilon_{ijl},
    \nonumber \\
    &&
     \hat{\CO}_{i}^{k} = \tr \tilde{\phi}^k (q_1)_i q_2,~~ 
     \hat{\CO}_{0}^{k} = \tr \tilde{\phi}^k q_2 q_2,
    \eea
  where $i, j$ label the $SU(2)_f$ adjoint representation.
  Due to (anti-)symmetric properties of $\tilde{\phi}^k$, we see that $\hat{\CO}_{s}^{2n} =\hat{\CO}_{0}^{2n} = \hat{\CO}_{i}^{' 2n+1} = 0$.
    The F-term conditions are
    \bea
    F_{\tilde{M}}
    &=&    \tr \tilde{\phi} q_2 q_2, ~~~
    F_{q_1}
     =     \tilde{\phi} q_1, ~~~
    F_{q_1}
    =     \tilde{M} q_2 \tilde{\phi}, 
          \nonumber \\
    F_{\tilde{\phi}}
    &=&    q_1 q_1 + \tilde{M} q_2 q_2 + Y \tilde{\phi}, ~~~
    F_Y
     =     \Tr \tilde{\phi}^2.
    \eea
  First of all, the last equation implies $\phi^2 = 0$ as above.
  Thus all $\hat{\CO}_s^{k}$, $\hat{\CO}_{i}^{k}$, $\hat{\CO}_{i}^{' k}$ and $\hat{\CO}_{0}^{k}$ vanish if $k>1$.
  The second condition implies that $\hat{\CO}_s^{k} \sim \hat{\CO}_{i}^{k} \sim \hat{\CO}_{i}^{' k} \sim 0$ for $k>0$.
  The third F-term condition $\tilde{M} q_2 \tilde{\phi} \sim 0$ implies that $\hat{\CO} \cdot \hat{\CO}_i^{k} \sim 0$ 
  and $\hat{\CO} \cdot \hat{\CO}_0^{k} \sim 0$ for  $k>0$.  
  One can further obtain the relation $\tilde{M} \Tr q_1 q_2 = 0$ from $\tilde{\phi}^2 = 0$, $\tilde{\phi} q_1 =0$ and $\tilde{M} q_2 \tilde{\phi} = 0$.
  
  As in the $N_f=1$ theory, 
  the moduli space of the Lagrange multiplier $Y$ is uplifted by the quantum effect as follows.
  By giving a vev to $Y$ the superpotential becomes $W = \sum_{i=1,2,3} (V^{i 4})^2 \tilde{M} + \sum_{i,j=1,2,3} V^{ij} V^{ij}$,
  where $V^{ij} = (q_1)^i (q_1)^j$ and $V^{i 4} =(q_1)^i q_2$.
  These $V$ satisfy the quantum chiral ring relation: ${\rm Pf} \,V \sim \Lambda^4$.
  However, this cannot be satisfied because of the F-term condition $V^{ij}=0$.
  
  Moreover, the D-term equation prevents the operator $\CO_{i}^{'0} = (q_1)^j (q_1)^k \epsilon_{ijk}$
  from having a vacuum expectation value. 
  Therefore this operator is also excluded from the chiral ring. 
  
  Now, the operators in the chiral ring identified with the $H_1$ theory operators are 
    \bea
    \tilde{M} = \CO_{{\rm CB}}, ~~~
    Y = \CO_{{\rm CB}}',~~~ 
    \hat{\CO}_i^{0} = \mu_i.
    \eea
  These satisfy the relation
    \bea
    \tilde{M} \cdot \hat{\CO}_{i}^{0}
     \sim     0.
    \eea

%%%%%%%%%%%%%%%%%%%%%%%%%%%%%%%%%%%%%%%%%%    
\paragraph{Duality}

  Since two theories flow to the same $H_1$ fixed point, these are dual to each other in the same sense as in $\CN=1$ SQCD in the conformal window   \cite{Seiberg:1994pq,Kutasov:1995ve,Kutasov:1995np} with or without adjoint. Since it involves the gauge-adjoint chiral multiplet, this reminds us of the Kutasov duality. However, a major difference from it is that the duality above does change the number of flavors. We are not aware of this kind of duality in the literature. 
  
  Indeed, this duality is valid even without some terms in the superpotentials. The $N_f=1$ theory with the superpotential (\href{https://qft.kaist.ac.kr/landscape/detail.php?id=1}{\#1})
    \bea
    W
     =    X \Tr \phi^2
    \eea
  flows to the fixed point $\hat{\CT}$. The central charges are $a_{\hat{\CT}} \simeq 0.452668$ and $c_{\hat{\CT}} \simeq 0.498618$. The same fixed point can be obtained from the $N_f=2$ theory \href{https://qft.kaist.ac.kr/landscape/detail.php?id=66}{\#66} with 
    \bea
    W
     =     \Tr \tilde{\phi}  q_1q_1 + Y \Tr \tilde{\phi}^2.
    \eea
  For both theories, the $SU(2) \times U(1)$ global (non-$R$) symmetry is manifest in the Lagrangian level. 
  At this fixed point, the operator $M \Tr q \tilde{q}$ in the former theory has dimension $\Delta \simeq 2.512$. 
  The corresponding operator in the latter theory is $\tilde{M} \Tr \tilde{\phi} q_2 q_2$.
  These do not break the global symmetry.
  Thus these operators are mapped by the duality.
  The resulting fixed points of the deformations by these operators are, therefore, the same. 
  
  At the $H_1$ fixed point, the identification of the operators are
    \bea
    M
     \leftrightarrow     
    \tilde{M}, ~~~
    \CO_i^{1}
     =     (\Tr \phi qq, \Tr \phi q \tilde{q}, \Tr \tilde{q}\tilde{q})
     \leftrightarrow     
    \hat{\CO}_{i}^{0}
     =     \Tr (q_1)_i q_2.
    \eea
  The strongest check of the duality is to see the equivalence of the superconformal indices of these two theories. 
  Indeed, the indices obtained from the Lagrangian descriptions pass this test.

\subsection{Lagrangian Duals for the Deformed Argyres-Douglas Theories} \label{subsec:deformAD}
\paragraph{Deformed $(A_1, A_4)$ theory}
It is known that the $(A_1,A_4)$ Argyres-Douglas theory has the following $\CN=1$ UV Lagrangian in $Sp(2)$ $N_f=1$ adjoint SQCD \cite{Maruyoshi:2016aim}:
\begin{align}
W=X_1\Tr\phi^2+M_1 \phi Q_1^2 + \phi Q_2^2 + X_2\Tr\phi^4+M_2\phi^3 Q_1^2\,
\end{align}
        {\renewcommand\arraystretch{1.4}
        \begin{table}[H]
            \centering
            \begin{tabular}{|c|c|c||c|c|c|c|}\hline
                id&$a$ & $c$ & $R_{X_i}$ &  $R_{M_i}$ & $R_{Q_i}$ & $R_\phi$ \\\hline\hline
             \href{https://qft.kaist.ac.kr/landscape/detail.php?id=8491}{\#8491} & $\frac{67}{84}$& $\frac{17}{21}$ & $\frac{38}{21}$, $\frac{34}{21}$ & $\frac{20}{21}$, $\frac{16}{21}$ & $\frac{10}{21}$, $\frac{20}{21}$  & $\frac{2}{21}$\\\hline
            \end{tabular}
        \end{table}}
\noindent The flip fields $M_1, M_2$ correspond to the Coulomb branch operators of dimension $10/7$ and $8/7$ respectively. 
When we deform the $(A_1,A_4)$ AD theory by the Coulomb branch operator of dimension $\Delta = 10/7$ or $M_1$, then we obtain $\CN=1$ theory with the central charges $(a,c)=(\frac{633}{2000},\frac{683}{2000})$ which is identical to the one studied in \cite{Xie:2021omd}. This deformed theory corresponds to the following fixed point of $Sp(2)$ theory with $N_f=1$ and an adjoint:
\begin{align} \label{eq:minimal Sp2} 
     W=X_1\Tr\phi^2+M_1 \phi Q_1^2 + \phi Q_2^2 + X_2\Tr\phi^4+M_2\phi^3 Q_1^2 + M_1 
\end{align}
        {\renewcommand\arraystretch{1.4}
        \begin{table}[H]
            \centering
            \begin{tabular}{|c|c|c||c|c|c|c|}\hline
                id&$a$ & $c$ & $R_{X_i}$ &  $R_{M_i}$ & $R_{Q_i}$ & $R_\phi$ \\\hline\hline
             $\cdot$ & $\frac{633}{2000}$& $\frac{683}{2000}$ & $\frac{8}{5}$, $\frac{6}{5}$ & $2$, $\frac{8}{5}$ & $-\frac{1}{10}$, $\frac{9}{10}$  & $\frac{1}{5}$\\\hline
            \end{tabular}
        \end{table}}
\noindent Notice there is a field with negative R-charge. This makes it technically more involved to calculate the superconformal index, so we did not include the theories having a chiral multiplet with negative R-charge in our database. Nevertheless, there is no real obstacle to computing the index.\footnote{~We refer to \cite{Agarwal:2014rua} for a discussion on the contour choice of the index integral when there exists $R\le 0$ fields.} Carefully computing the index for the deformed $(A_1, A_4)$ theory, we obtain
\begin{align}
   \CI_{\textrm{red}} = t^{2.4} + t^{3.6}(1-\chi_{\half}(y)) + t^{4.8} + t^{7.2} + O(t^9)\,.\label{eq:deformedA1A4}
\end{align}

Interestingly, we find that this fixed point obtained via deformation of $(A_1, A_4)$ has another Lagrangian dual description in terms of $SU(2)$ $N_f=2$ adjoint SQCD! It is given by the fixed point described by the superpotential
\begin{align}
 W=M_1 q_1q_2 + M_2 q_1\tilde{q}_1+\phi q_1\tilde{q}_2 + M_2\phi q_2^2 + M_1 M_2 + \phi  \tilde{q}_1^2 + X\Tr\phi^2 \ , 
\end{align}
and the R-charges and the central charges are given as the following table: 
        {\renewcommand\arraystretch{1.4}
        \begin{table}[H]
            \centering
            \begin{tabular}{|c|c|c||c|c|c|c|c|}\hline
                id&$a$ & $c$ & $R_{X}$ &  $R_{M_i}$ & $R_{q_i}$ & $R_{\tilde{q}_i}$& $R_\phi$ \\\hline\hline
             \href{https://qft.kaist.ac.kr/landscape/detail.php?id=46107}{\#46107} & $\frac{633}{2000}$& $\frac{683}{2000}$ & $\frac{8}{5}$& $\frac{6}{5}$, $\frac{4}{5}$ & $\frac{3}{10}$, $\frac{1}{2}$& $\frac{9}{10}$, $\frac{3}{2}$  & $\frac{1}{5}$\\\hline
            \end{tabular}
        \end{table}}
\noindent Notice that in this description, R-charges for all the chiral multiplets are positive, so there is no subtlety computing the index. We indeed reproduces the same result of \eqref{eq:deformedA1A4}. 
Therefore we establish the duality between $\CN=1$ deformed $\CN=2$ AD theory and an $\CN=1$ Lagrangian gauge theory. 
Since the $(A_1, A_4)$ AD theory also admits an $\CN=1$ Lagrangian description, we can also regard this result as a duality between $SU(2)$ and $Sp(2)$ gauge theories with different matter multiplets and superpotentials. 
 
\paragraph{Deformed $(A_2, A_3)$ theory}
We also find there is a fixed point in the landscape of $SU(3)$ $N_f=1$ adjoint SQCD that has the same central charges as the $\CN=1$ deformed $(A_2, A_3)$ Argyres-Douglas theory \cite{Cecotti:2010fi, Xie:2012hs}. The $(A_2, A_3)$ theory is a rank-3 $\CN=2$ SCFT with Coulomb branch operator dimensions $\Delta=12/7, ~9/7, ~8/7$. Consider deforming this theory via $W=u_1$, where $u_1$ is the Coulomb branch operator of dimension $12/7$. Assuming the deformed theory flows to an interacting SCFT, we obtain central charges 
\begin{align}
    a = \frac{231}{256} \ , \quad c = \frac{239}{256} \ . 
\end{align}
It turns out we get the same central charges from the fixed point of $SU(3)$ theory with 1 adjoint and $N_f=1$ with the superpotential 
\begin{align}
    W=X_1\Tr\phi^2+X_2\Tr\phi^3+q^2\tilde{q}^2 \ . 
\end{align}
The R-charges of the fields are given as in the table below:
        {\renewcommand\arraystretch{1.4}
        \begin{table}[H]
            \centering
            \begin{tabular}{|c|c|c||c|c|c|c|}\hline
                id&$a$ & $c$ & $R_{X_i}$ & $R_{q}$ & $R_{\tilde{q}}$ &$R_\phi$ \\\hline\hline
             \href{https://qft.kaist.ac.kr/landscape/detail.php?id=45756}{\#45756} & $\frac{231}{256}$& $\frac{239}{256}$ & $\frac{5}{3}$, $\frac{3}{2}$  & $\half$ & $\frac{1}{2}$  & $\frac{1}{6}$\\\hline
            \end{tabular}
        \end{table}}
\noindent The superconformal index of this theory (\href{https://qft.kaist.ac.kr/landscape/detail.php?id=45756}{\#45756}) passes all  consistency checks. 
Therefore we claim that this $SU(3)$ gauge theory provides a Lagrangian dual for the $\CN=1$ deformed $(A_2, A_3)$ theory. No $\CN=1$ UV Lagrangian flowing to $\CN=2$ $(A_{k-1}, A_{p-1})$ (with $(p, k)=1$) has been found. It is interesting to note  that $\CN=1$ deformed non-Lagrangian theories seem to be more amenable to  Lagrangian descriptions than the undeformed ones. See also \cite{Razamat:2019vfd, Zafrir:2019hps, Zafrir:2020epd, Razamat:2020gcc, Kang:2023dsa, Maruyoshi:2023mnv}. 

Similarly, it is possible to find a candidate fixed point in the landscape of $SU(2)$ $N_f=2$ adjoint SQCD that has the same central charges as the putative fixed point of the $(A_2,A_3)$ theory linearly deformed by the Coulomb branch operator of dimension $8/7$. Assuming there is no accidental symmetry, we obtain the central charges $(a, c) = (\frac{63}{2048}, \frac{159}{2048})$, which does not look sensible. Indeed, the putative fixed point described by $SU(2)$ $N_f=2$ adjoint theory with the superpotential
 \begin{align} 
 W=\phi q_1^2+M_1 q_2\tilde{q}_1+\phi^4+M_2\phi \tilde{q}_2^2+M_3\phi q_2\tilde{q}_2 + M_3 q_1\tilde{q}_1+M_1M_4+M_4 \ , 
 \end{align}
 gives the R-charges and central charges as follows:
{\renewcommand\arraystretch{1.4}
\begin{table}[H]
    \centering
    \begin{tabular}{|c|c|c||c|c|c|c|}\hline
        id &$a$ & $c$ & $R_{M_i}$ & $R_{q_i}$ & $R_{\tilde{q}_i}$ &$R_\phi$ \\\hline\hline
     $\cdot$ & $\frac{63}{2048}$& $\frac{159}{2048}$ & 0, 3, $\frac{3}{4}$, 2  & $\frac{3}{4}$, $\frac{3}{2}$ & $\frac{1}{2}$, $-\frac{3}{4}$  & $\frac{1}{2}$\\\hline
    \end{tabular}
\end{table}}
\noindent We can directly confirm via superconformal index that this putative fixed point indeed violates unitarity. Therefore there is no good interacting SCFT triggered by the linear deformation via dimension $8/7$ of $(A_2, A_3)$ theory. It is still interesting to note that this `bad' theory has a Lagrangian description. It would be interesting to reconstruct an $\CN=1$ Lagrangian of the $(A_2, A_3)$ from this observation.

\subsection{Non-Commuting RG flow: Ordering Matters} \label{subsec:non-commuting}
  Sometimes the deformations and the endpoint of RG flows can be different depending on the ordering. This happens when we integrate out some of the massive fields ``too early", before reaching another fixed point. The higher-dimensional operator generated upon integrating out massive fields can be irrelevant at one fixed point, but relevant at the other fixed point. 

  An illustrative example is the deformation of the $H_0$ Argyres-Douglas theory. The $H_0$ theory is in the landscape whose seed theory is $SU(2)$ $N_f=1$ adjoint SQCD:
    {\renewcommand\arraystretch{1.4}
        \begin{table}[H]
            \centering
            \begin{tabular}{|c|c|c||c|c|c|c|c|}\hline
                id &$a$ & $c$ & $R_{X_1}$ &  $R_{M_1}$ & $R_{q}$ & $R_{\tilde{q}}$ & $R_\phi$ \\\hline\hline
              \href{https://qft.kaist.ac.kr/landscape/detail.php?id=9}{\#9} & $\frac{43}{120}$& $\frac{11}{30}$ & $\frac{26}{15}$  & $\frac{4}{5}$  & $\frac{8}{15}$ & $\frac{14}{15}$ & $\frac{2}{15}$\\\hline
            \end{tabular}
        \end{table}}
  \noindent The superpotential is $W=X_1\Tr\phi^2 + M_1\phi q^2+ \phi \tilde{q}^2$. We can further deform this theory by adding $\delta W = M_1^2$ which leads to 
    {\renewcommand\arraystretch{1.4}
        \begin{table}[H]
            \centering
            \begin{tabular}{|c|c|c||c|c|c|c|c|}\hline
                id &$a$ & $c$ & $R_{X_1}$ &  $R_{M_1}$ & $R_{q}$ & $R_{\tilde{q}}$ & $R_\phi$ \\\hline\hline
              \href{https://qft.kaist.ac.kr/landscape/detail.php?id=16}{\#16} & $\frac{263}{768}$& $\frac{271}{768}$ & $\frac{5}{3}$  & $1$  & $\frac{5}{12}$ & $\frac{11}{12}$ & $\frac{1}{6}$\\\hline
            \end{tabular}
        \end{table}}
  \noindent In completing this flow, we assumed that we first arrived at the $H_0$ fixed point, and then deformed by $M_1^2$, {\it i.e.} the $M_1^2$ term is below the scale at which the $M_1 \phi q^2$ term becomes relevant. 
  
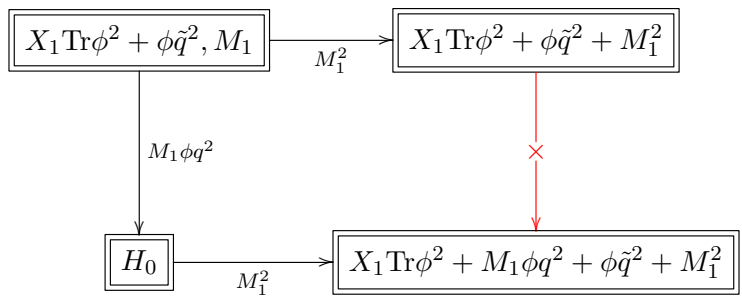
\begin{figure}[t]
  \begin{align*}
    \xymatrix @R=5pc {
  *++[F=]{X_1\Tr\phi^2+\phi\tilde{q}^2, M_1 } \ar[d]^{M_1\phi q^2} \ar[r]_-{M_1^2} & *++[F=]{X_1\Tr\phi^2+\phi\tilde{q}^2+ M_1^2 }\ar@[red][d]|-{\displaystyle \color{red}{\times}}  \\
  *++[F=]{H_0} \ar[r]_-{M_1^2} &*++[F=]{X_1\Tr\phi^2+M_1\phi q^2+\phi\tilde{q}^2+M_1^2}}
  \end{align*}
  \caption{``Non-commuting" RG flow of $H_0$ fixed point. }
  \label{fig:noncom}
  \end{figure}

  At some point when dialing the the scale of $M_1^2$ up, we hit a critical scale for which we no longer flow to the fixed point \href{https://qft.kaist.ac.kr/landscape/detail.php?id=16}{\#16}. In particular, let us integrate out $M_1$ first, then arriving at $W = X \Tr \phi^2 + \phi \tilde{q}^2 + (\phi q^2)^2$. The first two terms lead to the fixed point 
    {\renewcommand\arraystretch{1.4}
        \begin{table}[H]
            \centering
            \begin{tabular}{|c|c|c||c|c|c|c|c|}\hline
                id &$a$ & $c$ & $R_{X_1}$  & $R_{q}$ & $R_{\tilde{q}}$ & $R_\phi$ \\\hline\hline
              \href{https://qft.kaist.ac.kr/landscape/detail.php?id=5}{\#5} & $0.3451$& $0.3488$ & $1.6976$    & $0.4708$ & $0.9244$ & $0.1512$\\\hline
            \end{tabular}
        \end{table}}
  \noindent with the operator $(\phi q^2)^2$ which is {\it not} relevant. Thus we cannot flow to the fixed point \href{https://qft.kaist.ac.kr/landscape/detail.php?id=16}{\#16}, even though the superpotentials are equivalent at the classical level. 
  In summary, depending on the scale of the mass term $M_1^2$, the end point of the RG flow is different. In other words, the ordering of adding the terms in the superpotential really matters for the IR physics. See Figure \ref{fig:noncom} for an illustration of this argument. Such RG flow ``non-commutativity" is also be observed in the deformation of the $H_1$ Argyres-Douglas theory (\href{https://qft.kaist.ac.kr/landscape/detail.php?id=2}{\#2}). 

  The similar deformation of the $H_2$ theory produces  a slightly different situation, which however shares the non-commuting nature. The $H_2$ Argyres-Douglas theory is in the landscape of $SU(2)$ $N_f=2$ adjoint SQCD with the superpotential $W=M_1 q_1q_2 + \phi \tilde{q}_1\tilde{q}_2 + X_1\Tr\phi^2$ (\href{https://qft.kaist.ac.kr/landscape/detail.php?id=54}{\#54}). At this fixed point the deformation by $M_1^2$ is marginally irrelevant. Therefore the deformation does not drive the theory to a new fixed point. On the other hand, when the scale of $M_1^2$ is large enough, again this term is simply the mass term for $M_1$, and then integrating it out gives the superpotential $W=(q_1q_2)^2 + \phi \tilde{q}_1\tilde{q}_2 + X_1\Tr\phi^2$. At the fixed point with $\phi \tilde{q}_1\tilde{q}_2 + X_1\Tr\phi^2$ (\href{https://qft.kaist.ac.kr/landscape/detail.php?id=51}{\#51}) the deformation $(q_1 q_2)^2$ is irrelevant, so the RG flow ends up at the fixed point (\href{https://qft.kaist.ac.kr/landscape/detail.php?id=51}{\#51}). 

  Such RG flow ``non-commutativity" was first observed in $\CN=1$ singlet SQCD in \cite{Barnes:2004jj, Amariti:2012wc}.
  It would be interesting to study it further.

\subsection{Duality: Not in the $N_f=1$ Landscape but in the $N_f=2$ Landscape} \label{subsec:nf1nf2}
  As stated in section \ref{sec:su2nf1}, our deformation procedure differs slightly from the one in \cite{Maruyoshi:2018nod}. In particular, we check the consistency of a fixed point at each step of deformation procedure, whereas in \cite{Maruyoshi:2018nod}, a consistency check was done only after completing all possible deformations. For the landscape of $SU(2)$ $N_f=1$ adjoint SQCD we obtain 26 fixed points, rather than 34 fixed points obtained in the previous work. This means that some seemingly consistent fixed points were obtained by deforming inconsistent fixed points.

  For example, consider the following consistent fixed point (\href{https://qft.kaist.ac.kr/landscape/detail.php?id=18}{\#18}):
  {\renewcommand\arraystretch{1.4}
  \begin{table}[H]
    \centering
    \begin{tabular}{|c|c|c|c|c|c|c|c|}\hline
    $W$&\multicolumn{7}{c|}{$X_1\Tr\phi^2+M_1\phi\tilde{q}^2+M_2 q\phi\tilde{q}+M_1^2$}\\\hline
        id&$a$ & $c$ & $R_{X_1}$& $R_{M_i}$ & $R_{q}$ & $R_{\tilde{q}}$ &$R_\phi$ \\\hline
     \href{https://qft.kaist.ac.kr/landscape/detail.php?id=18}{\#18} & $0.4487$& $0.5156$ & $1.4646$  & 1, $0.8031$ & $0.3662$  & $0.5631$& $0.2677$\\\hline
    \end{tabular}
  \end{table}}
  \noindent At this fixed point (\href{https://qft.kaist.ac.kr/landscape/detail.php?id=18}{\#18}), we can flip the operators $q\tilde{q}$ and $M_2$ since they have R-charges smaller than $4/3$. If we flip the operator $q\widetilde{q}$, then we obtain:
  {\renewcommand\arraystretch{1.4}
  \begin{table}[H]
    \centering
    \begin{tabular}{|c|c|c|c|c|c|c|c|}\hline
    $W'$&\multicolumn{7}{c|}{$X_1\Tr\phi^2+M_1\phi\tilde{q}^2+M_2 q\phi\tilde{q}+M_1^2+M_3q\tilde{q}$}\\\hline
        id&$a$ & $c$ & $R_{X_1}$& $R_{M_i}$ & $R_{q}$ & $R_{\tilde{q}}$ &$R_\phi$ \\\hline
     $\cdot$ & $\frac{57}{128}$& $\frac{65}{128}$ & $\frac{3}{2}$  & 1, $\frac{3}{4}$, 1 & $\frac{3}{8}$  & $\frac{5}{8}$& $\frac{1}{4}$\\\hline
    \end{tabular}
  \end{table}}
  \noindent At this deformed fixed point, the superconformal index with $v$ as the fugacity for the $U(1)$ flavor symmetry is given by:
  \begin{align}
  \begin{split}
    \mathcal{I}_{{\rm red}}&=t^{\frac{9}{4}}v^3+t^3(1+v^4)-t^{\frac{15}{4}}\chi_{\half}(y)v+t^{\frac{9}{2}}(v^{-6}+v^{-2}+v^6)+t^{\frac{21}{4}}v^{7}\\
    &\quad+t^6(-1+v^8-\chi_{\half}(y)v^4)+t^{\frac{27}{4}}(v^9-\chi_{\half}(y)v^{-3}+\chi_{\half}(y)v^5)+O(t^{\frac{15}{2}})\,.
  \end{split}
  \end{align}
  The last term, $+t^{27/4}\chi_{1/2}(y)v^5$, violates unitarity. Also, if we flip the operator $M_2$, then we obtain: 
  {\renewcommand\arraystretch{1.4}
  \begin{table}[H]
    \centering
    \begin{tabular}{|c|c|c|c|c|c|c|c|}\hline
    $W''$&\multicolumn{7}{c|}{$X_1\Tr\phi^2+M_1\phi\tilde{q}^2+M_2 q\phi\tilde{q}+M_1^2+M_2M_3$}\\\hline
        id&$a$ & $c$ & $R_{X_1}$& $R_{M_i}$ & $R_{q}$ & $R_{\tilde{q}}$ &$R_\phi$ \\\hline
     $\cdot$ & $0.4332$& $0.4919$ & $1.4462$  & 1, $0.8307$, 1.1693 & $0.3616$  & $0.5309$& $0.2769$\\\hline
    \end{tabular}
  \end{table}}
  \noindent with the superconformal index
  \begin{align}
    \mathcal{I}_{{\rm red}}=t^{2.68}+t^3+t^{3.51}-t^{3.83}\chi_{\half}(y)+t^{4.02}+t^{4.34}+t^{5.35}-t^{5.49}+O(t^{5.86})\,.
  \end{align}
  The last term, $-t^{5.49}$, again violates unitarity.

  These two deformed fixed points have inconsistent superconformal indices, so we stop further deformations from them. However, in \cite{Maruyoshi:2018nod}, a seemingly consistent fixed point was found by further deforming; or in other words, by turning on the superpotential $M_3q\tilde{q}$ and $M_2M_3$ simultaneously from $W$:
  \begin{align}
    W'''=X_1\Tr\phi^2+M_1\phi\tilde{q}^2+M_2 q\phi\tilde{q}+M_1^2+M_3q\tilde{q}+M_2M_3\,.\label{eq:non-commuting}
  \end{align}
  This fixed point (\ref{eq:non-commuting}) is not captured if we deform superpotential terms one by one and check consistency at every step, so it does not appear in this paper. On the other hand, if we turn on multiple superpotential terms simultaneously, such fixed points can appear.

  Additionally, there is no different superpotential providing an equivalent fixed point to (\ref{eq:non-commuting}) in our procedure. This indicates that the $N_f=1$ landscape has ``non-commuting" RG flows, meaning that we cannot obtain an equivalent fixed point by changing the order of superpotential terms.

  Note that we cannot definitely say that such seemingly consistent fixed points, obtained by deforming inconsistent fixed points, are truly consistent and exist. The presence of an inconsistent index suggests that there may be spontaneous supersymmetry breaking or accidental symmetry enhancement, which could invalidate our $a$-maximization results. Consequently, with our technique, we cannot accurately track what really happens near the inconsistent fixed points along the RG flow and cannot make definitive statements about their consistency or existence of the seemingly consistent points.

  We list all cases of non-commuting RG flows in the $SU(2)$ $N_f=1$ landscape in Table \ref{tab:non-commuting}. In our landscape, the fixed points in the third column do not appear. The No.2 in Table \ref{tab:non-commuting} is the fixed point (\ref{eq:non-commuting}), which is the example we have just discussed. The No.3, 4, and 5 fixed points in the Table are deformed theories from (\ref{eq:non-commuting}), and other cases also exhibit similar behavior. 

  {\renewcommand\arraystretch{1.4}
  \begin{table}[t]
    \centering
    \begin{tabular}{|c|c|c|}
    \hline
        No. & Superpotential at the inconsistent fixed point & \scriptsize{Deformation to a consistent fixed point} \\\hline\hline
        1 & $W_0 + M_2 q\phi\tilde{q} + M_1q\tilde{q}$ & $+M_1M_3$\\\hline
        2&$W_0 +M_2 q\phi\tilde{q}+M_1^2+M_3q\tilde{q}$ & $+M_2M_3$\\\hline
        3&$W_0 +M_2 q\phi\tilde{q}+M_1^2+M_3q\tilde{q}$ & $+M_2M_3+M_3M_4$\\\hline
        4&$W_0 +M_2 q\phi\tilde{q}+M_1^2+M_3q\tilde{q}$ & $+M_2M_3+M_4 \phi\tilde{q}^2$\\\hline
        5&$W_0 +M_2 q\phi\tilde{q}+M_1^2+M_3q\tilde{q}$ & $+M_2M_3+M_4 \phi\tilde{q}^2+M_3M_5$\\\hline
        6&$W_0 +M_2q\phi\tilde{q}+M_1M_2+M_2M_3$ &  $+q\phi\tilde{q}^3+M_1M_4$\\\hline
        7&$W_0 +M_2q\phi\tilde{q}+M_1M_2+M_2M_3$ &  $+q\phi\tilde{q}^3+M_1M_4+M_5q^2\phi$\\\hline
        8&$W_0 +M_2\phi\tilde{q}^2+M_3q\phi\tilde{q}+M_1M_3+M_1M_4+M_5q\tilde{q}$&  $+M_2M_5$ \\\hline
        9&$W_0 +M_2\phi\tilde{q}^2+M_3q\phi\tilde{q}+M_1M_3+M_1M_4+M_5q\tilde{q}$&  $+M_2M_5+M_5M_6$ \\\hline
    \end{tabular}
    \caption{The second column lists superpotentials that give fixed points with inconsistent superconformal indices. The $W_0$ is given by $W_0=X_1\Tr\phi^2+M_1 \phi q^2$. The third column shows the deformation of superpotential from the inconsistent fixed point to the consistent fixed point.}
    \label{tab:non-commuting}
  \end{table}}

  Interestingly, the seemingly consistent fixed points in the third column of Table \ref{tab:non-commuting}, which disappear in our $N_f=1$ landscape, arise in the $N_f=2$ landscape. From the viewpoint of the $N_f=2$ landscape, these fixed points are obtained by deforming the superpotential terms one by one and do not exhibit any peculiar behaviour at each step, unlike in the $N_f=1$ case. In other words, the seemingly consistent but peculiar theories in the $N_f=1$ landscape that can only be obtained by turning on multiple superpotential terms simultaneously can be naturally achieved by increasing the dimension of the theory space. We list the corresponding fixed points in the $N_f=2$ landscape in Table \ref{tab:nf2commuting}. Note that the fixed point equivalent to the first entry of Table \ref{tab:non-commuting} has not yet been found in the $N_f=2$ landscape. We expect that this fixed point will appear if we enumerate the $N_f=2$ landscape further. 
  {\renewcommand\arraystretch{1.4}
  \begin{table}[t]
    \centering
    \begin{tabular}{|c|c|c|}
        \hline
        No. & id & Corresponding fixed points in the $N_f=2$ landscape \\\hline\hline
        1 &  & not found\\\hline
        2 & \href{https://qft.kaist.ac.kr/landscape/detail.php?id=1711}{\#1711}&$\phi q_1q_2+ M_1q_1\tilde{q}_1+M_1^2+\phi\tilde{q}_1^3\tilde{q}_2+\phi\tilde{q}_2^2+X_1\Tr\phi^2$ \\\hline
        3& \href{https://qft.kaist.ac.kr/landscape/detail.php?id=2707}{\#2707} &$\phi q_1q_2+ M_1q_1\tilde{q}_1+M_1^2+\phi\tilde{q}_1^3\tilde{q}_2+\phi\tilde{q}_2^2+X_1\Tr\phi^2+M_2\tilde{q}_1\tilde{q}_2$\\\hline
        4& \href{https://qft.kaist.ac.kr/landscape/detail.php?id=2708}{\#2708} &$\phi q_1q_2+ M_1q_1\tilde{q}_1+M_1^2+\phi\tilde{q}_1^3\tilde{q}_2+\phi\tilde{q}_2^2+X_1\Tr\phi^2+M_2q_2\tilde{q}_1$\\\hline
        5&\href{https://qft.kaist.ac.kr/landscape/detail.php?id=3215}{\#3215}&$\phi q_1q_2+ M_1q_1\tilde{q}_1+M_1^2+\phi\tilde{q}_1^3\tilde{q}_2+\phi\tilde{q}_2^2+X_1\Tr\phi^2+M_2q_2\tilde{q}_1+M_3\tilde{q}_1\tilde{q}_2$ \\\hline
        6&\href{https://qft.kaist.ac.kr/landscape/detail.php?id=3721}{\#3721}& $\phi q_1q_2+ M_1q_1\tilde{q}_1+M_2\phi\tilde{q}_2^2+M_3\phi\tilde{q}_1^2+M_1M_3+X_1 M_2+M_4q_2\tilde{q}_1+M_5q_1\tilde{q}_2$ \\
        && $+q_2\tilde{q}_2+X_2M_4+X_3\Tr\phi^2+M_6 \tilde{q}_1\tilde{q}_2+M_5M_6$ \\\hline
        7&\href{https://qft.kaist.ac.kr/landscape/detail.php?id=5340}{\#5340}& $\phi q_1q_2+ M_1q_1\tilde{q}_1+M_2\phi\tilde{q}_2^2+M_3\phi\tilde{q}_1^2+M_1M_3+X_1 M_2+M_4q_2\tilde{q}_1+M_5q_1\tilde{q}_2$ \\
        && $+q_2\tilde{q}_2+X_2M_4+X_3\Tr\phi^2+M_6 \tilde{q}_1\tilde{q}_2+M_5M_6+M_1M_7$ \\\hline
        8&\href{https://qft.kaist.ac.kr/landscape/detail.php?id=5339}{\#5339}&$\phi q_1q_2+M_1q_1\tilde{q}_1+M_2\phi\tilde{q}_2^2+M_3\phi\tilde{q}_1^2+M_1M_3+X_1M_2+M_4q_2\tilde{q}_1+M_5 q_1\tilde{q}_2$\\
        & &$ +q_2\tilde{q}_2+ X_2M_4+X_3\Tr\phi^2+M_6\tilde{q}_1\tilde{q}_2+M_5M_6+M_5M_7$\\\hline
        9&\href{https://qft.kaist.ac.kr/landscape/detail.php?id=6853}{\#6853}&$\phi q_1q_2+M_1q_1\tilde{q}_1+M_2\phi\tilde{q}_2^2+M_3\phi\tilde{q}_1^2+M_1M_3+X_1M_2+M_4q_2\tilde{q}_1+M_5 q_1\tilde{q}_2$\\
        && $+q_2\tilde{q}_2 + X_2M_4+X_3\Tr\phi^2+M_6\tilde{q}_1\tilde{q}_2+M_5M_6+M_5M_7+M_1M_8$\\\hline
    \end{tabular}
    \caption{Fixed points in the $SU(2)$ $N_f=2$ landscape corresponding to the fixed points in the third column in Table \ref{tab:non-commuting}.}
    \label{tab:nf2commuting}
  \end{table}}

\subsection{Possible New SUSY Enhancing Fixed Point} \label{subsec:newN2}

\paragraph{Possible new $\CN=2$ SCFT of rank 1?}
  In the landscape of SCFTs we explored, we find one particular case which can potentially give rise to a new $\CN=2$ SCFT. We find a fixed point in the %(extended) 
  landscape of $SU(2)$ $N_f=2$ adjoint SQCD, which satisfies tbe $\CN=2$ SUSY enhancement condition and passes the unitarity criterion. It is given as follows:
  \begin{align} \label{eq:WsusyN2}
    W=\phi q_1^2+ M_1 q_2\tilde{q}_1 + M_2 q_2\tilde{q}_2 + \phi \tilde{q}_1\tilde{q}_2 + X_1\Tr\phi^2 + \phi q_2\tilde{q}_1+ X_2 M_1\,.
  \end{align}
  {\renewcommand\arraystretch{1.4}
  \begin{table}[H]
    \centering
    \begin{tabular}{|c|c||c|c|c|c|c|}\hline
        $a$ & $c$ & $R_{X_i}$ &  $R_{M_i}$ & $R_{q_i}$ & $R_{\tilde{q}_i}$ & $R_\phi$ \\\hline\hline
        $\frac{19}{48}$ & $\frac{5}{12}$ & $\frac{5}{3}$, $\frac{11}{6}$  & $\frac{1}{6}$, $\frac{5}{6}$  & $\frac{11}{12}$, $\frac{7}{12}$ & $\frac{5}{4}$, $\frac{7}{12}$ & $\frac{1}{6}$\\\hline
    \end{tabular}
  \end{table}}
  \noindent In fact, this fixed point is not accessible by our general procedure. This is because when we construct our database of fixed points, we only deform by one (super)relevant operators at once. If we proceed in this way, we find that at the fixed point described by the first 5 terms in \eqref{eq:WsusyN2} (\href{https://qft.kaist.ac.kr/landscape/detail.php?id=104}{\#104}), the operator $\phi q_2 \tilde{q}_1$ turns out to be removed from the chiral ring from the F-term for $\tilde{q}_2$.
  However if we start with the fixed point for the first 3 terms in \eqref{eq:WsusyN2} (\href{https://qft.kaist.ac.kr/landscape/detail.php?id=69}{\#69}), which has the relevant operators of the form $\phi \tilde{q}_1 \tilde{q}_2$ and $\phi q_2 \tilde{q}_1$, and then add both terms simultaneously, we do find a candidate non-trivial fixed point of the RG flow. 
  Therefore this comprises another example of `non-commuting RG flow' and there is no obvious reason not to consider this fixed point. Notice that flips by $X_1$ and $X_2$ are the `passive flips' that are done to get rid of the decoupled operator along the flow. 

  Let us look at the spectrum of this theory in detail. The low-lying gauge-invariant operators of this theory are given as
  \begin{align}
    M_2,\quad \phi q_2\tilde{q}_2,\quad \phi\tilde{q}_2^2,\quad q_1q_2,\quad M_2^2,\quad X_1 \ , 
  \end{align}
  whose scaling dimensions are
  \begin{align}
    M_2: \frac{5}{4},\quad \phi q_2\tilde{q}_2:2,\quad q_1q_2:\frac{9}{4},\quad M_2^2:\frac{5}{2},\quad X_1: \frac{5}{2} \ . 
  \end{align}
  The reduced index of this fixed point is given by
  \begin{align} \label{eq:N2candidateIdx}
  \begin{split}
    \mathcal{I}_{\textrm{red}}&=t^{5/2}v^{5/4}-t^{7/2}\chi_{\half}(y)v^{1/4}+t^4 v^{-1}+t^{9/2}v^{-3/4}+t^5v^{5/2}-2t^6-t^6\chi_{\half}(y)v^{3/2}\\
    &\quad +t^7v^{1/2}+t^7\chi_{\half}(y)v^{-1}+t^{15/2}v^{15/4}+t^8(v+v^{-2})-t^{17/2} (v^{5/4}-\chi_{\half}(y)v^{11/4}) \\
    &\quad - t^9 (\chi_{\half}(y)+\chi_1 (y) v^{3/2}) + t^{19/2}(v^{7/4}+2\chi_{\half}(y)v^{1/4})\\
    &\quad +t^{10}(-v^{-1}+v^5+\chi_{\half}(y)v^{1/2}+\chi_{\frac{3}{2}}(y)v^{1/2}) \\
    &\quad + t^{21/2}(-3v^{-3/4}+v^{9/4}-\chi_{\half}(y)v^{-3/4}) \\
    &\quad +t^{11}(-v^{5/2}-\chi_{\half}(y)v^4+\chi_{1}(y)(v^{5/2}-v^{-1/2}))\,.
  \end{split}
  \end{align}
  We find that the superconformal index of this theory satisfies all the necessary and sufficient conditions to exhibit $\CN=2$ supersymmetry enhancement discussed in section \ref{subsec:SUSYenhancement}. To rewrite the index in the more familiar form of $\CN=2$ superconformal index $I(\fp, \fq, \ft)$ \cite{Gadde:2011uv}, let us write $\fp=t^3y$, $\fq=t^3/y$, and $\ft={t}^4v^{-1}$. Then, the index in the Macdonald limit ($\fp \goto 0$) is given as
  \begin{align}
    \mathcal{I}_{M}(q, \ft) =1+ \ft + 2 \fq \ft + 2\fq^2\ft + 2\fq^3\ft+\ft ^2 + 3\fq^2\ft^2+\ft^3+\fq\ft^3+\cdots \ . 
  \end{align}
  In the Hall-Littlewood limit ($\fp,\fq\goto0$), the index reads
  \begin{align}
    \mathcal{I}_{HL} (\ft) =1+\ft+ \ft^2 + \ft^3 +\cdots = \frac{1}{1-\ft} \,  \ .
  \end{align}
  The Schur limit of the index ($\fp\goto0$, $\ft \goto \fq$) is given as 
  \begin{align}
    \mathcal{I}_S (\fq) =1+ \fq+3 \fq^2+4\fq^3+6\fq^4+\cdots\,,
  \end{align}
  and the Coulomb branch index ($\fp \fq/\ft \goto u$, $\fp,\fq,\ft \goto 0$) is given as 
  \begin{align}
    \mathcal{I}_C(u) =1+u^{5/4}+ u^{5/2}+u^{15/4}+u^5+u^{25/4}+\cdots = \frac{1}{1-u^{5/4}}\,.
  \end{align}
  Therefore, it seems that it is a rank-1 $\CN=2$ SCFT (which has 1-dimensional Coulomb branch) with Coulomb branch operator dimension $\Delta = \frac{5}{4}$. Such scaling dimension is not allowed in the geometric classification of 4d $\CN=2$ SCFT \cite{Argyres:2015ffa, Argyres:2015gha, Argyres:2016xmc, Argyres:2016xua, Caorsi:2018zsq, Argyres:2018urp}. If this fixed point indeed exists as we stated above, then there must be a loophole in the geometric classifications. 

  However, the Hall-Littlewood index we computed above, which is identical to the Higgs branch Hilbert series, sums up to $\frac{1}{1-\mathfrak{t}}$. This means that the Higgs branch is simply given by a complex plane. This is inconsistent with the expectation that the Higgs branch for a scale-invariant theory has to be a cone. Hence, it is likely that there is some accidental symmetry along the RG flow which modifies our R-symmetry. The resulting theory can still be a product of interacting theory and a decoupled field. However, we do not have the proper tool to analyze the ultimate fate of this fixed point theory. It would be interesting to definitively answer this question.

\paragraph{Satisfying necessary condition for $\CN=2$ but not sufficient}
  Another candidate of new $\CN=2$ SCFT is found in the $Sp(2)$ with adjoint and 2 fundamentals landscape discussed in section \ref{subsec:Sp2nf2}. As stated there, the fixed point is described by the superpotential
  \begin{align}
    W=S^3+X_1 Q_1 Q_2 \  
  \end{align}
  have the following central charges and R-charges:
    {\renewcommand\arraystretch{1.4} 
        \begin{table}[H]
            \centering
            \begin{tabular}{|c|c|c||c|c|c|}\hline
               id & $a$ & $c$ & $R_{X_1}$ & $R_{Q_{i}}$ & $R_S$ \\\hline\hline
               \href{https://qft.kaist.ac.kr/landscape/detail.php?id=45748}{\#45748} & $\frac{95}{48}$& $\frac{47}{24}$ & $\frac{4}{3}$ & $\frac{1}{3}$, $\frac{1}{3}$  & $\frac{2}{3}$\\\hline
            \end{tabular}
        \end{table}
    }
  \noindent The superconformal index of this theory is given as 
  \begin{align}
  \begin{split}
    \mathcal{I}_{\textrm{red}}& =2t^4-t^6\left( a^2+\frac{1}{a^2} \right)-t^7\chi_{\half}(y) \left(1+a^2+\frac{1}{a^2} \right) 
     +t^8 \left(5+a^4+a^2+\frac{1}{a^2}+\frac{1}{a^4}\right) + \cdots \,.
  \end{split}
  \end{align}
  Here $a$ is the fugacity for the $U(1)$ symmetry rotating $Q_1$ and $Q_2$ by opposite phases. 
  Unlike the case of previous example in equation \eqref{eq:N2candidateIdx}, this index cannot be rewritten in terms of $\CN=2$ fugacities $(\fp, \fq, \ft)$, because we do not have an access to the $U(1) \subset SU(2)_I \times U(1)_r$ part of the $\CN=2$ R-symmetry. Notice that there was a term $t^4/v$ in \eqref{eq:N2candidateIdx}, which maps to the moment map operator in the current multiplet. Here, $a$ appears as a fugacity for the conserved current. This is only possible if the conserved current is non-abelian, since abelian conserved current is not charged under the symmetry. 
  Therefore, there must be at least one more conserved current that contributes to $- 1 \times t^6$ in the index so that we have a proper $SU(2)$ non-abelian flavor symmetry (as it can be seen from the Lagrangian), which gets cancelled by a contribution by a marginal operator. We do find a marginal operator given by $S Q_1^2 Q_2^2$, which contributes to the $t^6$ term. 
  
  In order for this theory to have enhanced $\CN=2$ supersymmetry, we also need an extra $U(1)$ flavor symmetry (from the $\CN=1$ viewpoint) which becomes part of the $\CN=2$ R-symmetry. This requires the existence of an additional marginal operator which we do not find evidence for. Therefore, we conjecture that this theory does not enhance to $\CN=2$ supersymmetry in the infrared.

\subsection{Distinct Theories with Identical Central Charges}\label{subsec:iden}
Some fixed points, despite having different superconformal indices, can share the same values for the central charges $a$ and $c$.
We find 26 pairs of such inequivalent fixed points in the landscape of $SU(2)$ $N_f=2$ adjoint SQCD. For example, the following two fixed points with superpotentials $W_1$ (\href{https://qft.kaist.ac.kr/landscape/detail.php?id=58}{\#58}) and $W_2$ (\href{https://qft.kaist.ac.kr/landscape/detail.php?id=2301}{\#2301}),
\begin{align}
\begin{split}
    W_1&=\phi q_1^2 + \phi^2\tilde{q}_1\tilde{q}_2+\phi^4\,,\\
    W_2&=\phi q_1 q_2+M_1 q_1\tilde{q}_1+M_2\phi \tilde{q}_2^2+M_3\phi \tilde{q}_1^2 + M_1M_3\\
     &\quad +X_1M_2 + M_4 q_2\tilde{q}_1 + M_4^2+q_2\tilde{q}_1 + \tilde{q}_2^2\,,
\end{split}
\end{align}
have the identical central charges
\begin{align}
    a=\frac{75}{128},\quad c=\frac{95}{128}\,.
\end{align}
However, they have different superconformal indices:
\begin{align}
\begin{split}
    W_1:\quad \mathcal{I}_{\textrm{red}}&=2t^{9/4}+4t^3+4t^{15/4}+\cdots\,,\\
    W_2:\quad \mathcal{I}_{\textrm{red}}&=2t^{9/4}+2t^{5/2}+t^{11/4}+t^3+t^{13/4}+\cdots\,.\\
\end{split}
\end{align}
Therefore these two fixed points are distinct SCFTs but having the same central charges. 

The full list of distinct fixed points that have identical central charges are available at \url{https://qft.kaist.ac.kr/landscape/identicalAC.php}.

\subsection{The Theory with Largest $a/c$} \label{subsec:acRecord}
Consider $SO(5)$ supersymmetric gauge theory with one traceless symmetric tensor $(S)$ and one vector $(q)$ chiral multiplets, with the superpotential \begin{align}
    W=M_1S^2+M_1^2\label{eq:maxratio}
\end{align}
It flows to an SCFT(\href{https://qft.kaist.ac.kr/landscape/detail.php?id=10701}{\#10701}), which 
has the largest ratio of central charges $a/c$ among the entire landscape of SCFTs we explore, with value of approximately 1.21. 
This fixed point has the following central charges, R-charges, and relevant/marginal operator spectrum:
{\renewcommand\arraystretch{1.4}
\begin{table}[H]
    \centering
    \begin{tabular}{|c|c||c|c|c||c|c|}\hline
        $a$ & $c$ &  $R_{M_1}$ & $R_{q_i}$& $R_S$ & Relevant Operators& Marginal Operators\\\hline\hline
        $\frac{507}{256}$ & $\frac{419}{256}$ & $1$  & $\frac{3}{2}$ & $\frac{1}{2}$& $M_1$, $\Tr S^3$ &$\Tr S^4$\\\hline
    \end{tabular}
\end{table}}
 \noindent The previously known largest value of $a/c$ was $183/158 \simeq 1.158$, which is realized by the ISS model \cite{Intriligator:1994rx}, which is an $SU(2)$ theory with a chiral multiplet in the spin-$3/2$ representation. 
 Our $SO(5)$ fixed point (\ref{eq:maxratio}) sets a new record for $a/c$. 
 Let us make a remark that for 4d $\CN=1$ SCFTs, the bound of central charge ratio is given by $\frac{1}{2}\leq \frac{a}{c}\leq \frac{3}{2}$ \cite{Hofman:2008ar}. However, no known example has $a/c$ close to $3/2$. Moreover, notice that our maximal value is still lower than $5/4=1.25$, which is the upper bound for the $\CN=2$ SCFT. It is tempting to conjecture that there is no interacting $\CN=1$ SCFT with $a/c \ge 5/4$. 

The superconformal index of this theory is given as
\begin{align}
    \mathcal{I}_{{\rm red}}=t^3+t^{9/2}+t^6-t^{15/2}(1-\chi_{\half}(y))+\cdots\,.
\end{align}
It passes all the unitarity tests we perform. 
We find one marginal operator $\Tr S^4$. Since there is no flavor symmetry this operator can break, it is also exactly marginal and we have a one-dimensional conformal manifold. This is consistent with the coefficient of the $t^6$ term in the index, which counts the number of marginal operators minus the dimension of the flavor symmetry.

%%%%%%%%%%%%%%%%%%%%%%%%%%%%%%%%%%%%%%%%%%%%
\section{Discussion}
\label{sec:discussion}
  We have discovered a large landscape of SCFTs reached by the RG flows triggered by various deformations of simple $\CN=1$ supersymmetric gauge theories. In this way, we gave a full classification of superconformal fixed points that can be obtained via superpotential deformations of certain small gauge theories. We found these SCFTs in the landscape exhibit curious phenomena like dualities, enhancement of  (super)symmetry, and non-commutativity of the RG flow. Moreover, we performed various statistical analysis of this large set of SCFTs on the distribution of the central charges or the ratio $a/c$ in relation with other quantum numbers.  

  The $a$ versus $c$ plot in Figure \ref{fig:acplots} of the entire SCFTs we have obtained shows that the allowed region of the central charges tends to be smaller than the Hofman-Maldacena bounds \cite{Hofman:2008ar} ${\frac{1}{2} \le \frac{a}{c} \le \frac{3}{2}}$. It is reasonable that the theories that saturate the Hofman-Maldacena bounds are free theories, and interacting CFTs are a bit further from the free theories. Such a phenomenon is proven to be true for $\CN=2$ SCFTs, for the central charge $c$: $c = \frac{1}{24}$ is realized by the free (half)-hypermultiplet, whereas the interacting theory requires $c\ge \frac{11}{30}$ \cite{Liendo:2015ofa}. It would be interesting to continue our program further and give evidence for a more stringent constraint on the allowed bound for existing SCFTs. 

  We have seen a remarkable correlation between the central charge ratio $a/c$ against the dimension of the lightest operator. There is a pattern that when $a/c$ increases the dimension also increases, as we have seen in Figure \ref{fig:combining dimtoratio}. In the AdS/CFT context, it is conjectured that a large-$N$ CFT with a parametrically large gap in the higher-spin operator \cite{Heemskerk:2009pn} is dual to a local gravity in the bulk. The higher-spin gap is also constrained by the difference of the central charges $c-a$ \cite{Camanho:2014apa} , which also controls the correction to  Einstein gravity in the bulk. The correlation between the central charge ratio and the dimension of the lightest operator is in good accordance with this holographic argument. 

  We have also found an interesting correlation between $a/c$ and the number of relevant operators. There is a sense in which the number of relevant operators measures the degrees of freedom of the SCFT. This is because a system with more degrees of freedom can have more ways to deform the system, compared to the ones with fewer degrees of freedom. This notion has been proposed and explored in \cite{Gukov:2015qea, Gukov:2016tnp}. Thus our result suggests the $a/c$ ratio is also potentially related to the degrees of freedom of the SCFT. It would be interesting to clarify this correlation qualitatively, along the lines of \cite{Gukov:2015qea, Gukov:2016tnp}.

  Furthermore, we found an SCFT with the smallest central charge $a$ in the $SU(2)$ $N_f=2$ landscape. This SCFT is identified with the one discovered in \cite{Xie:2021omd} as the deformation of the $(A_1, A_4)$ Argyres-Douglas theory. We provide a ``Lagrangian description" of this minimal theory. It would be interesting to clarify the chiral ring of the minimal theory via our Lagrangian description. 

  We have found  several fixed point with  enhanced $\CN=2$ supersymmetry, including the $H_0=(A_1, A_2)$, $H_1=(A_1, A_3)$, $H_2=(A_1, D_4)$, $(A_1, A_4)$, and $(A_1, A_5)$ theories, which reproduces the results of \cite{Maruyoshi:2016tqk, Maruyoshi:2016aim, Agarwal:2016pjo}. So far we have not been able to produce other non-Lagrangian theories as fixed points of the Lagrangian seed theories by our method. 
  Even though we do not find a previously unknown Lagrangian description for the AD theory itself, we found a new Lagrangian description for the deformed $(A_2, A_3)$ theory.
  It would be interesting to push the investigation up to higher-rank gauge groups and larger number of matters to see if all the known Argyres-Douglas theories appear as fixed points. 
  Also, we found candidate fixed points with enhanced supersymmetry whose central charges are not in the list of the classification done through the Coulomb branch geometry \cite{Argyres:2015ffa, Argyres:2015gha, Argyres:2016xua, Argyres:2016xmc}. As seen in section \ref{subsec:newN2}, the index of one of the candidate theories, which is in the landscape of $SU(2)$ $N_f=2$ adjoint SQCD, satisfies all the necessary and sufficient conditions for $
  \CN=2$ supersymmetry. We are not aware of any definite argument which excludes this fixed point as a sensible fixed point. It is very interesting to study this fixed point and its enhancement of supersymmetry.

One of the lessons we learned through our classification program is that the landscape of 4d $\CN=1$ SCFT is extremely rich with non-trivial coherency. Even though we unveiled a large set of SCFTs, it is obvious that this covers only tiny part of the SCFT landscape. We have only considered a dozen `seed theories,' from which we obtained more than 7,000 SCFTs. We can start with any theories in the list of \cite{Agarwal:2020pol} as a seed theory to further expand this SCFT landscape. Let us make a remark, however,  that our method does not depend on the existence of a  Lagrangian description. It would be interesting to perform our analysis systematically for non-Lagrangian SCFTs as well. 

As we have seen through statistical analysis of the SCFT landscape, we find interesting and non-trivial patterns. This may give hints on further constraining the SCFT theory space, which in turn can be used to improve the superconformal bootstrap. Also, by invoking AdS/CFT, we can hope to convert  universal constraints on the CFT theory space to a Swampland conjecture in AdS quantum gravity.

Finally, let us remark again that our method crucially depends on identification of all the symmetries, especially the ones that can be mixed with R-symmetry. We have seen a few cases where unidentifiable accidental symmetries do seem to occur. In such cases, our matching of anomalies itself can be accidental. So it is possible that some of our candidate fixed points might be invalidated, or that  ones we dropped survive. It is extremely important to develop a toolkit to further test or exclude such cases.

\begin{acknowledgments}
We thank Prarit Agarwal, Jacques Distler, Dongmin Gang, Ken Intriligator, Monica Kang, Seunggyu Kim, Craig Lawrie, Ki-Hong Lee, Kimyeong Lee, Sungjay Lee, Matteo Lotito, Shlomo Razamat, Yuji Tachikawa, Piljin Yi for helpful discussions. 
We also thank the various institutions that have provided us with an environment for collaboration over the years: Caltech, Simons Center for Geometry and Physics, KAIST, Kavli IPMU, KIAS, Seikei University, UCLA, and UCSD. 
The work of KM is supported in part by JSPS Grant-in-Aid for Scientific Research No. 20K03935. The work of EN is supported in part by World Premier International Research Center Initiative (WPI), MEXT, Japan. The work of MC and JS is supported in part by National Research Foundation of Korea Grants No. RS-2023-00208602 and RS-2024-00405629. The work of JS is also supported by POSCO Science Fellowship of POSCO TJ Park Foundation. 

\end{acknowledgments}

\appendix

\section{Operators and Deformations of $\CN=1$ SCFTs}%%%%%%%
\label{app:ops}

In this appendix, we review some aspects of the representation theory of the 4d $\CN=1$ and $\CN=2$ superconformal algebras, which are pertinent to the classes of theories we consider. 
We label representations of the $\CN=1$ and $\CN=2$ superconformal algebras by their quantum numbers as follows,\footnote{~We follow conventions from \cite{Cordova:2016emh}, except that $j_1,j_2$ are half-integers.}
\ba{
\CN=1:\quad [j_1,j_2]_{\Delta}^{(R)}\,,\qquad\quad\CN=2:\quad [j_1,j_2]_{\Delta}^{(2I_3,r)}\,.
} 
Here $j_1$ and $j_2$ are the half-integer-valued spins for the Lorentz group; $\Delta$ is the scaling dimension; and the superscripts are labels for the R-symmetry---$R$ denotes the $\CN=1$ $U(1)_R$ charge, while $(2I_3,r)$ label the charges of the $\CN=2$ $SU(2)_R\times U(1)_r$ symmetry, with $2I_3$ the integer-valued Dynkin label for $SU(2)_R$. The $\CN=1$ R-symmetry algebra can be embedded in the $\CN=2$ algebra with the following identification of charges,
\ba{
R = \frac{1}{3} (r+ 4I_3)\,,\qquad J = \frac{1}{2} (r - 2I_3)\,,
\label{RJ}
}
where $J$ denotes the $\CN=1$ flavor (non-$R$) symmetry in the embedding. 

\paragraph{Deformations}%%%%%%%%%%%

This work pertains to the possible relevant superpotential deformations of $\CN=1$ SCFTs. These arise from $\CN=1$ chiral multiplets with scalar chiral primaries $\CO$,
\ba{
\label{n1chiral}
\CO = L\overline{B}_1[0;0]^{(R)}_{\Delta}\,,\qquad \Delta(\CO) = \frac{3}{2} R(\CO)\,.
}
(When we say $``\CO = \dots"$ we mean that $\CO$ is the primary of the multiplet, so that the quantum numbers labeling the multiplet coincide with those of $\CO$.)  This labeling follows conventions from \cite{Cordova:2016emh} and is interpreted as follows: $L$ (for {\it long}) indicates that there is no shortening condition with respect to the $Q$ supercharges, while $\overline{B}_1$ indicates that the  multiplet satisfies a shortening condition with respect to the $\overline{Q}$ supercharges (in this case, $\bar{Q} \CO=0$).
%In Dolan-Osborn naming conventions, $L\overline{B}_1[0;0]^{(R)}_{\Delta}=\CB_R(0,0)$. 
Then, the possible relevant supersymmetry-preserving F-term deformations of 4d $\CN=1$ SCFTs are of the form, 
\ba{
\label{deform}
\delta \CL = Q^2 \CO\,,\qquad \CO = L\overline{B}_1[0;0]^{(R)}_{\Delta}\,,\qquad \Delta(\CO) < 3\,\  \leftrightarrow\, \  R(\CO) < 2
}
(or similarly, $\delta \CL = \overline{Q}^2 \CO$ for $\CO$ an anti-chiral multiplet). 

The conformal manifold is parameterized by marginal deformations \eqref{deform} with $\Delta(\CO)=3$. In an $\CN=1$ SCFT, marginal deformations are exactly marginal if and only if they do not break any flavor symmetries; while the operators \eqref{deform} with $R<2$ are absolutely protected, the chiral multiplet with $\Delta=3$ can pair up with a flavor current multiplet to form a long multiplet.\footnote{~On the other hand, since the $\CN=2$ flavor current multiplet and Coulomb branch multiplet are absolutely protected, all marginal $\CN=2$ preserving deformations must be exactly marginal.}

\paragraph{$\CN=1$ Multiplets}

Besides \eqref{n1chiral}, some other important $\CN=1$ multiplets are as follows:

\begin{itemize}

\item The multiplets $A_\ell \overline{B}_1[j;0]^{( \frac{2}{3}(j+1))}_{j+1}$ with $\ell=2$ if $j=0$ and $\ell = 1$ if $j>0$ contain free spin-$j$ fields. In particular, the multiplet whose primary is a free chiral scalar with $\Delta  = 1$ belongs to  $A_2\overline{B}_1[0,0]^{(\frac{2}{3})}_1$, while the case $j=\frac{1}{2},\ell=1$ with $\Delta = \frac{3}{2}$ is a free vector multiplet.

\item The flavor current multiplet is $A_2\overline{A}_2[0,0]^{(0)}_2$, whose top component of dimension 3 is the conserved flavor current $j_\mu$. 

\item The stress tensor multiplet is $A_1\overline{A}_1[\frac{1}{2},\frac{1}{2}]_3^{(0)}$, whose top component of dimension $\Delta =4$ is the conserved traceless stress tensor $T_{\mu\nu}$. This multiplet also contains the conserved traceless supersymmetry current of dimension $\Delta = 7/2$, and the R-symmetry current $j_\mu^R$ of dimension 3.  This multiplet is unique and always exists in any SCFT.

\item The multiplet $A_1\overline{A}_2[\frac{1}{2},0]_{\frac{5}{2}}^{(\frac{1}{3})}$ is an extra supersymmetry current multiplet, which is necessary if the $\CN=1$ theory has enhanced $\CN=2$ supersymmetry.

\end{itemize}

\paragraph{$\CN=2$ Multiplets}

As the landscape of fixed points we consider also includes $\CN=2$ SCFTs, let us  briefly also review some important $\CN=2$ multiplets. 

\begin{itemize}

\item The $\CN=2$ multiplet whose primary is a Coulomb branch operator $\CO_{\text{CB}}$ is,
\ba{
\CO_{\text{CB}} = L\overline{B}_1[0,0]^{(0;r)}_{\frac{r}{2}}\,,\qquad \Delta(\CO_{\text{CB}}) = \frac{r}{2}\,,\qquad r>2\,.
}
%$(=\overline{\CE}_{r(0,0)})$. 
The level-2 descendent of this multiplet with quantum numbers $\CO'_{\text{CB}}=[0,0]_{\frac{r+2}{2}}^{(2,r-2)}$ is itself a primary of an $\CN=1$ chiral multiplet. In particular, the Coulomb branch multiplet decomposes into $\CN=1$ multiplets as, 
\ba{
 L\overline{B}_1[0,0]^{(0;r)}_{\frac{r}{2}}= L\overline{B}_1[0,0]^{(\frac{r}{3})}_{\frac{r}{2}}\oplus  L\overline{B}_1[0,0]^{(\frac{r+2}{3})}_{\frac{r+2}{2}}\oplus  L\overline{B}_1[\frac{1}{2},0]^{(\frac{r+1}{3})}_{\frac{r+1}{2}}
}

When $r=2$, the CB operator just lives in the vector multiplet $A_2\overline{B}_1[0,0]_1^{(0,2)}$, whose primary is the free vector with charge $R=2/3$. 

\item The multiplet $A_2\overline{B}_1[0,0]_2^{(1,2)}$ is an extra supersymmetry current multiplet.

\item The $\CN=2$ multiplet $B_{1}\overline{B}_1[0,0]^{(1+i;0)}_{1+i}$ for $i=0,1$ contains Higgs branch operators. For $i=0$ this is a free half-hypermultiplet $(=\hat{\CB}_1)$, whose primary has charge $R=2/3$. For $i=1$ this is a flavor current multiplet, whose primary is the moment map operator $\mu$ in the triplet of $SU(2)_R$,
\ba{
\mu = B_1\overline{B}_1[0,0]_2^{(2;0)}
}
and whose top component is the conserved flavor current $j_\mu$.

\item The stress tensor multiplet $A_2\overline{A}_2[0,0]_2^{(0,0)}$, whose top component is the conserved traceless stress tensor $T_{\mu\nu}$, and which also contains the supersymmetry currents $S_{\mu\alpha}$ of dimension $7/2$, and $SU(2)_R\times U(1)_r$ R-symmetry currents.  

\end{itemize}

\section{Implementing the Landscape Database} \label{app:database}

When implementing the landscape, two major calculations are required. The first is determining the R-charges through $a$-maximization, and the second is calculating the superconformal index to perform a consistency check and to read the operator spectrum.

The $a$-maximization procedure, which is to find a local maximum of a cubic equation in terms of R-charges, is tantamount to simply solving (coupled) quadratic equations. \textit{Mathematica} is an optimal tool for this, and the calculations are quick and easy to complete. The bottleneck lies in calculating the superconformal indices.

Consider a gauge theory with a gauge group $G$ of rank $N$ withchiral multiplets $\Phi$ in representations $\textbf{R}_{\Phi}$. Then, the superconformal index (\ref{eq:Idx}) can be calculated by the residue integral
\begin{align}
    I=\oint \prod_{i=1}^N \frac{dz_i}{2\pi i z_i}\prod_{\a\in\a^+} (1-z^\a) I_V(t,y;z_i)\prod_{\Phi}I_{\Phi}(t,y,;x,z_i)\,,\label{eq:integral}
\end{align}
where $z_i$ are the gauge holonomy variables, $\a=(\a_1,\dots\a_N)$ represents a vector in the set $\a^+$ of positive roots of $G$, and $z^\a$ denotes $z^\a=\prod_i z_i^{\a_i}$. The contribution of the vector multiplet and chiral multiplets, $I_V$ and $I_{\Phi}$ are given by
\begin{align}
    I_V(t,y;z_i)=\PE\left[\frac{-t^3y-t^3/y+2t^6}{(1-t^3y)(1-t^3/y)}\chi_{adj}(z_i)\right]\,,
\end{align}
and
\begin{align}
    I_{\Phi}(t,y;x,z_i)=\PE\left[\frac{t^{3R_i}x^f\chi_{\textbf{R}_{\Phi}}(z_i)-t^{6-3R_i}x^{-f}\chi_{\bar{\textbf{R}}_{\Phi}}(z_i)}{(1-t^3y)(1-t^3/y)}\right]\,,\label{eq:chiral}
\end{align}
respectively. Here, $\PE$ denotes the plethystic exponential defined by
\begin{align}
    \PE\left[f(x_i)\right]=\exp\left(\sum_{n=1}^\infty \frac{1}{n}f(x_i^n)\right)\,,\label{eq:PE}
\end{align}

It is important for us to calculate the index at least up to order $t^8$ to ensure consistency regarding spin-1 operators. With the integral representation for the index, \textit{Mathematica} may not be the ideal tool to implement our Landscape database. This is because \textit{Mathematica} is not efficient enough for computing the series expansion of arbitrarily long expressions. In our Landscape, a fixed point may have a lot of matter fields. For example, one of the IR fixed points of $SU(2)$ $N_f=2$ adjoint SQCD (\href{https://qft.kaist.ac.kr/landscape/detail.php?id=5704}{\#5704}) has a total of 16 chiral multiplets. Such a large number of matter fields makes the numerator in (\ref{eq:chiral}) very long, and their plethystic exponential in a series expansion becomes too large to deal with.

Additionally, a fixed point may have chiral multiplets with very small R-charges. For instance, at the fixed point (\href{https://qft.kaist.ac.kr/landscape/detail.php?id=45786}{\#45786}) of $SU(3)$ $N_f=1$ adjoint SQCD, the R-charge of the fundamental chiral multiplet is $1/12$. Therefore, in this case, the parameter $n$ in the plethystic exponential (\ref{eq:PE}) must run at least up to $n=28$. 
Note that this example is one of the consistent fixed points, but there are candidate fixed points having chiral multiplets with much smaller R-charges that give inconsistent indices. The small value of R-charge also makes the evaluation of the plethystic exponentials very computationally challenging. 

To calculate the index more efficiently, we utilize \textit{LiE} \cite{LiE1992}. It is a computer algebra system specialized in computations regarding Lie groups and their representations. When calculating the index, we find it particularly useful for determining the number of singlets in the products of representations.

For example, consider the previous example (\href{https://qft.kaist.ac.kr/landscape/detail.php?id=45786}{\#45786}). When expanding the plethystic exponential, we find the following expression, which involves products of the character $\ch_\textbf{3}$ of the fundamental representation with fugacities of different powers:
\begin{align}
    \ch_{\textbf{3}}(z_1,z_2)^7\ch_{\textbf{3}}(z_1^3,z_2^3)\ch_{\textbf{3}}(z_1^4,z_2^4)^2\ch_{\textbf{3}}(z_1^6,z_2^6)\,.\label{eq:chars}
\end{align}
Using \textit{LiE}, we do not need to multiply the Haar measure with (\ref{eq:chars}) or expand them to obtain gauge-singlets. Instead, we can directly obtain the number of singlets from the expression (\ref{eq:chars}) using the functions \textit{tensor}, \textit{p\_tensor}, and \textit{Adams} in \textit{LiE}. The function \textit{tensor} gives the decomposition of a tensor product of two representations into a sum of the highest-weight representations. The function \textit{p\_tensor} provides the decomposition of the $n$-th tensor power of a representation. The function \textit{Adams} returns the decomposition of the $n$-th Adams operation, which maps to $\ch_{\textbf{R}}(z_i)\mapsto \ch_{\textbf{R}}(z_i^n)$, of a representation $\textbf{R}$. Thus, we do not expand the products of characters and instead keep their form symbolic. This greatly reduces the computational time since we do not have to directly perform contour integration. 

Since we are calculating indices for thousands of fixed points, repetitive calculations of products of characters occur frequently. Therefore, we save the results of \textit{LiE} calculations and read them as needed to avoid redundant computations. This helps to reduce computational time for calculating indices.

In addition, we use \textit{FORM} \cite{Kuipers:2012rf}, a powerful software program for manipulating arbitrarily long mathematical expressions.
To check consistency for a given fixed point, we often expand the plethystic exponential to high orders in $t$. Additionally, to examine the operator spectrum via the superconformal index, we introduce fictitious fugacities to track how each operator contributes to the index. 
For example, we write the index contribution from chiral multiplets with fictitious fugacities $\phi$ and $\psi$ as:
\begin{align}
    I'_{\Phi}(t,y;x,z_i,q)=\PE\left[ \frac{ \phi\,t^{3R_i}x^f\chi_{\textbf{R}_{\Phi}}(z_i)- \psi\,t^{6-3R_i}x^{-f}\chi_{\bar{\textbf{R}}_{\Phi}}(z_i)}{(1-t^3y)(1-t^3/y)}\right]\,
\end{align}
The fugacity $\phi$ indicates the contribution coming from the scalar in the chiral multiplet, while $\psi$ denotes the contribution coming from its superpartner. This allows us to extract the candidates for gauge-invariant operators that contribute to the index.
\textit{Form} demonstrates excellent performance in calculating higher-order indices having many fugacities. 

One drawback of \textit{FORM} is that the exponents must be integers, while the R-charges are usually irrational numbers. As a result, it is often hard to keep the exact values of these irrational numbers, so we need to round them with sufficient decimal precision. Then, we need to appropriately redefine the fugacity $t$ so that the exponents become integers. For practical reasons, we rescale the exponents by a factor of 500, take the integer part, then multiply the remaining decimal places by 5000 and round to the nearest integer. Store these two integers in the new fugacities $t$ and $s$:
\begin{align}
    t^a\goto t^{\lfloor500a\rfloor}s^{ \lfloor 5000(500a-\lfloor500a\rfloor)+0.5\rfloor}
\end{align}

Here is a simple example, which is part of the code used to calculate the index of (\href{https://qft.kaist.ac.kr/landscape/detail.php?id=2}{\#2}). Note that $g_i$'s denote the fugacities for $U(1)$ flavor charges.
{\footnotesize
\begin{verbatim}
  * The maximum term size should be sufficiently large.
  #: maxtermsize 600000
  Off statistics;

  * Declaration symbols.
  * The last 9 symbols correspond to the fugacities in the index.
  S m, n, i, j, z, s, t(: 4500), y, X1, g1, g2, M1, q1, qb1, phi1;

  * q, qb, phi indicates the characters of each representation.
  CF c, d, q, qb, phi;

  * Declaration functions, which correspond to numerators of single-letter indices.
  Function K1, Kb1, K2, Kb2, K3, Kb3, K4, Kb4;

  * c and d correspond to the descendants.
  Polyratfun c, d;
  L G = c(1,(1-m));
  .sort
  
  Polyratfun c(expand,m,3);
  L H = d(1,(1-n));
  .sort
  
  Polyratfun d(expand,n,3);
  id c(m?)=m;
  id d(n?)=n;
  .sort
  
  L J = G*H;
  id m=t^1500*y;
  id n=t^1500/y;
  .sort

  * Summation of a single-letter index of the fundamental chiral multiplet.
  L I1 = sum_(j,1,6,(K1(t^j,s^j,y^j,q1^j,g1^j,g2^j)*q(j)
                     +Kb1(t^j,s^j,y^j,q1^j,g1^j,g2^j)*qb(j))/j);
  id K1(t?,s?,y?,q1?,g1?,g2?)=J*(+1*g1^-1*g2^4*q1*t^833*s^1667);
  .sort
  id Kb1(t?,s?,y?,q1?,g1?,g2?)=J*(-1*g1^1*g2^-4*q1^(-1)*t^2166*s^3333);
  .sort

  * Exponentiate
  L P1 = z;
  id z= z*I1;
  #do i=2, 6
  id z= 1+z*I1/`i';
  .sort:step `i';
  #enddo
  .sort

  * Other chiral multiplets are omitted.
  ...

  * Combining contributions of all representations
  L result1 = 1;
  .sort
  L result2 = result1*(1+P1);
  .sort
  L result3 = result2*(1+P2);
  .sort

  ...

  Print result5;
  .end
\end{verbatim}
}

We calculate the R-charges via $a$-maximization using \textit{Mathematica}. Then, keeping the characters symbolic, we use \textit{FORM} to perform the series expansion of $I_V$ and $I_{\Phi_i}$. Finally, we use \textit{LiE} to extract gauge-singlets from the products of characters and obtain the index. In analyzing the resulting index, such as performing consistency checks or reading the operator spectrum, \textit{Mathematica} is again useful. Finally, the results of the calculations are directly sent to and stored in a \textit{MySQL} database.
All these processes can be organically controlled through \textit{Python}. 

By using the \textit{multiprocessing} module in \textit{Python}, we can parallelize these processes for multiple fixed points. Note that if there are equivalent fixed points, we keep the one with the shorter superpotential as the representative. Therefore, it is preferable to enumerate the landscape in order of increasing superpotential length to avoid redundant calculations. For parallel computations, tasks are divided into `levels' based on superpotential length and processed sequentially.

%%%%%%%%%%%%%%%%%%%%%%%%%%%%%%%%%%%%%%%%%%
\bibliographystyle{jhep}
\bibliography{MNSbib}

\end{document}